\documentclass{aa}  

\usepackage{graphicx}
\usepackage[percent]{overpic}
\usepackage{lipsum}
\usepackage{multirow}
\usepackage{color}
\usepackage{rotating}
\usepackage{mwe,tikz}
\usepackage[percent]{overpic}
\usepackage{mathtools}
\usepackage{physics}
\usepackage{amsmath}
\usepackage{longtable}
\usepackage{wasysym}
\usepackage[normalem]{ulem}
\usepackage{CJK}
\usepackage{soul}
\usepackage{pdflscape}
\usepackage{longtable}
\usepackage{booktabs,tabularx}
\usepackage{graphicx}	
\usepackage{amsmath}	
\usepackage{amssymb}	
\usepackage{latexsym}
\usepackage{afterpage}

\usepackage{txfonts}
\newcommand{\lta}{$\; \buildrel < \over \sim \;$}
\newcommand{\simlt}{\lower.5ex\hbox{\ltsima}}
\newcommand{\gta}{$\; \buildrel > \over \sim \;$}
\newcommand{\simgt}{\lower.5ex\hbox{\gtsima}}

\def\FeH{{\rm[Fe/H]}}

\def\kms{{\rm\,km\,s^{-1}}}

\def\deg{^\circ}

\def\s{\ifmmode \widetilde \else \~\fi}
\def\={\overline}

\def\spose#1{\hbox to 0pt{#1\hss}}

\def\eg{{e.g.,\ }}
\def\ie{{i.e.,\ }}
\def\lta{\mathrel{\spose{\lower 3pt\hbox{$\mathchar"218$}}
     \raise 2.0pt\hbox{$\mathchar"13C$}}}
\def\gta{\mathrel{\spose{\lower 3pt\hbox{$\mathchar"218$}}
     \raise 2.0pt\hbox{$\mathchar"13E$}}}
\def\Dt{\spose{\raise 1.5ex\hbox{\hskip3pt$\mathchar"201$}}}    
\def\dt{\spose{\raise 1.0ex\hbox{\hskip2pt$\mathchar"201$}}}    

\def\dotsfill{\leaders\hbox to 1em{\hss.\hss}\hfill}

\def\FeH{{\rm[Fe/H]}}

\newcommand{\logg}{\ensuremath{\log g}}
\newcommand{\teff}{\ensuremath{T_{\mathrm{eff}}}\xspace}

\newcommand{\Gaia}{{\emph{Gaia}}}

\def\FeH{{\rm[Fe/H]}}

\newcommand*\samethanks[1][\value{footnote}]{\footnotemark[#1]}
\newcommand{\gkai}[1]{\begin{CJK*}{UTF8}{gkai}\raisebox{.1em}{(}#1\raisebox{.1em}{)}\end{CJK*}}

\begin{document}

   \title{The Pristine survey -- XXIII.}

   \subtitle{Data Release 1 and an all-sky metallicity catalogue based on \Gaia\ DR3 BP/RP spectro-photometry\thanks{Catalogues described in Tables~\ref{table:cahk}, \ref{table:FeHphot_CaHKsyn}, and \ref{table:FeHphot_Pr} are only available in electronic form at the CDS via anonymous ftp to cdsarc.u-strasbg.fr (130.79.128.5) or via \url{http://cdsweb.u-strasbg.fr/cgi-bin/qcat?J/A+A/}.}}

   \author{Nicolas F. Martin\inst{1,2}\fnmsep\thanks{Both authors contributed equally.\\ \email{nicolas.martin@astro.unistra.fr;\\estarkenburg@astro.rug.nl}}
	\and Else Starkenburg\inst{3}\fnmsep\samethanks[2]
	\and Zhen Yuan \gkai{袁珍}\inst{1}
	\and Morgan Fouesneau\inst{2} 
	\and Anke Ardern-Arentsen\inst{4}
	\and Francesca De Angeli\inst{4}
	\and Felipe Gran\inst{5}
	\and Martin Montelius\inst{3}
	\and Samuel Rusterucci\inst{1,3}
	\and Ren\'e Andrae\inst{2}
	\and Michele Bellazzini\inst{6}
	\and Paolo Montegriffo\inst{6}
	\and Anna F. Esselink\inst{3}
	\and Hanyuan Zhang\inst{4}
	\and Kim A. Venn\inst{7}
	\and Akshara Viswanathan\inst{3}	
	\and David S. Aguado\inst{8,9}
	\and Giuseppina Battaglia\inst{8,9}
	\and Manuel Bayer\inst{3}
	\and Piercarlo Bonifacio\inst{10}
	\and Elisabetta Caffau\inst{10}
	\and Patrick C\^ot\'e\inst{11}
	\and Raymond Carlberg\inst{12}
	\and S\'ebastien Fabbro\inst{11} 
	\and Emma Fern\'andez-Alvar\inst{8,9}
	\and Jonay I. Gonz\'alez Hern\'andez\inst{8,9}
	\and Isaure Gonz\'alez Rivera de La Vernhe\inst{5}
	\and Vanessa Hill\inst{5}
	\and Rodrigo A. Ibata\inst{1} 
	\and Pascale Jablonka\inst{13,10}
	\and Georges Kordopatis\inst{5}
	\and Carmela Lardo\inst{14}
	\and Alan W. McConnachie\inst{11}
	\and Camila Navarrete\inst{5}
	\and Julio Navarro\inst{7}
	\and Alejandra Recio-Blanco\inst{5}
	\and Rub\'en S\'anchez Janssen\inst{15}
	\and Federico Sestito\inst{7}
	\and Guillaume F. Thomas\inst{8,9}
	\and Sara Vitali\inst{16}
	\and Kristopher Youakim\inst{17}
          }

   \institute{Universit\'e de Strasbourg, CNRS, Observatoire astronomique de Strasbourg, UMR 7550, F-67000 Strasbourg, France
         \and
         Max-Planck-Institut f\"{u}r Astronomie, K\"{o}nigstuhl 17, D-69117 Heidelberg, Germany
         \and
         Kapteyn Astronomical Institute, University of Groningen, Landleven 12, 9747 AD Groningen, The Netherlands
         \and
         Institute of Astronomy, University of Cambridge, Madingley Road, Cambridge CB3 0HA, UK
         \and
         Universit\'e C\^ote d'Azur, Observatoire de la C\^ote d'Azur, CNRS, Laboratoire Lagrange, Nice, France
	\and
	INAF - Osservatorio di Astrofisica e Scienza dello Spazio di Bologna, via Piero Gobetti 93/3, 40129 Bologna, Italy
	\and
	Dept. of Physics and Astronomy, University of Victoria, P.O. Box 3055, STN CSC, Victoria BC V8W 3P6, Canada 
	\and
	Instituto de Astrof{\'\i}sica de Canarias, E-38205 La Laguna, Tenerife, Spain
	\and
	Universidad de La Laguna, Dept. Astrof{\'\i}sica, E-38206 La Laguna, Tenerife, Spain
	\and
	GEPI, Observatoire de Paris, Universit\'e PSL, CNRS, 5 Place Jules Janssen, 92195, Meudon, France
	\and
	NRC Herzberg Astronomy and Astrophysics, 5071 West Saanich Road, Victoria, BC V9E 2E7, Canada
	\and
	Department of Astronomy \& Astrophysics, University of Toronto, Toronto, ON M5S 3H4, Canada
	\and
	Institute of Physics, Laboratory of Astrophysics, Ecole Polytechnique F\'ed\'erale de Lausanne (EPFL), Observatoire de Sauverny, 1290 Versoix, Switzerland
	\and
	Dipartimento di Fisica e Astronomia, Universit\`a degli Studi di Bologna, Via Gobetti 93/2, I-40129 Bologna, Italy
	\and
	UK Astronomy Technology Centre, Royal Observatory, Blackford Hill, Edinburgh, EH9 3HJ, UK
	\and
	N\'ucleo de Astronom\'ia, Facultad de Ingenier\'ia y Ciencias Universidad Diego Portales, Ej\'ercito 441, Santiago, Chile
	\and
	Department of Astronomy, Stockholm University, AlbaNova University Centre, SE-106 91 Stockholm, Sweden
             }

\date{Received XXX; accepted XXX}

 
  \abstract{We used the spectro-photometric information of $\sim219$ million stars from \Gaia's Data Release 3 (DR3) to calculate synthetic, narrow-band, metallicity-sensitive $CaHK$ magnitudes that mimic the observations of the Pristine survey, a survey of photometric metallicities of Milky Way stars that has been mapping more than 6,500\,deg$^2$ of the northern sky with the Canada-France-Hawaii Telescope since 2015. These synthetic magnitudes were used for an absolute recalibration of the deeper Pristine photometry and, combined with broadband \Gaia\ information, synthetic and Pristine $CaHK$ magnitudes were used to estimate photometric metallicities over the whole sky. The resulting metallicity catalogue is accurate down to $\FeH\sim-3.5$ and is particularly suited for the exploration of the metal-poor Milky Way ($\FeH<-1.0$). We make available here the catalogue of synthetic $CaHK_\mathrm{syn}$ magnitudes for all stars with BP/RP information in \Gaia\ DR3, as well as an associated catalogue of more than $\sim30$~million photometric metallicities for high signal-to-noise FGK stars. This paper further provides the first public data release of the Pristine catalogue in the form of higher quality recalibrated Pristine $CaHK$ magnitudes and photometric metallicities for all stars in common with the BP/RP spectro-photometric information in \Gaia\ DR3. We demonstrate that, when available, the much deeper Pristine data greatly enhance the quality of the derived metallicities, in particular at the faint end of the catalogue ($G_\mathrm{BP} \gta 16$). Combined, both photometric metallicity catalogues include more than two million metal-poor star candidates ($\FeH_\mathrm{phot}<-1.0$) as well as more than 200,000 and $\sim$8,000 very and extremely metal-poor candidates ($\FeH_\mathrm{phot}<-2.0$ and $<-3.0$, respectively). Finally, we show that these metallicity catalogues can be used efficiently, among other applications, for Galactic archaeology, to hunt for the most metal-poor stars, and to study how the structure of the Milky Way varies with metallicity, from the flat distribution of disk stars to the spheroid-shaped metal-poor halo.}


   \maketitle
   
\section{Introduction}

In any given closed environment, the lowest metallicity stars are also the oldest ones. As such, they are often thought of as the ultimate targets that allow us to look back in time to the infancy of our Galaxy through the methods of Galactic archaeology \citep[see, \eg][for detailed reviews]{tinsley80,beers05,belokurov13,frebel15,helmi20}. It is likely that most, if not all, of the first (metal-free) stars are inaccessible to us, because they are very massive and, therefore, short-lived \citep[for a recent review, see][and references therein]{klessen23}. As a consequence, the most metal-poor stars we can still observe today not only contain information on the earliest buildup of the proto-galaxy that would later become the Milky Way (MW), but they also hold invaluable clues about the properties of the first stars and their mass function. They give us unique information on times long gone and complement studies that focus on the high-redshift Universe through observations. For example, recent results from the \emph{James Webb} Space Telescope include observations of damped Lyman-alpha systems \citep[\eg][]{welsh23}, high-redshift, magnified stars \citep{welch22a,welch22b}, or the possible direct observation of the impact of the first stars on the photoionization of gas \citep{maiolino24}.

However, these important tracers of the early Galaxy are very rare among the more metal-rich and younger populations of the MW. In the Solar neighborhood, looking out of the Galactic plane through the dense foreground of predominantly young and metal-rich disk stars, it has been estimated that stars with less than one-thousandth of the solar metallicity represent, at most, one star in a thousand. Below this metallicity value, the metallicity distribution function (MDF) may decline even more steeply than at higher metallicities \citep[\eg][]{youakim20, bonifacio21,yong21}, leading to a very challenging search for ``extremely metal-poor'' stars (EMP stars; $\FeH<-3.0$) or ``ultra-metal-poor'' stars (UMP stars, following the terminology from \citealt{beers05}; $\FeH<-4.0$). At the moment, databases collecting detailed chemical abundance studies of such stars \citep[see, \eg the SAGA database,][and online updates\footnote{\url{http://sagadatabase.jp}}]{suda08} contain a few hundred EMP stars and significantly fewer UMP stars, with only $\sim40$ such stars currently known \citep[\eg][]{sestito19}.

Historically, one of the most successful avenues to isolate the low-metallicity end of the MDF, starting with the ``very metal-poor'' regime (VMP stars; $\FeH<-2.0$), has been to employ objective-prism spectroscopy in the region of the metallicity-sensitive Ca H\& K lines \citep{bond70,bidelman73,bessel77}. The two most recent such endeavors, the HK survey \citep{beers92} and the Hamburg-ESO survey \citep{christlieb08}, led to the buildup of the first significant samples of EMP stars via the spectroscopic follow-up of apparently metal-deficient stars identified in the surveys. In parallel, they helped discover the first confirmed UMP stars \citep{christlieb02,frebel05}. Large, low-resolution spectroscopic surveys such as the Sloan Digital Sky Survey \citep[SDSS; ][]{york20} and the Large Sky Area Multi-Object Fibre Spectroscopic Telescope \citep[LAMOST; ][]{zhao06} have also proven to be a major source of stars at the low-metallicity end of the MDF. Although the survey strategy was not designed to specifically target EMP and UMP candidates, samples of millions of stellar spectra naturally included a sizable number of such stars. These were discovered through the application of specifically developed pipelines and analyses (\eg the Turn-Off Primordial Stars survey --- TOPoS --- \citealt{caffau13}; and see also \citealt{aguado16} and \citealt{li18b}).

The community has now mainly shifted toward using narrow-band photometric surveys to identify these exceptional stars, thanks to both the preponderance of wide-field CCD imagers and advances in filter technology that allow for the construction of well-behaved medium- and narrow-band filters (\ie near top-hat and providing stable photometry over a wide focal plane). Two such surveys have been driving the field over the last decade: the SkyMapper panoptic survey \citep{wolf18,onken19,chiti21}, in the south, that includes a Str\"omgren $v$ filter that covers the metallicity-sensitive Ca H \& K lines in addition to the more classical $ugriz$ broadband filters; and the Pristine survey \citep{starkenburg17b}, in the north, that relies on a specifically tailored narrow-band filter centered on the Ca H \& K lines and is mounted on the MegaCam imager at the Canada-France-Hawaii Telescope (CFHT). The combination of the metallicity-sensitive medium- or narrow-band photometry with broadband photometry has led to large samples of VMP and EMP candidates. Dedicated spectroscopic follow-up surveys have confirmed that both surveys produce samples of VMP stars with a high purity and that they are efficient at isolating true EMP stars, with success rates of $\sim20$\% when quality flags are carefully considered  \citep{youakim17,aguado19,dacosta19, marino19}. Both surveys have also led to the discovery of a handful of new UMP stars \citep{starkenburg18,kielty21,nordlander19,lardo21}, with the most iron-deficient star known to date being among them \citep{keller14}. Very recently, additional new surveys that follow similar strategies, such as J-PLUS \citep{cenarro19}, S-PLUS \citep{almeida-fernandes22}, and SAGES \citep{zhou23}, have also started to produce results \citep{galarza22,placco22,yang22,yang23}.

The data provided by the recent \Gaia\ Data Release 3 \citep[DR3;][]{vallenari23} promise to bridge the two techniques --- very low-resolution spectroscopy and narrow-band photometry --- with the release of the spectro-photometric observations conducted with the Blue Prism (BP) and the Red Prism (RP) on board the spacecraft \citep{carrasco21,deangeli23,montegriffo23b}. Specifically, the BP prism includes the region of the Ca H \& K lines so the \Gaia\ spectro-photometry is expected to perform well at constraining the metallicity of a star, even in the EMP regime \citep{witten22,xylakis-dornbusch22}. For more generic MW stars, the BP/RP information is already proving invaluable to derive accurate stellar parameters, including $\FeH$ \citep{andrae22,andrae23,bellazzini23,zhang23,xylakis-dornbusch24}, despite being currently limited to the fairly bright sky in DR3 ($G\lta17.65$).

The release of the DR3 BP/RP information is particularly exciting in the context of the Pristine survey. As shown by \citet{montegriffo23}, the BP/RP information --- distributed by the consortium as coefficients from the projection of the observed low resolution spectra on a set of basis functions --- can be used to reproduce synthetic photometry for any filter that overlaps the large wavelength coverage of the observations (330--1,050\,nm). In particular, they show that synthetic Pristine $CaHK_\mathrm{syn}$ magnitudes compare favorably with the actual Pristine observations (see their Figure~32). This brings a unique opportunity: (1) to supersede the previous, relative calibration of the Pristine survey and calibrate all of the narrow-band $CaHK$ photometry on an absolute scale provided by the $CaHK_\mathrm{syn}$ magnitudes\footnote{A similar effort was also successfully undertaken by the J-PLUS collaboration to calibrate their multi-narrowband photometry, including photometry from their own CaHK filter \citep{lopez-sanjuan24}.}, and (2) to build a catalogue of \Gaia-based synthetic $CaHK_\mathrm{syn}$ Pristine-like photometry over the whole sky. The latter can be pushed through an updated version of the Pristine photometric metallicity model to yield photometric metallicities that are accurate down to the EMP regime.

This paper describes both these efforts. In addition to information based on the shallower $CaHK_\mathrm{syn}$ magnitudes, we assemble and distribute the first public Data Release (DR1) of the Pristine survey that now includes the newly recalibrated photometry and metallicities from the updated photometric metallicity model. This data release includes all sources observed in the Pristine survey by the end of March 2024 that have BP/RP information in \Gaia\ DR3.

The paper is organized as follows. Section~\ref{sec:CaHKsyn} presents the catalogue of \Gaia-based synthetic $CaHK_\mathrm{syn}$ photometry based on the BP/RP information. Section~\ref{sec:Pristine} provides a detailed update to the data reduction of the now $\sim$11,500\,images of the Pristine survey and, in particular, explains the calibration of the survey photometry onto the absolute scale provided by the \Gaia\ synthetic magnitudes. In Section~\ref{sec:extinction}, we describe how we handle the complex issue of correcting observed magnitudes from extinction, while Section~\ref{sec:variability} presents a probabilistic model to isolate likely variable stars that can have erroneous photometric metallicities. Section~\ref{sec:model} describes the updated Pristine $(CaHK,G,G_\mathrm{BP},G_\mathrm{RP})\rightarrow\FeH_\mathrm{phot}$ model that is now based on \Gaia\ information alone to complement the Pristine photometry. The catalogues of photometric metallicities generated from pushing the \Gaia-based $CaHK_\mathrm{syn}$ and the Pristine $CaHK$ magnitudes through the model are detailed in Section~\ref{sec:met_catalogues}, in which we also compare the resulting data sets with spectroscopic catalogues of metal-poor stars and other metallicity catalogues based on \Gaia\ DR3. In the same section, we also provide advice on how to best use the photometric metallicity catalogues. In Section~\ref{sec:tests}, we show two possible applications of the catalogues, one to study Galactic globular clusters and another to map the MW as a function of metallicity. Finally, Section~\ref{sec:summary} summarizes the contents of the paper.

Accompanying this paper, we provide a number of large catalogues, which will be accessible through the CDS. 

The $CaHK_\mathrm{syn}$ photometric catalogue the contains the synthetic, Pristine-like $CaHK_\mathrm{syn}$ magnitudes for all stars with BP/RP coefficients information in \Gaia\ DR3. It is described in Section~\ref{sec:CaHKsyn}. The Pristine DR1 photometric catalogue contains the Pristine $CaHK$ magnitudes for all stars in the Pristine footprint as of the end of March 2024 that also have BP/RP information in \Gaia\ DR3. It is described in Section~\ref{sec:Pristine}. Both photometric catalogues are merged into a single file\footnote{219.2~million rows, 6.4~million of which also include Pristine DR1 photometry, 23\,GB in csv format.}, whose content is described in Table~\ref{table:cahk}.

The Pristine-\Gaia\ synthetic metallicity catalogue contains the photometric metallicities created from pushing relevant stars from the $CaHK_\mathrm{syn}$ catalogue through the Pristine $(CaHK,G_\mathrm{BP},G,G_\mathrm{RP})\rightarrow\FeH_\mathrm{phot}$ model. It contains $\FeH_\mathrm{CaHKsyn}$ photometric metallicities over the whole sky and is most accurate in the bright regime ($G_\mathrm{BP}\lta16.0$). The catalogue is available online\footnote{52.3~million rows, 21\,GB in csv format.} and described in Section~\ref{sec:met_catalogues} and Table~\ref{table:FeHphot_CaHKsyn}.

The Pristine DR1 metallicity catalogue contains the photometric metallicities created from pushing the relevant stars from the Pristine $CaHK$ DR1 catalogue through the Pristine $(CaHK,G,G_\mathrm{BP},G_\mathrm{RP})\rightarrow\FeH_\mathrm{phot}$ model. It retains high S/N photometry for the full magnitude range of the \Gaia\ BP/RP data, and therefore provides more accurate $\FeH_\mathrm{Pristine}$ photometric metallicities but over a smaller footprint of $\sim$6,500\,deg$^2$. The catalogue is available online\footnote{6.4 million rows, 2.7\,GB in csv format.} and described in Section~\ref{sec:met_catalogues} and Table~\ref{table:FeHphot_Pr}.

\section{The G\lowercase{aia} synthetic $C\lowercase{a}HK$ catalogue}
\label{sec:CaHKsyn}

\begin{figure*}
\sidecaption
\includegraphics[width=12cm]{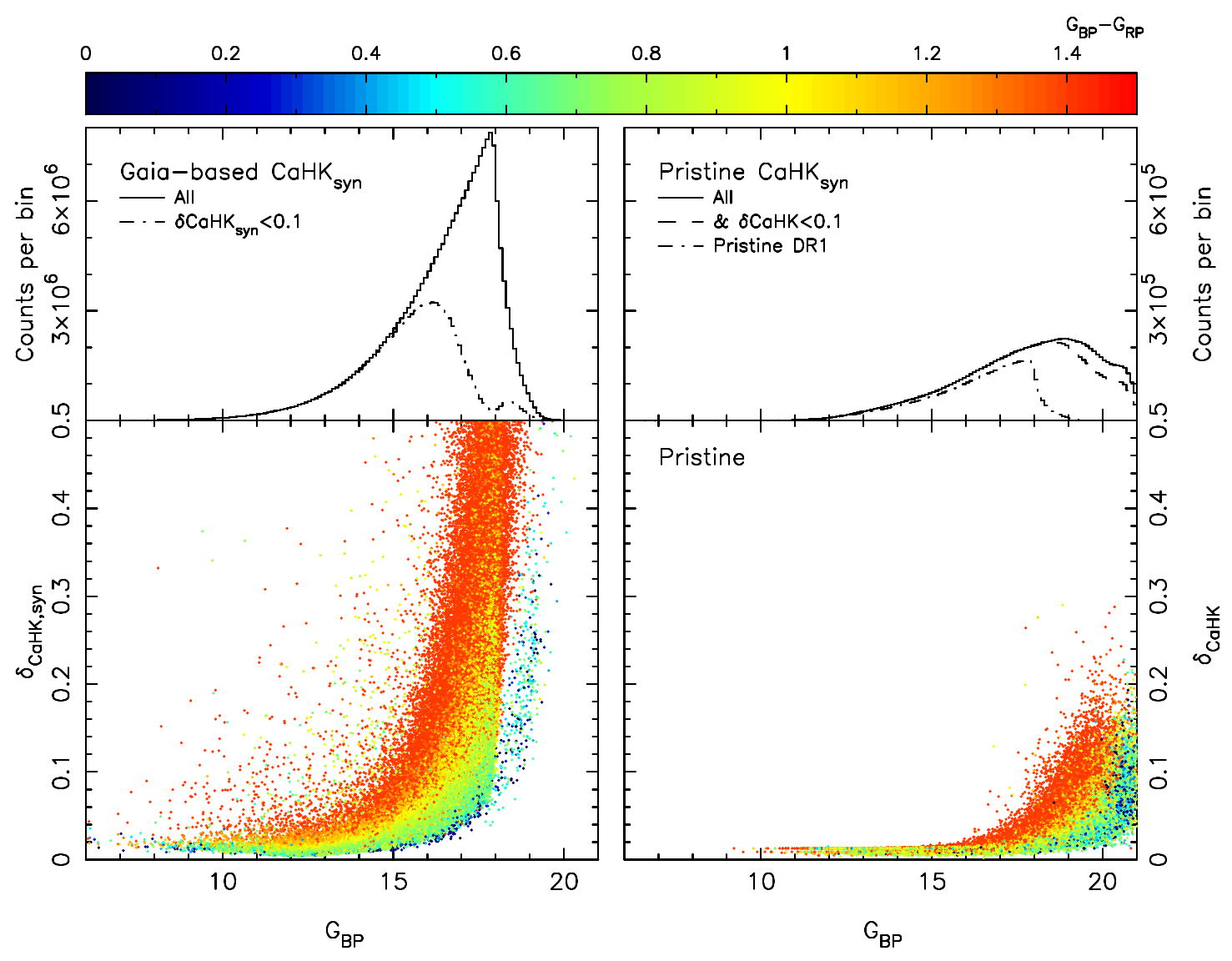}
\caption{Quality of the $CaHK$ magnitudes. \emph{Bottom panels:} $CaHK$ magnitude uncertainties for stars in a representative sample of the \Gaia-based $CaHK_\mathrm{syn}$ (left) and Pristine (right) catalogues. The colors code the $G_\mathrm{BP}-G_\mathrm{RP}$ color of a given star and show the clear impact of a star's color on the $CaHK$ signal-to-noise in both samples. The sharp faint edge in the \Gaia-based synthetic catalogue is driven by the DR3 $G<17.65$ cut and the small number of fainter sources correspond to specific fainter targets included in the catalogue by the \Gaia\ collaboration \citep{deangeli23}. The Pristine survey, on the other hand, has significantly higher signal-to-noise and only reaches photometric uncertainties $\delta CaHK=0.1$ at $19.0 \lta G \lta 21.0$. \emph{Top panels:} histograms of star counts as a function of magnitude, with the line style tracking different samples, as labeled in the panels.\label{d_CaHK}}
\end{figure*}

\Gaia-based synthetic $CaHK$ magnitudes, $CaHK_\mathrm{syn}$, are calculated using \texttt{GaiaXPy}\footnote{\url{https://pypi.org/project/GaiaXPy/}} for all 219.2 million objects for which the \Gaia\ DR3 includes BP/RP coefficients. We follow the procedure presented in \citet{montegriffo23}, integrating the BP/RP spectra under the curve of the MegaCam CaHK\footnote{In what follows, CaHK refers to the filter and $CaHK$ refers to the magnitudes obtained using this filter.} filter, available on the CFHT website\footnote{\url{https://www.cfht.hawaii.edu/Instruments/Filters/megaprimenew.html}}. This narrow-band filter has the significant advantage of presenting a near-top-hat transmission curve and of being centered on the Calcium H \& K lines (near 395\,nm) that are very sensitive to the metallicity of a MW FGK star (see Figure~1 of \citealt{starkenburg17b}). By construction, this catalogue covers the fairly bright sky as the \Gaia\ catalogue only includes the BP/RP information for objects detected with sufficiently high signal-to-noise (S/N). The added difficulty of working toward the blue end of the BP spectrum, which has lower S/N than most of the BP/RP wavelength coverage, means that uncertainties on the $CaHK_\mathrm{syn}$ magnitudes can be very large. We nevertheless include them in the catalogue, but the reader should be aware that quality cuts and a careful treatment of uncertainties are necessary and essential to use this catalogue. For instance, in the main case presented in this paper, that of calculating photometric metallicities based on the Pristine $(CaHK,G,G_\mathrm{BP},G_\mathrm{RP})\rightarrow\FeH_\mathrm{phot}$ model, we only consider stars with $CaHK_\mathrm{syn}$ uncertainties $\delta CaHK_\mathrm{syn}<0.1$. The left-hand panel of Figure~\ref{d_CaHK} shows that this limit is reached for $G_{BP}\sim$15--17. This large range stems from the broad distribution of colors of the catalogue stars: since the CaHK filter is very blue the resulting S/N of a blue or very red star observed under the same conditions by \Gaia\ leads to much larger $CaHK_\mathrm{syn}$ uncertainties for the latter.

\begin{figure*}
\begin{center}
\includegraphics[width=\hsize]{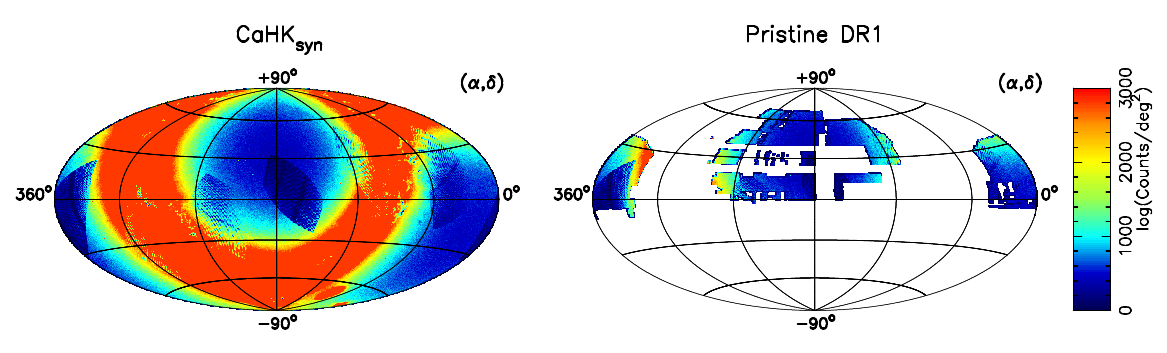}
\caption{\label{map_CaHK} Maps of the source density in the $CaHK_\mathrm{syn}$ catalogue (left) and in its cross-match with the Pristine catalogue (right). The localized artifacts in the density values, particularly visible in the map of $CaHK_\mathrm{syn}$ sources are a consequence of the \Gaia\ scanning law.}
\end{center}
\end{figure*}

From Figure~\ref{d_CaHK}, it is evident that the \Gaia-based $CaHK_\mathrm{syn}$ is most useful at the bright end and that, when available, the Pristine data usually has much higher quality (right-hand panel). The new catalogue is nevertheless extremely valuable for two main reasons: it gives us a ground truth against which we can finally calibrate the Pristine survey to a very high level of photometric accuracy, and it provides an all-sky (bright) equivalent to Pristine on which we can apply the Pristine model to build an all-sky catalogue of MW metal-poor stars. It is however important to keep in mind that this catalogue is impacted by the \Gaia\ scanning law \citep[\eg][]{cantat-gaudin23}. In particular, the S/N and minimum number of visits required for the publication of the BP/RP coefficients in the \Gaia\ DR3 archive translates to a nonuniform inclusion of stars in the catalogue. More generically, the varying number of visits at different locations of the sky that are imparted by the \Gaia\ scanning law translates into a nonuniform S/N over the sky, even for stars with the same properties. This is visible in the left-hand panel of Figure~\ref{map_CaHK}, which presents the density of stars in the $CaHK_\mathrm{syn}$ catalogue. Patches of the sky with significantly lower densities of stars are clearly visible and result in a complicated footprint and inhomogeneous densities of stars in the resulting $CaHK_\mathrm{syn}$ catalogue.

The full catalogue that contains $CaHK_\mathrm{syn}$ values and their uncertainties are be made available online at CDS 
for the 219.2~million sources with BP/RP coefficients from \Gaia\ DR3 and its content is described in Table~\ref{table:cahk}.

\begin{table*}
\begin{center}
  \caption{Description of the columns of the photometry catalogue that includes both the synthetic $CaHK_\mathrm{syn}$ and the Pristine $CaHK$ photometric information.\label{table:cahk}}
  \begin{tabular}{@{}lp{10cm}ll@{}}
Column & Description & Unit & Type\\
\hline
\texttt{RA} & \Gaia\ DR3 right ascension & ICRS (J2016) & \texttt{float}\\
\texttt{Dec} & \Gaia\ DR3 declination & ICRS (J2016) & \texttt{float}\\
\texttt{source\_id} & \Gaia\ DR3 \texttt{source\_id} & --- & \texttt{longint}\\
\texttt{CaHK\_flux} & CaHK mean flux & e-/s & \texttt{float}\\
\texttt{d\_CaHK\_flux} & uncertainty on the CaHK mean flux & e-/s & \texttt{float}\\
\texttt{CaHK\_syn} & $CaHK_\mathrm{syn}$, synthetic Pristine-like $CaHK$ magnitude & mag & \texttt{float}\\
\texttt{d\_CaHK\_syn} & $\delta CaHK_\mathrm{syn}$, uncertainty on $CaHK_\mathrm{syn}$ & mag & \texttt{float}\\
\texttt{Pvar} & Probability for the source to be variable, as defined in section~\ref{sec:variability} & --- & \texttt{float}\\
\texttt{RA\_Pr} & Pristine right ascension& ICRS (J2015.5) & \texttt{float}\\
\texttt{Dec\_Pr} & Pristine declination & ICRS (J2015.5) & \texttt{float}\\
\texttt{CaHK\_Pr} & $CaHK$ Pristine magnitude & mag & \texttt{float}\\
\texttt{d\_CaHK\_Pr} & uncertainty on the $CaHK$ Pristine magnitude & mag & \texttt{float}\\
\texttt{merged\_CASU\_flag} & morphology and quality flag\footnotemark & --- & \texttt{int}\\
\afterpage{\footnotetext{Merged flag based on the \texttt{CASU\_flag} of individual detections, as mentioned in sub-section~\ref{sec:merging}. The flag discriminates between saturated stars (\texttt{-9}), likely point sources (\texttt{-2}), very likely point sources (\texttt{-1}), extended sources (\texttt{+1}), discrepant flags from individual detections (\texttt{+2}), merged source from individual detections with inconsistent photometry (\texttt{+3}).}}
\end{tabular}
\end{center}
\end{table*}

\section{The Pristine \emph{C\lowercase{a}HK} survey}
\label{sec:Pristine}
\subsection{The Pristine photometric data}
\label{sec:Pristine_phot}
The design of the Pristine survey, along with its core science goals, are presented in detail in \citet{starkenburg17b}. In a nutshell, the Pristine survey is a narrow-band photometric survey that relies on the CaHK narrow-band filter that was procured in 2014 by the CFHT for its wide-field imager MegaCam \citep{boulade03}. Thanks to the large $\sim1\times1$\,deg$^2$ field of view of the MegaCam camera and the dedication of the CFHT engineering and service observing crew, we have now gathered $\sim$11,500 images with the CaHK filter. These cover more than 6,500\,deg$^2$ of the north and south Galactic caps. The current coverage of the Pristine survey is shown in the right-hand panel of Figure~\ref{map_CaHK}. The observational set-up and the data reduction of Pristine remain in general the same as those described in \citet{starkenburg17b} for the first two semesters of observations. In what follows, we mainly focus on the updates to the data reduction pipeline.

Since semester 2016B, Pristine is set up as a CFHT snapshot (\ie poor-weather) program and we enforce no restriction on observing conditions. While this means that, over time, some of the fields had to be reobserved because the initial observing conditions proved too challenging, it also ensures that a sizable number of images are regularly observed. This strategy also leads to a footprint that is not contiguous and is instead driven by poor weather or gaps in the observing queue of normal CFHT programs. While we have been striving, over the years, to fill in the holes in the coverage, some of these remain, especially in the region of the northern Galactic cap (right-hand panel of Figure~\ref{map_CaHK}). Our agreement with the \emph{William Herschel} Telescope Enhanced Area Velocity Explorer (WEAVE) Galactic Archaeology spectroscopic survey for the follow-up of EMP star candidates \citep{jin24} leads us to favor covering the expected footprint of the ``WEAVE Galactic Archaeology Low Resolution high latitude'' (WEAVE-GA LR-highlat) subsurvey that focuses on halo regions ($|b|\gta25\deg$ and $\delta>0\deg$).

The initial survey strategy, when Pristine was observed as a normal program at CFHT and that we presented in \citet{starkenburg17b}, included exposures of $1\times100$\,s or $2\times100$\,s, depending on the rankings of our programs. Since 2016B, we have settled on a single 200\,s exposure for all fields. After the images are observed, they are preprocessed by the CFHT staff with the \texttt{elixir} software  to remove most of the instrumental signatures (de-biasing, flat-fielding; \citealt{magnier04}). All individual preprocessed exposures are retrieved from the Canadian Astronomy Data Center (CADC) archive at this stage and we use the pipeline from the Cambridge Astronomy Survey Unit \citep[CASU; ][]{irwin01,irwin04} for the next steps of data reduction. Initially, in \citet{starkenburg17b}, only the central 36 CCDs of the images were used, but we now treat all 40 CCDs of the imager in our analysis, including the ``ears'' that are formed by the 4 CCDs to the east and west of the central square-degree of the field-of-view\footnote{\url{https://www.cfht.hawaii.edu/Instruments/Imaging/Megacam/specsinformation.html}}.

We refine the astrometry of the images downloaded from the archive and we now use \Gaia\ DR2 \citep{lindegren18}\footnote{Switching to \Gaia\ (E)DR3 leads to no significant improvement of the astrometry compared to \Gaia\ DR2. To avoid having to reprocess the thousands of images observed before December 2020, we continue using \Gaia\ DR2 as a reference. The resulting rms on the astrometry of stars is consistently better than 0.1".} as an astrometric reference instead of 2MASS. The small exposure times, combined with the sometimes low density of stars toward the MW halo, the limited number of photons that pass through the very blue and narrow filter, and the sometimes challenging observing conditions mean that a small number of images fail at this astrometry stage. The corresponding fields are then sent back in the queue to be reobserved later.

After the astrometry is refined, we also use the CASU pipeline to perform aperture photometry on the images. This step can be made difficult by the low S/N of some observations and the absence of many high S/N, bright stars from which the pipeline can build a reliable point-spread function (PSF) and, then, a reliable aperture correction. This step, which is conducted independently for each CCD of each image, sometimes fails and generates obviously wrong aperture corrections. In such cases, the average aperture correction of successful CCDs on a given image is propagated to the CCDs of the same image that failed the photometry step. In total, this is necessary for a few percent of all the CCDs. We show below that, after calibration, the resulting photometry is very stable despite these necessary corrections.

The CASU pipeline automatically assigns to each source a flag that tracks its morphology: whether it is point-source-like ($\mathtt{CASU\_flag=-1}$ for very likely stars, \texttt{-2} for likely stars, or \texttt{-9} for saturated stars with extrapolated photometry), appears extended (\texttt{+1} for a source that is likely extended), or resembles noise (\texttt{0}). However, the difficulty to sometimes determine a good PSF from a limited number of high S/N stars on poorly populated CCDs means that this flag is not always reliable. From experience and numerous tests, we find that this step is still reliable to discriminate between spurious and real astrophysical objects but does not always work efficiently to separate point sources from extended sources. This is not dramatic since Pristine is a fairly bright survey tailored to the \Gaia\ depth and all of the envisaged Pristine science requires the contribution of broad-band photometry (SDSS before, now typically \Gaia), whose star/galaxy discriminator can be used to assign a morphological class to a given source. Consequently, we keep in the catalogue sources that are not defined as noise by the CASU pipeline ($\mathtt{CASU\_flag\neq0}$).

\subsection{Data calibration}
Previous versions of the Pristine photometry were calibrated relatively to each other using the stellar locus of red dwarf stars \citep[using SDSS photometry, $(g-i)_0>1.2$;][]{starkenburg17b} because, at the time, there was no absolute photometric scale to calibrate against. With the availability of the all-sky, beautifully and absolutely calibrated \Gaia-based $CaHK_\mathrm{syn}$ catalogue, we now rework our procedure to calibrate against this catalogue. In \citet{starkenburg17b}, we showed that two different effects needed to be taken into account for the calibration: an offset in the calibration of a given field that depends on the observing conditions (clouds, dust in the atmosphere, reflectivity of the mirror) and a variation of the zero point of the $CaHK$ magnitudes as a function of the location on the field of view (FoV). This latter effect is well known and not subtle, and can produce magnitude differences larger than 0.1 for the same star depending on whether it is observed at the center of the field or its outskirts (see \citealt{starkenburg17b} and, \eg Figure~3 of \citealt{ibata17b}; see also \citealt{regnault09}).

Until now, the contribution of both effects to the calibration was determined separately by comparing the median location of the red part of the stellar locus, first dealing with the zero-point offset before building a smooth model of magnitude offsets as a function of the location on the FoV by combining all the observations of a given semester, once again using the red part of the stellar locus. There are, however, reasons to believe that the shape and location of these offsets subtly changes each time the camera is placed on the telescope, that is for every MegaCam ``run'' that happens once or twice a month \citep{ibata17b}. A quick comparison between the catalogue generated using the previous calibration and the $CaHK_\mathrm{syn}$ catalogues shows that, in general, the previous zero-point offsets are good, with only a small fraction of fields showing significant offsets. Similarly, the FoV correction was adequate, but nevertheless shows some artifacts when compared to the $CaHK_\mathrm{syn}$ catalogue. We also find that the systematic floor on the $CaHK$ uncertainties, which was previously determined to be 18\,mmag from a mosaic of overlapping fields observed under good conditions during the same semester, was significantly underestimated. This number was in fact closer to 40\,mmag once taking into account all survey fields observed under very different conditions over a period of 8~years.

\subsubsection{Updated calibration model}
Taking the $CaHK_\mathrm{syn}$ magnitudes as the truth to calibrate against, we now perform the two corrections simultaneously for all images of a given run and allow for more flexible and detailed, run-specific FoV models. To achieve this goal, we build a simple neural network model, \texttt{PhotCalib}\footnote{\url{https://github.com/zyuan-astro/PhotCalib}}, that we apply to the independent photometric catalogues of all $\sim$11,500\,Pristine images observed during 85~MegaCam runs between semesters 2015A and 2023B (until run 23Bm03 in October 2023; no additional data where gather after this run until April 2024).

In practice, \texttt{PhotCalib} is fed with all the photometric catalogues of Pristine images observed in run $j$, to which it assigns a zero-point offset, $zp(i)$, as a free parameter for each image $i$, and models the FoV correction of this run, $FOV_j(X,Y)$, as a smooth function that depends on the $(X,Y)$ position in the field. For a star of uncalibrated magnitude $CaHK_\mathrm{uncalib}$, the calibrated magnitude, $CaHK_\mathrm{calib}$, is therefore

\begin{equation}
\centering
CaHK_\mathrm{calib} =  CaHK_\mathrm{uncalib} + zp(i) + FOV_j(X,Y).
\label{eq_calib}
\end{equation}

Specifically, we use a free $n$-D parameter tensor to describe the $zp(i)$ offsets and the $FOV_j$ model uses three fully connected neural layers, with each layer having 200 neurons, followed by an activation and a normalization layer. The input data for calibration are the positions of the stars $(X, Y)$ in the FoV, the image number $i$, and both the $CaHK_{\rm uncalib}$ and  $CaHK_{\rm syn}$ catalogues for a given run. The loss function is simply the uncertainty-weighted chi-square between the $CaHK_{\rm calib}$ values, as defined in the equation above, and the corresponding $CaHK_{\rm syn}$ values. For fast convergence, we use the stochastic gradient descent optimizer, which remains fast even for the largest run of almost 1,000 images and 404,144 stars.

\subsubsection{Setup and application}

To ensure that we determine the calibration models on reliable stars, we apply the following quality cuts to the Pristine data set:
\begin{itemize}
  \item $\texttt{CASU\_flag = -1}$;
  \item $\delta CaHK_\mathrm{syn} <0.1$;
  \item no star within radius $r_\mathrm{max}$ from the center of globular clusters, with $r_\mathrm{max}$ the rough estimate of the size of a cluster defined by \citet{vasiliev21};
  \item no star with $|CaHK_\mathrm{uncalib}  - CaHK_\mathrm{syn,med}|>0.2$, with $CaHK_\mathrm{syn,med}$ the median $CaHK_\mathrm{syn}$ value of the considered field (these stars are likely variable stars\footnote{Applying a more specific variability cut that relies on the $P_\mathrm{var}$ quantity defined later, in Section~\ref{sec:variability}, only mildly impacts the calibration, with differences of less than 3\,mmag on the $zp(i)$ values and significantly less than the systematic floor of 13\,mmag that results from the new calibration process (see Sub-section~\ref{sec:final_catalogue} below).} or catastrophic failures).  
  \end{itemize}

\begin{figure*}
\begin{center}
\includegraphics[width=\hsize]{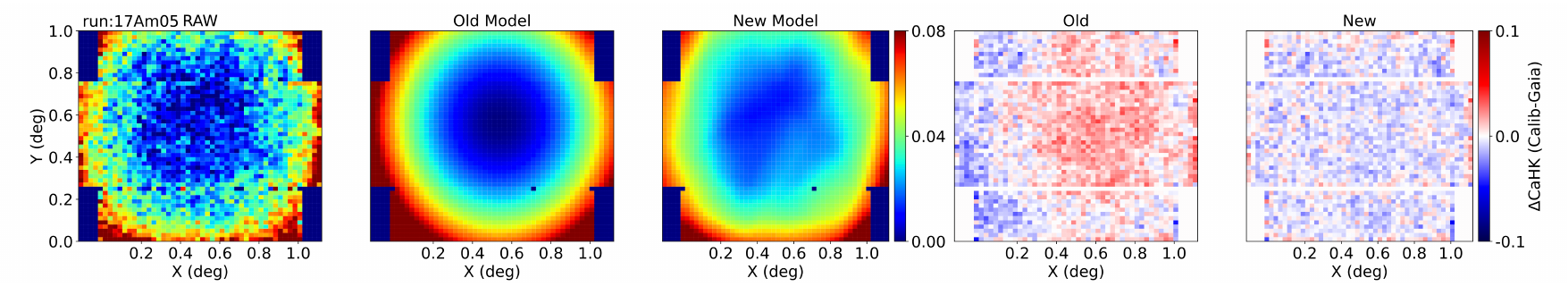}
\caption{Field-of-view corrections of run 17Am05. From left to right: the un-calibrated data after the field zero-point offset corrections ($CaHK_{\rm uncalib} + zp(i) - CaHK_{\rm syn}$); the old analytic model; the new FoV model from \texttt{PhotCalib}; the median residual over the field of view between the calibrated and the synthetic \Gaia\ values ($\Delta CaHK = CaHK_\mathrm{calib} - CaHK_\mathrm{syn}$) from the old and the new model.\label{fig_fov}}
\end{center}
\end{figure*}

Figure~\ref{fig_fov} visualizes the results of the FoV model for a randomly chosen run, 17Am05 (89,994 stars with $\delta CaHK_\mathrm{syn} <0.1$, a number that is typical of most runs, although see below for the special handling a particularly small runs). In the leftmost panel, we show the median magnitude offsets  $\Delta CaHK = CaHK_\mathrm{uncalib} + zp(i) -  CaHK_{\rm syn}$, after correcting the photometry of all images by the corresponding zero-point offset. It makes the need for the FoV correction particularly evident. The old FoV model is shown in the second panel of the figure and, while it reproduced the global shape of the offsets, it was clearly not perfect. The FoV model determined by \texttt{PhotCalib} is shown in the third panel. Its increased flexibility more subtly captures the irregularities in the FoV offsets. This is confirmed in the fourth and fifth panels of the figure that present the residuals after applying the two FoV models. While median magnitude offsets with the $CaHK_\mathrm{syn}$ magnitudes remain small with the previous model (usually within $\pm0.03$) and do not invalidate the past Pristine catalogues, the application of \texttt{PhotCalib} yields significantly flatter photometry.

We achieve similar results to those presented in Figure~\ref{fig_fov} for most runs. Five of the runs contain fewer than 3,000~good quality stars each and yield visually poorer results for their FoV models. For those, we combine two small neighboring runs (21Bm03 and 21Bm04) with 1,073 and 2,889 stars, respectively, to obtain a single model for both runs. For the other three small runs (16Bm02, 19Am05, and 22Am04), the number of stars is always less than 10\% of that of the neighboring runs. For those, we chose to use the FoV model from the two neighboring runs and only let \texttt{PhotCalib} determine the zero-point offsets of each image. From these two models, we compare the distribution of the residuals after the calibration and pick the one with the smallest intrinsic dispersion as the best model.

Overall, contiguous runs show subtle variations in their FoV models, with a range of $\pm0.02$\,mag. It justifies the choice of building a model for each run when possible, but these variations are small enough that using the FoV model of adjoining runs for those with a small number of stars is still a reasonable choice and much better than not applying a FoV correction.

\subsubsection{Dealing with a sharp feature in runs 19Am05, 19Am06 \& 19Bm01}

\begin{figure*}
\begin{center}
\includegraphics[width=\hsize]{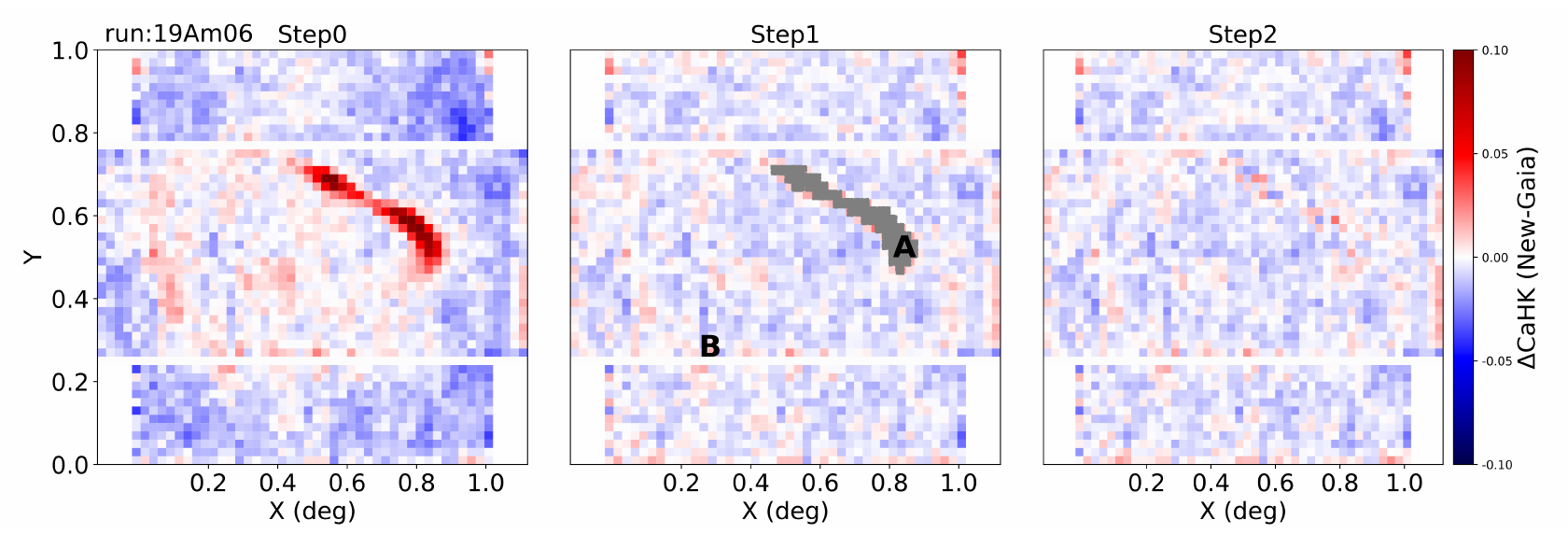}
\caption{Median residuals per pixel over the field of view between the calibrated and the synthetic \Gaia\ values ($CaHK_{\rm calib} - CaHK_{\rm syn}$) for run 19Am06 that is affected by the sharp feature. Step~0: results after applying only the regular model. Step~1: the masked region of the feature is defined as region~A (gray) and the rest is region~B, which has been corrected by applying the regular model only on region~B. Step~2: after applying the modified model on region~A, the sharp feature is removed. The corrections in region~B are the same as in Step~1.\label{fig_hair}}
\end{center}
\end{figure*}

Three of the runs --- 19Am05, 19Am06, and 19Bm01 --- show a problematic sharp feature in the map of residuals, as shown in the left-hand panel of Figure~\ref{fig_hair}. The CFHT staff tracked this feature to the presence, during the summer of 2019, of a hair on one of the MegaCam lenses. They removed it after run 19Bm01 but the direct consequence is that all the photometric catalogues observed during a period of about 3~months suffer from this additional source of localized absorption. It causes the $CaHK$ values to get fainter by up to $\sim$0.15\,mag. \texttt{PhotCalib} under-corrects the magnitudes at the location of the feature and over-corrects them around it as the jump in the data are locally so sharp  that the gradient descent method fails when reaching the boundaries of the feature. Fortunately, two of these three runs (19Am06, 19Bm01) are the largest runs in the survey and contain hundreds of images (300,000--400,000 good quality stars for each run). We are therefore in the fortunate position to be able to build a refined version of the calibration for this specific situation.

We first use the FoV model from the closest large run (19Am03) to determine temporary zero-point offsets for all the images in run 19Am06 (Step~0, left-hand panel of Figure~\ref{fig_hair}). We then determine the boundaries of the region of the feature (region~A) by grouping together neighboring pixels that show median $\Delta CaHK$ deviations larger than the 97.5th percentile of the entire distribution. The pixels outside region~A define region~B. We then apply \texttt{PhotCalib} only to stars in region B, which yields the final zero-point offsets as well as a reliable $FOV_\mathrm{19Am06}$ model in regions not affected by the feature (Step~1; central panel of Figure~\ref{fig_hair}). After Step~1, the $zp(i)$ values are frozen for all the stars in this run and, to learn the behavior of the feature, we set a different neural network containing one convolutional and one linear layer, with each accompanied by an activation and a normalization layer like for the regular model. We apply this new neural network only to region B with fixed $zp(i)$ values. We then simply tile together the models from the two regions in Step~2 (right-hand panel of Figure~\ref{fig_hair}), which successfully removes the feature while, at the same time, leaves its surroundings un-affected. We follow a similar approach to calibrate 19Bm01, with the only difference being that the FoV model used in Step~0 is the model from region B of 19Am06. The third run affected by this feature is 19Am05, which contains only 5~images and 1,142~good-quality stars. For this run, we apply the recipe we designed above to deal with small runs and use the FoV model from its large neighboring run, 19Am06.

\subsubsection{Final results}
\begin{figure*}
\begin{center}
\includegraphics[width=\hsize]{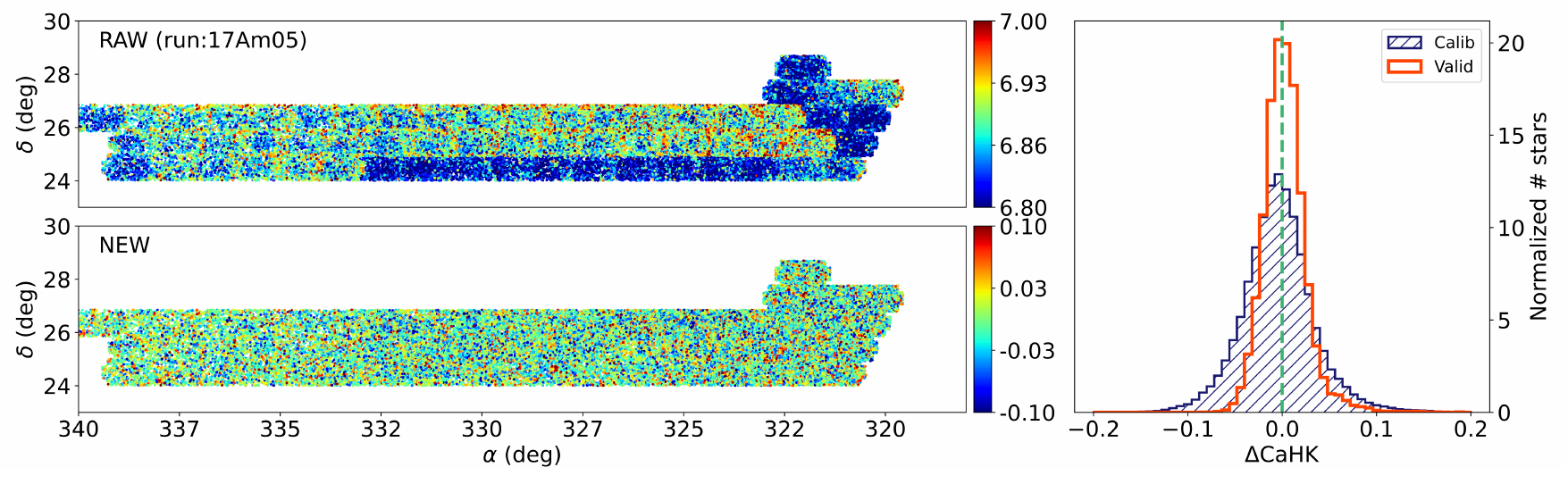}
\caption{Effect of the calibration procedure on the $CaHK$ magnitudes. \textit{Left:} maps of the $CaHK$ offsets between the uncalibrated (top) and calibrated (bottom) Pristine catalogues and the \Gaia\ $CaHK_\mathrm{syn}$ magnitudes (the color codes $CaHK_\mathrm{uncalib} - CaHK_\mathrm{syn}$ and $CaHK_\mathrm{calib} - CaHK_\mathrm{syn}$, respectively) for run 17Am05. \textit{Right:} histogram of a random sampling of the residual ($\Delta CaHK = CaHK_\mathrm{calib} - CaHK_\mathrm{syn}$) for 1\% of stars in the calibration ($\delta CaHK < 0.05$; blue) and validation ($\delta CaHK< 0.015$; orange) samples. \label{fig_calib}}
\end{center}
\end{figure*}

The left-hand panels of Figure~\ref{fig_calib} give a visual representation of the success of the calibration by showing a map of the differences between the Pristine $CaHK$ magnitudes and the \Gaia\ synthetic $CaHK_\mathrm{syn}$ magnitudes before and after calibration for stars of good-quality synthetic data ($\delta CaHK_\mathrm{syn}< 0.05$). The chosen 17Am05 run combines data that are affected by different amounts of absorption from the observing conditions, which produce the obvious field-to-field offsets in the top-left panel. The magnitude variations with the location on the FoV are also visible. After calibration, both of these effects are handled and the data are very flat. This is summarized for the full data set in the right-hand panel: the blue histogram represents a random sampling of 1\% of the calibration sample with $\delta CaHK_\mathrm{syn}< 0.05$ over the full data set, and the orange histogram shows a random sample of 1\% of the validation sample ($\delta CaHK_\mathrm{syn}< 0.015$), still over the full data set. After calibration, the mean difference is $0.006$\,mag and the distribution, which looks Gaussian-like, has a dispersion that is barely larger than the \Gaia\ photometric-uncertainty cut used to build the sample ($\sim$18\,mmag).

\subsection{The Pristine \emph{CaHK} DR1}
\label{sec:final_catalogue}

\begin{figure}
\begin{center}
\includegraphics[width=1.05\hsize]{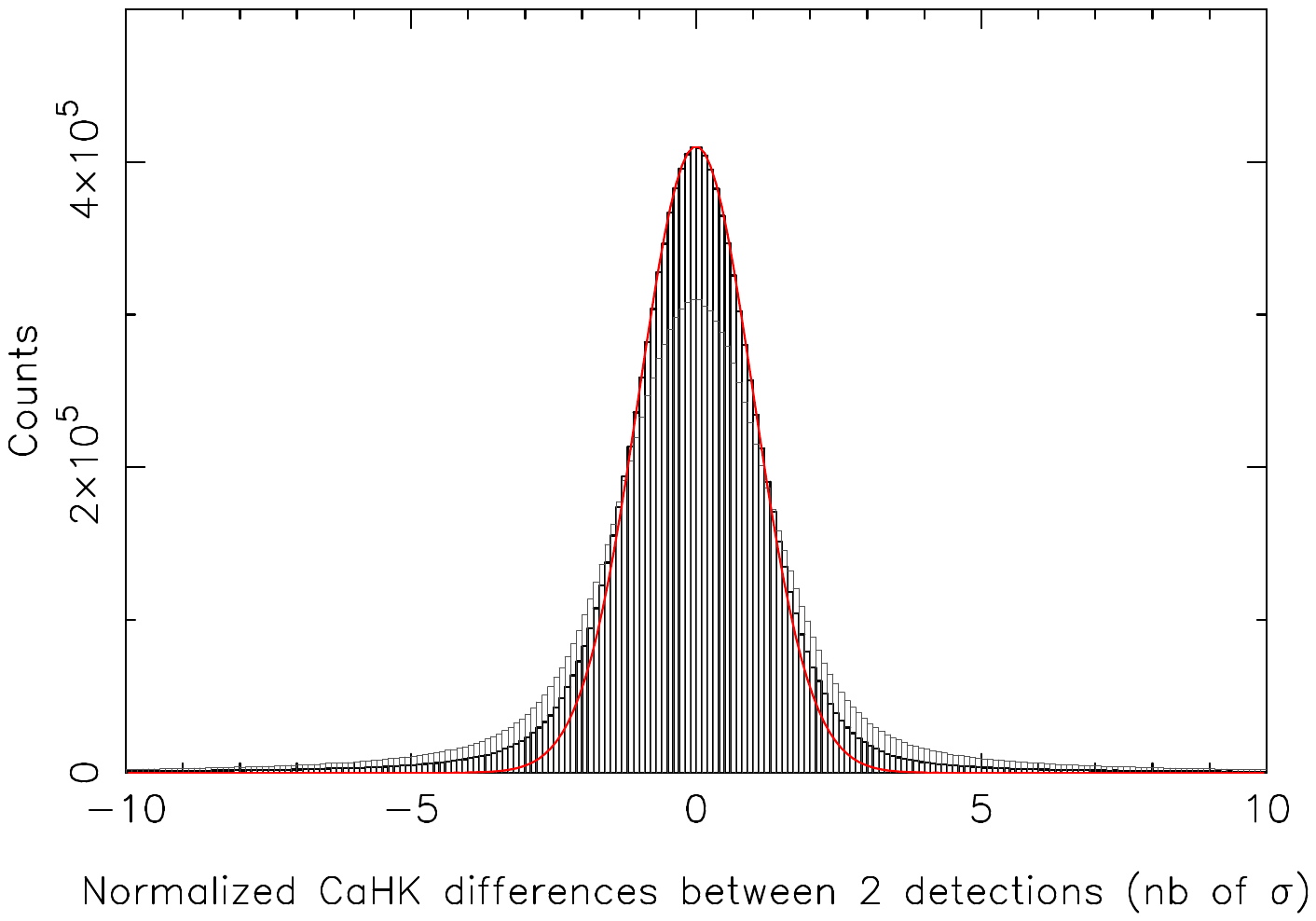}
\caption{Distribution of normalized magnitude differences between repeat observations in Pristine. The gray data corresponds to the reduced, calibrated data, while the black histogram also includes a systematics uncertainty floor of 13\,mmag, added in quadrature to the photometric uncertainties coming out of the pipeline. The black histogram is a good approximation of the expected distribution (red line) but shows a small fraction of objects in the tails of the distribution produced by incompatible repeat observations (variable stars, artifacts, etc.). These represent $\sim7.5$\% of repeats.\label{fig:merges}}
\end{center}
\end{figure}

\subsubsection{Merging of repeat individual detections}
\label{sec:merging}
While the Pristine survey was designed as a single-exposure survey, there is nevertheless a significant fraction of survey stars that are observed multiple time, either from the repeat observations of low-quality fields that could be salvaged, from the $2'$ overlap between the central, square portion of neighboring fields, or from the 4 side CCDs (the ``ears'') that were not assumed to be available when the survey tiling was constructed. In total $\sim$20\% of sources in the detection catalogue correspond to repeat observations. To increase the signal-to-noise of those objects, we merge their individual detections using a $0.5''$ search radius. The magnitudes of the individual detections are combined into the magnitude and corresponding uncertainty of the merged detection using an arithmetic mean weighted by the uncertainties of the individual measurements.

Figure~\ref{fig:merges} shows the resulting magnitude differences between repeat measurements, normalized by their uncertainties (gray histogram). Perfect data would result in a Gaussian centered on zero, with width unity (red line). It is not the case here, which shows that there remains some level of systematic uncertainties in the photometric calibration, as is expected in any data set. The black histogram, whose central distribution is a good match to the red model has a systematic uncertainty of 0.013\,mag added in quadrature to the photometric uncertainties of individual detections. This value is entirely compatible with the width of the orange histogram in Figure~\ref{fig_calib} and allows us to conclude that the Pristine magnitudes are calibrated at the 13\,mmag level. The tails of the back histogram in Figure~\ref{fig:merges} correspond to variable stars or stars with problematic photometry. In the catalogue of merged detections, we flag with \texttt{merged\_CASU\_flag=+2} merged detections whose \texttt{CASU\_flag} is not consistent between the different individual detections ($\sim20\%$ of merges). The \texttt{merged\_CASU\_flag} otherwise duplicates the consistent value of \texttt{CASU\_flag} from the different individual detections. Finally, we flag with \texttt{merged\_CASU\_flag=+3} merged objects that have two individual detections that are incompatible at more than the $5\sigma$ level (2.9\% of repeats for the full Pristine survey, including faint sources).

\subsubsection{Catalogue}

With this paper, we make public the part of the Pristine data that overlaps with the $CaHK_\mathrm{syn}$ catalogue; that is, we provide the Pristine information on the merged detections for all 6,311,676~stars with \Gaia\ DR3 BP/RP coefficients that overlap the Pristine footprint. The density of stars in this catalogue is shown in the right-hand panel of Figure~\ref{map_CaHK}. When available, the Pristine photometric information for these sources is added to the photometric catalogue described earlier in Table~\ref{table:cahk} and temporarily available on the private repository before paper acceptance.

\section{Extinction correction}
\label{sec:extinction}
The \Gaia\ broad-band filters that we use in our photometric metallicity model (see Section~\ref{sec:model}) are so broad that the extinction correction in a given filter depends on the underlying properties of the star (\teff, \logg, \FeH), as well as the extinction toward that star. The \Gaia\ team provides a relation to derive the extinction coefficients as a function of \teff (or $G_\mathrm{BP} - G_\mathrm{RP}$) and $A_0$ (the monochromatic extinction at 541.4\,nm)\footnote{\url{https://www.cosmos.esa.int/web/gaia/edr3-extinction-law}}. We follow a similar methodology as described there to re-derive this relation including a dependence on \FeH, using the \texttt{dustapprox} package \citep{fouesneau22} with Kurucz synthetic spectra \citep{castelli03}, adopting the \citet{fitzpatrick99} extinction law with $R_V = 3.1$, the \citet{riello21} \Gaia\ passbands and the MegaCam CaHK passband. 

The synthetic grid covers $4000\leq\teff\leq10\,000$\,K with step of 250\,K, $0.0\leq\logg\leq0.5$ with steps of 0.5, and $-2.5\leq\FeH\leq+0.5$ with steps of 0.5\,dex, with the addition of spectra at $\FeH=+0.2$ and $\FeH=-4.0$. We adopt [$\alpha$/Fe] = +0.4 for $\FeH \leq -1.0$ and [$\alpha$/Fe] = 0.0 for $\FeH \geq -0.5$. We fit separate relations for giants (all $\FeH$, $4000 \leq \teff \leq 7500$\,K, $\logg = 4.0$ for $\teff > 5250$\,K and $\logg = -8.3000 + 0.0023\,\teff$ for cooler stars) and dwarfs (all $\teff$, $\logg = 4.5$ and all $\FeH$ for $\teff \leq 7500$\,K but only $\FeH \geq -0.5$ for hotter stars). 

\begin{figure}
\centering
\includegraphics[width=1.0\hsize]{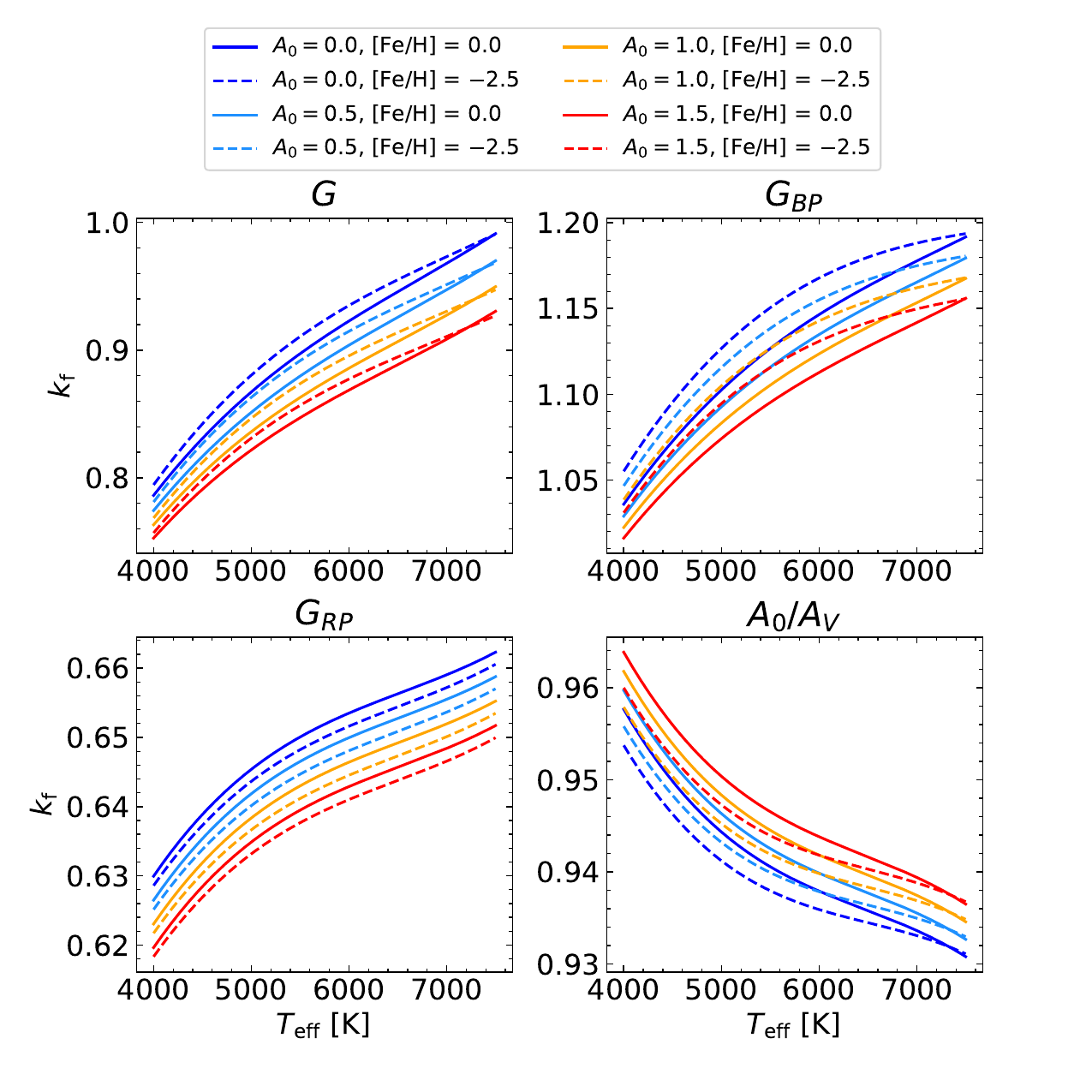}
\caption{Extinction coefficients for the three \Gaia\ filters and the ratio $A_0/A_V$ as a function of the effective temperature \teff, for different values of $A_0$ and \FeH\ (see legend).}
\label{fig:extcoeff} 
\end{figure}

We compute synthetic photometry for $0.01\leq A_0\leq 2.99$ with steps of 0.02 and we then fit polynomial relations of the following form to the synthetic photometry from \texttt{dustapprox}: 
\begin{multline} \label{eq:kf}
k_{f} = a_0 + a_1 T + a_2 A_0 + a_3 \FeH + a_4 T^2 + a_5 A_0^2 + a_6 \FeH^2 \\ 
+ a_7 T^3 + a_8 A_0^3 + a_9 \FeH^3 + a_{10} T A_0 + a_{11} T \FeH + a_{12} A_0 \FeH \\ 
+ a_{13} A_0 T^2  + a_{14} T A_0^2 + a_{15} \FeH T^2 + a_{16} T \FeH^2 + a_{17} \FeH A_0^2 \\
+ a_{18} A_0 \FeH^2 + a_{19} T A_0 \FeH,
\end{multline}

\noindent with $T = \teff/5040$\,K and $k_{f}$ the extinction coefficients in a given filter. The best fit polynomial coefficients are listed in Table~\ref{tab:extinction_giant} (giants) and ~\ref{tab:extinction_dwarf} (dwarfs) in the Appendices. The giant solution as a function of \teff for different $A_0$ and \FeH\ is presented in Figure~\ref{fig:extcoeff} for the three \Gaia\ filters. The sensitivity to the stellar parameters for the narrow CaHK filter is negligible ($< 0.3\%$) and not shown in the Figure. 

In this work we use the \citet[][hereafter SFD]{schlegel98} extinction map, which provides $E(B-V)$ instead of $A_0$. We derive ${A_0}/{A_V}$ as a function of the stellar parameters following the same methodology as above\footnote{This ratio is not equal to one because $A_V$ is for a broad filter and depends on the stellar parameters, whereas $A_0$ does not.}, as shown in the bottom-right panel of Figure~\ref{fig:extcoeff}. Finally, the extinction in filter $f$ is given by $A_f = R_f E(B-V)$ with $R_f = k_f \frac{A_0}{A_V} 3.1$, which is multiplied by an additional factor of 0.86 when using $E(B-V)$ from the SFD map (from the recalibration of the SFD map by \citealt{schlafly11}). 

We iteratively derive the best extinction coefficients for each individual star from the photometry, together with \teff and \FeH. For \teff, we adopt the \citet{casagrande21} \Gaia\ EDR3 photometric \teff relation that depends on $(G_{BP}-G_{RP})_0$, \FeH, and \logg\ (we fix the latter to 3.0 as it only has a minor effect). An estimate of $\FeH$ is then derived using our photometric metallicity model (see Section~\ref{sec:model}). If a star falls outside the metallicity grid, we assign it $\FeH = -4.0$ if the star is above the edge of the metallicity grid and $\FeH = 0.0$ otherwise. If it falls outside the validity range of the \citet{casagrande21} relation (3500--9000\,K), \teff is set to the closest valid \teff. Initial guesses for $A_{(BP-RP)}$, $A_{G}$ and $A_{CaHK}$ are $E(B-V)_{SFD} \times 1, 2$ and $4$, respectively, and the initial guess for $A_0/A_V = 1$. We find that the extinction coefficients converge quickly and that five iterations are sufficient. 

For a typical halo giant star ($\teff = 5000$\,K, $\FeH = -1.0$, $E(B-V) = 0.05$), we find $R_{CaHK} = 3.918$ for the SFD map, which is very similar to the $CaHK$ extinction coefficient of 3.924 that our team has been using previously \citep{starkenburg17b}.

\section{Probabilistic variability model}
\label{sec:variability}
Variable stars are one of the main sources of contamination in low-metallicity catalogues based on photometry \citep{starkenburg17b,lombardo23}. Previously, we relied on the repeat observations of sources in Pan-STARRS~1 to quantify the variability of a source \citep{hernitschek16}. As we strive to rely solely on Pristine and \Gaia\ information to characterize observed sources over the full sky coverage, we have moved toward using a single \Gaia-based diagnostic of variability \citep{fernandez-alvar21,lombardo23}, which we update here with a probabilistic approach.

\begin{figure}
     \begin{center}
     \includegraphics[width=\hsize,]{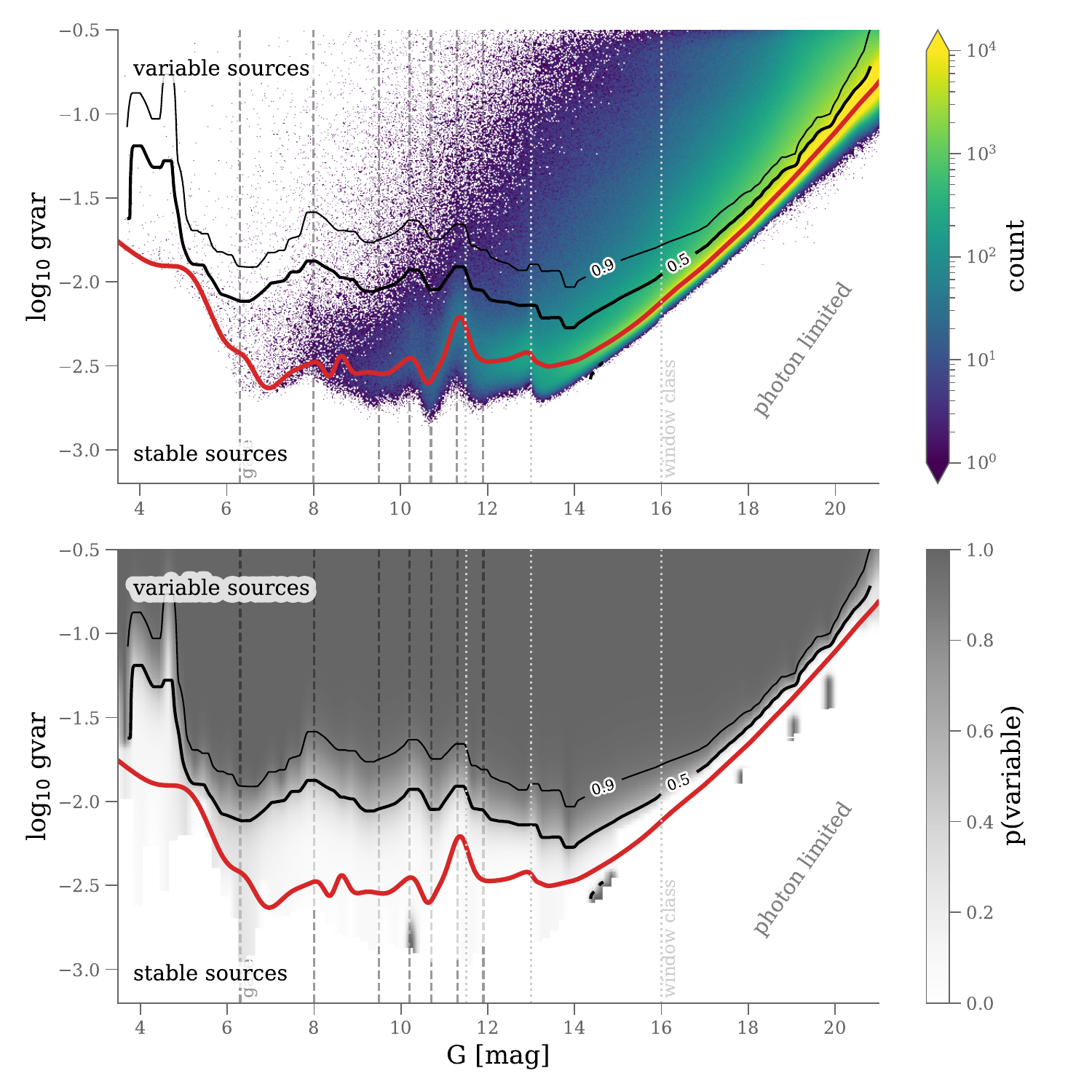}
     \end{center}
     \caption{
          Variability distribution as a function of the $G$ magnitude for 1,806,185,593 sources from \Gaia\ DR3.  The red line indicates our B-spline model that tracks the mean location of non-variable sources. We note that for $G<13$, the photometry is affected by the gating and windowing system onboard \Gaia. We indicate the approximate $G$ magnitudes of the eleven gates with the dashed gray lines and the three window class changes resulting in overall offsets in the photometric calibration (details in \citealt{riello21}). Between these positions, the configuration affects the photometry in various manners, generating apparent wiggles and discontinuities. In DR3, the reported uncertainties include an additional calibration error on a single measurement of 2\,mmag corresponding to the apparent floor ($\log_{10}g_\mathrm{var}\sim-2.7$) at the bright end.}
          \label{fig_gvar_vs_gmag}
     \end{figure}

\Gaia\ observes repeatedly 1.8 billion sources quasi-simultaneously in $G$, $G_{BP}$, and $G_{RP}$ integrated photometric bands and the \Gaia\ DR3 contains the mean photometry for these sources based on $34$ months of multi-epoch observations. The repetition of the measurements provides precise mean photometric values but, additionally, the associated uncertainties can tell us about the photometric variations of a source (see also \citealt{belokurov17}). Assuming Poisson variation of the measurements, we can define a variability metric as the fractional flux uncertainty scaled by the square root of the number of observations. For the $G$ band, we defined this variability metric by

\begin{equation}
    \text{gvar} = \sqrt{N_{obs, G}} \cdot \frac{\sigma_G}{f_G},
\end{equation}

\noindent where $N_{obs, G}$ is the number of observations contributing to the photometric value, and $\frac{\sigma_G}{f_G}$ is the fractional flux uncertainty. This quantity corresponds to \texttt{sqrt(phot\_g\_n\_obs)/phot\_g\_mean\_flux\_over\_error} in the \texttt{gaia\_source} \Gaia\ tables\footnote{\url{https://gea.esac.esa.int/archive/}}. The top panel of Fig.~\ref{fig_gvar_vs_gmag} shows the complex distribution of this metric as a function of apparent $G$-magnitude. The majority of sources at any given magnitude resides close to the photon-noise limit of the instrument, i.e., small values of \texttt{gvar}. Complexity arises at the bright end, where the gating and windowing system on board \Gaia\ affects the photometry in various manners generating apparent wiggles and discontinuities in that distribution.

Defining which \Gaia\ sources are intrinsically variable ones requires finding the photon-limited sequence\footnote{Technically, systematics and instrumental effects dominate the noise budget of sources at the bright end.} and modeling the dispersion around it. We adapt the B-Spline approach of \citet{riello21} to derive a mean uncertainty model (see the \Gaia\ science performance pages\footnote{\url{https://www.cosmos.esa.int/web/gaia/science-performance\#photometric\%20performance}}). We set the spline knots such that we capture the main features (gates) and the sharp transitions due to window class changes (\eg $G=13.0$). We assume that the spline location identifies the ``stable'' sources and that they follow an almost Gaussian distribution around it (law of large numbers). In contrast, we expect the ``variable'' sources (strictly, the non-stable sources) to be rare and thus distribute more closely to a Poisson distribution toward larger values of \texttt{gvar}. In practice, we model the distributions at a given $G$-magnitude by the mixture of two $t$-distributions using the spline as a prior on the location of the peaks. A $t$-distribution is a continuous equivalent to a Poisson distribution, which encompasses a Gaussian distribution. The mixture ratio, $P_\mathrm{var}$, defines the probability of a source to be variable. The lower panel of Figure~\ref{fig_gvar_vs_gmag} shows our model predictions as a function of \texttt{gvar} and $G$-magnitude. Because we assume the distributions at different $G$ magnitudes are independent, our model can show artifacts at the lower bound of the data. However, it captures the complexity of the distribution, especially for magnitudes brighter than $G=15.0$.

We report $P_\mathrm{var}$ values in the $CaHK_\mathrm{syn}$ and Pristine DR1 photometric catalogues of Table~\ref{table:cahk}. An aggressive cut on the probability of being variable, $P_\mathrm{var}<0.3$, can be used to produce a sample of stars whose \Gaia\ magnitudes are unlikely to be affected by variability. This cut flags out 20\% of the 219.2 million stars with BP/RP information.

\section{The Pristine $(C\lowercase{a}HK,G,G_\mathrm{BP},G_\mathrm{RP})\rightarrow\FeH_\mathrm{phot}$ model}
\label{sec:model}
We use both the Pristine $CaHK$ magnitudes as well as the \Gaia-based synthetic $CaHK_\mathrm{syn}$ magnitudes to determine photometric metallicity estimates, $\FeH_\mathrm{phot}$. The method used here is in essence very similar to the method described in \citet{starkenburg17b}, but has been updated and adapted to use \Gaia\ broadband photometry ($G$, $G_\mathrm{BP}$, $G_\mathrm{RP}$) rather than SDSS broadband photometry.

\subsection{Spectroscopic training sample}
Our method is limited to FGK stars as, for hotter stars, the strength of the Ca H \& K absorption lines is too weak to be a reliable estimator of metallicity. At the other end of the temperature range, very cool M stars contain very prominent molecular bands that significantly drop the level of the pseudo-continuum in the relevant wavelength regime, making the measurement of the Ca H \& K absorption features challenging. We therefore restrain our analysis to the range $0.5<(G_\mathrm{BP}-G_\mathrm{RP})_0<1.5$, which approximately covers evolutionary stages between the upper main sequence, the turn-off, and the tip of the red giant branch for an old, (very) metal-poor stellar population. This color interval corresponds to a temperature range of approximately $3900 < T_{\rm{eff}} < 7000$\,K. 

Following \citet{starkenburg17b}, we use a training sample to provide a mapping from the de-reddened ($CaHK$, $G$, $G_\mathrm{BP}$, $G_\mathrm{RP}$) color space onto $\FeH_\mathrm{phot}$. A first, major, constituent of this sample are the SDSS/SEGUE stars \citep{smee13,yanny09} that are in the Pristine footprint. We restrict ourselves to those spectra with an average signal-to-noise ratio per pixel larger than 25 over the wavelength range 400--800\,nm. We further require that the SDSS pipeline provides a value for \logg, an adopted T$_{\rm eff} < 7000$\,K, a radial velocity with an uncertainty $< 10\kms$, and an adopted spectroscopic metallicity, $\FeH_\mathrm{adop}$, with an uncertainty $<0.2$\,dex. We also limit the sample to stars with nominal \texttt{n} flags, except for stars that also show the \texttt{g'} or \texttt{G} flag indicating a G-band feature. For outer halo science cases we are particularly interested in red giant stars and because these are less numerous in the sample, we complement our SDSS sample with APOGEE giants, stars that have APOGEE derived $\logg< 3.9$ and $\teff < 5800$\,K \citep{majewski17,wilson19,abdurrouf22}. The added APOGEE stars are furthermore restricted to spectra with a signal-to-noise ratio larger than 50, $\mathtt{FE\_H\_FLAG=0}$, and no $\mathtt{STARFLAG}$ raised. The APOGEE $\FeH$ values are shifted down by 0.1 dex to match the SDSS/SEGUE scale we have adopted. On the Pristine side, we restrict the sample to high-quality data: we use $\mathtt{CASU\_flag=-1}$ and $CaHK$ photometric uncertainties $\delta CaHK< 0.05$.

As mentioned earlier, the Pristine footprint has grown enormously since the publication of \citet{starkenburg17b}, from about 1,000\,deg$^{2}$ to $\sim6,500$\,deg$^{2}$ and, naturally, this commensurately enlarges the training sample, with a total of now over 66,000 stars that also have a \Gaia\ DR3 counterpart. Nevertheless, the range of EMP and UMP stars is not well sampled as these stars are exceedingly rare \citep[see, \eg][]{youakim17,bonifacio21,yong21}. We attempt to mitigate the sparsity of this important region of parameter space in two ways. First, we complement the SDSS sample with known VMP, EMP, and UMP stars. We add overlapping high-resolution observations of stars in the  Bo\"{o}tes I dwarf galaxy \citep{feltzing09,gilmore13,ishigaki14,frebel16}, as well as VMP stars from the third data release of the LAMOST survey, taken from \citet{li18b}. For this latter sample, we correct for spurious stars in the low effective temperature range, as described in \citet{sestito20}, but then use the most recent LAMOST DR8 catalogue. We add stars from the PASTEL sample \citep{soubiran16}, as also used by \citet{huang22}. Finally, we add targets followed up from the Pristine survey and confirmed to be VMP, EMP, or UMP stars \citep{youakim17,aguado19,venn20,kielty21,lardo21,lucchesi22}. We note that we use the same training sample, with $CaHK$ magnitudes derived from the Pristine survey rather than the synthetic $CaHK_\mathrm{syn}$, for the training of both catalogues presented in this paper. The distribution of these stars in the Kiel diagram is shown in the left-hand panel of Figure~\ref{fig_colorspace}.

\subsection{The Pristine color-color space}
\label{sec:ccspace}
To further sample the extremely low-metallicity regime, we calculate a grid of stellar spectra with $\FeH = -3.0$ and an $[\alpha/\textrm{Fe}]=+0.4$ using MARCS (Model Atmospheres in Radiative and Convective Scheme) stellar atmospheres and the Turbospectrum code \citep{alvarez98,gustafsson08,plez08}. To derive magnitudes, these spectra are integrated under the \Gaia\ EDR3 filter curves and that of the CaHK MegaCam filter. We also add a synthetic spectral grid calculated with the lowest metallicity atmospheres publicly available from MARCS and no metals at all in the step to create the spectra using Turbospectrum. 

\begin{figure*}
\begin{center}
\includegraphics[width=0.3\hsize]{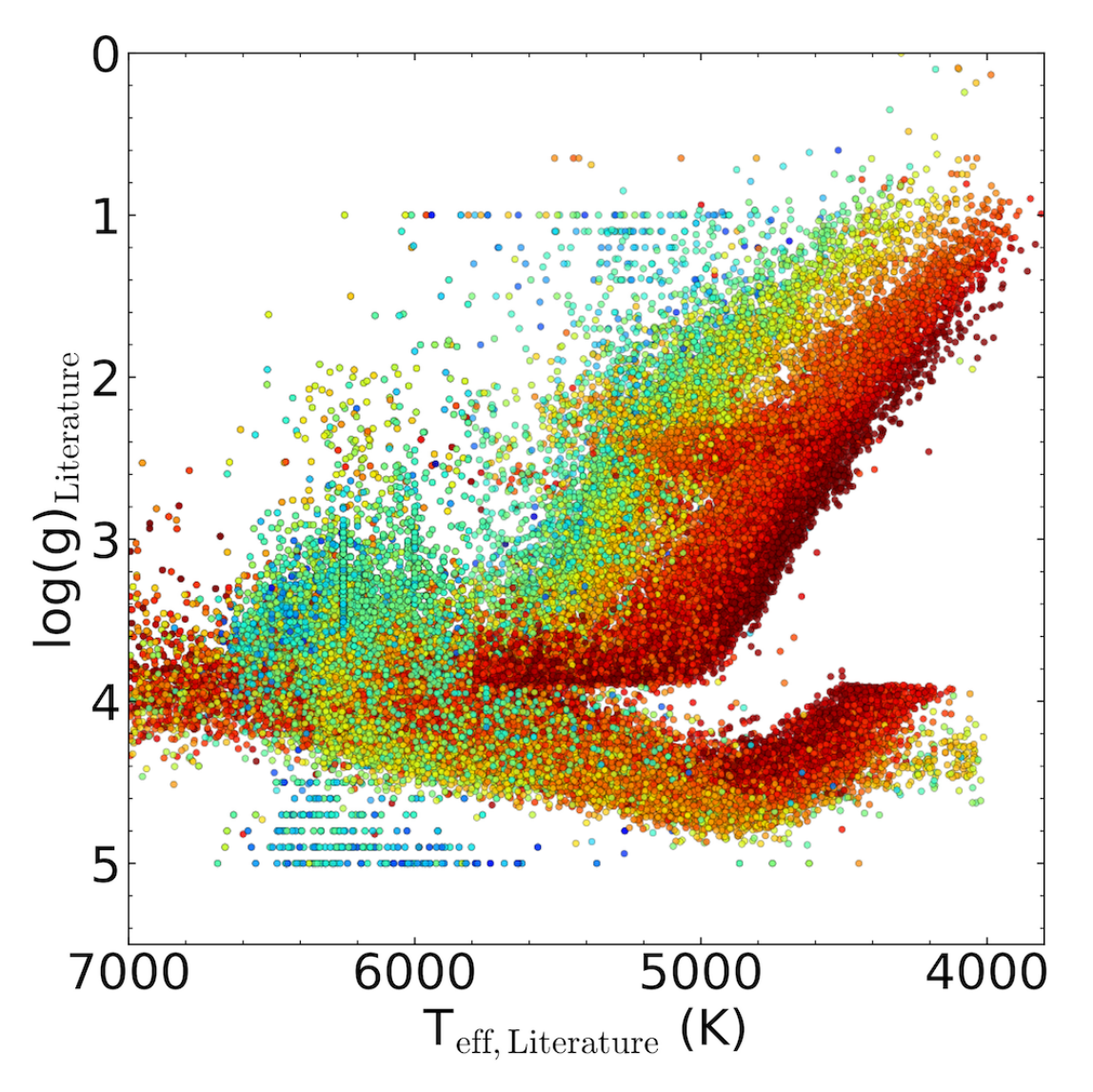}
\includegraphics[width=0.68\hsize]{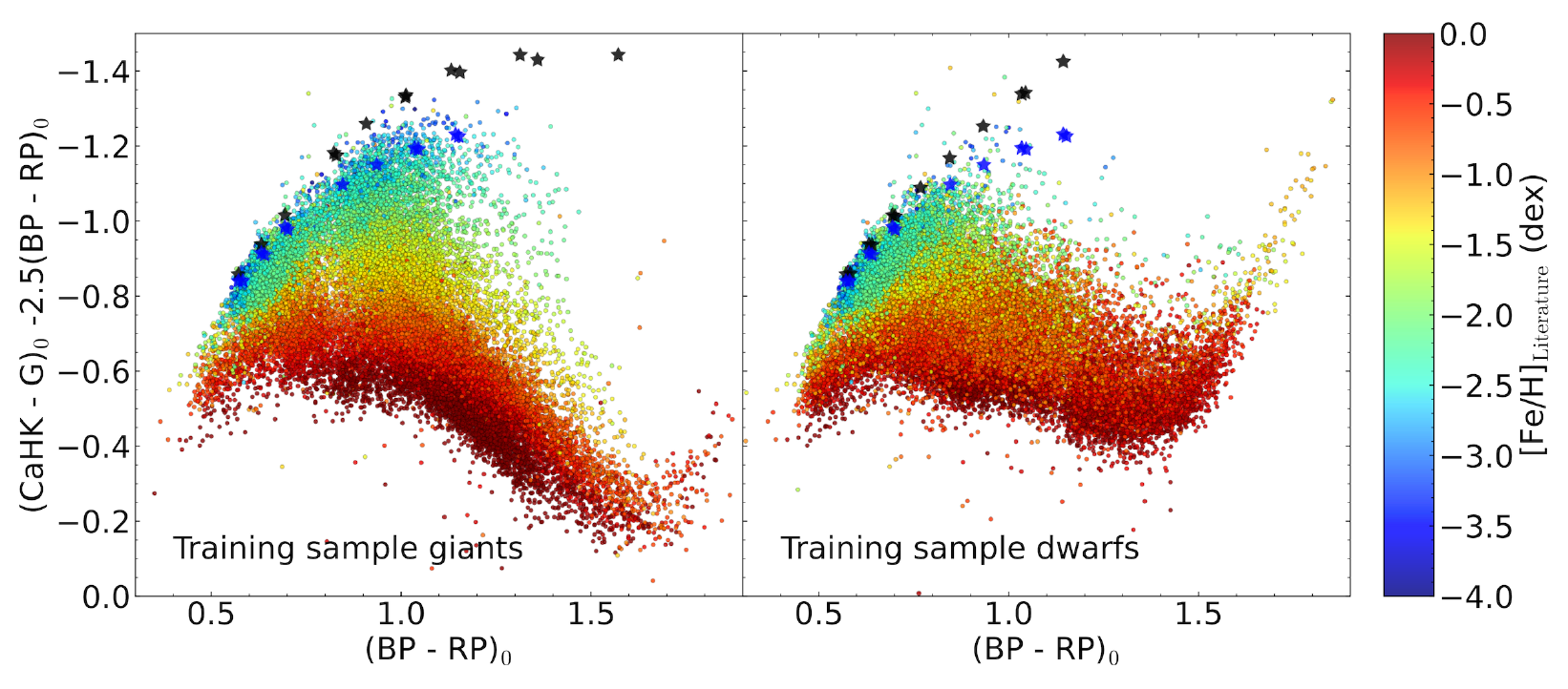}
\caption{Temperatures and surface gravities from our training sample (left) and their behavior in the Pristine color-color space for giants ($\logg< 3.9$; middle panel) and dwarf stars ($\logg> 3.9$ or $\teff>5,800$\,K; right-hand panel). The synthetic color resulting from synthetic spectra for stars with $\FeH = -3.0$ (blue) and (hypothetical) stars with no heavy metals (black) are overplotted with larger star symbols. Stars are plotted from the most metal-rich to the most metal-poor, so as to emphasize the metal-poor region. The sharp edge at $\logg\sim$ 3.9 in the left panel is due to the APOGEE sample that is limited to giants only (see text). The very low-metallicity stars from \citet{aguado19} have less physical $\logg$ determinations, showing up as horizontal lines in this panel. However, this does not significantly affect their metallicity, or location in the color-color space \citep[see][for a detailed discussion]{aguado19}. The magnitudes shown here are de-reddened following the procedure described in Section~\ref{sec:extinction}.\label{fig_colorspace}}
\end{center}
\end{figure*}

The color-color space used to determine $\FeH_\mathrm{phot}$ combines \Gaia\ $(G_\mathrm{BP}-G_\mathrm{RP})_0$ information as a temperature proxy and a combination of the \Gaia\ broad-band filters and the Pristine narrow-band $CaHK$, as illustrated in the two right-most panels of Figure~\ref{fig_colorspace}. From left to right, these include dwarf ($\logg> 3.9$ or $\teff>5,800$\,K) and giant stars ($\logg< 3.9$), respectively, and the two panels show both the training sample of observed stars (dots) and the synthetic results mentioned above (star symbols). Here, the magnitudes of stars are extinction corrected as described in Section~\ref{sec:extinction} by assuming the spectroscopic parameters of the stars. As expected, the stars arrange themselves in an orderly fashion so that, at fixed color, more metal-poor stars are higher up along the y axis of the figure as the $CaHK$ magnitude becomes brighter with the lower absorption of the smaller Ca H \& K spectral features of a more metal-poor star.

\subsection{The algorithm}
\label{sec:algorithm}
The algorithm that builds the model linking a location in color-color space to a metallicity value is described in detail in \citet{starkenburg17b} and \citet{fernandez-alvar21} and is available online\footnote{\url{https://github.com/ankearentsen/Pristine_code}}. In short, the color-color space is divided into a grid and, for each cell of the grid, the mean metallicity of the training sample is determined. The main difficulty in this determination is that the smaller number of metal-poor stars compared to more metal-rich stars means that any scattering from measurement uncertainties always scatters more metal-rich stars into the metal-poor regime rather than the other way around, biasing the metallicity upwards. We aim to mitigate this effect by sigma-clipping outliers beyond $2\sigma$ from the mean metallicity of a given pixel and, additionally, by imposing that, at a given $(G_\mathrm{BP}-G_\mathrm{RP})_0$ (the x axis in Figure~\ref{fig_colorspace}), the metallicity decreases monotonically with decreasing $[(CaHK-G)_0-2.5(G_\mathrm{BP}-G_\mathrm{RP})_0]$ (the y axis in Figure~\ref{fig_colorspace}) for $\FeH< -1.0$. In other words, stars that show less $CaHK$ absorption for similar broadband colors are only allowed to get a more metal-poor value. The synthetic colors calculated from the modeled spectra of EMP stars are used to anchor the model in the lowest metallicity regime. Grid points closer to the ``no-metals'' line than to the synthetic $\FeH = -3.0$ models are set to $\FeH_\mathrm{phot} = -4.0$. In addition, we account for possible systematics in the color of these spectra by setting cells to $\FeH = -4.0$ down to 0.2~magnitudes below the ``no-metals'' line (i.e, 0.2~magnitudes above the black symbols in Figure~\ref{fig_colorspace})\footnote{A clear consequence of this choice is that the model cannot return a metallicity $\FeH_\mathrm{phot} < -4.0$, but this regime is so poorly sampled that any model would only be a wild extrapolation beyond this limit.}. Finally, the resulting grid is smoothed with a kernel of 2-pixel dispersion to remove outlying grid points caused by a few metal-rich stars scattered into a metal-poor region. We assign no metallicity value to stars outside of the $0.5<(G_\mathrm{BP} - G_\mathrm{RP})_0<1.5$ color range, more than 0.2~magnitudes above the ``no-metals'' line, or more than 0.1~magnitudes below a polynomial fit to the solar metallicity stars (the last two restrictions are near the top and bottom of the color-color plots of Figure~\ref{fig_colorspace}, respectively).

As mentioned above, and since the extinction correction of the magnitudes used by the model depends on the stellar parameters of a star, including its metallicity, we cannot simply de-redden the magnitudes and push them through the photometric metallicity model as we did before when we relied on SDSS broad-band magnitudes. We therefore determine both the extinction coefficients and the metallicity of the stars iteratively, as mentioned at the end of Section~\ref{sec:extinction}.

This framework is used to build both a model for dwarfs ($\logg> 3.9$ or $\teff>5,800$\,K) and for giants ($\logg<3.9$). As the Pristine survey does not provide distance information for the observed stars, one does not a priori know whether a given object is a dwarf or a giant. However, since the Pristine survey and model are mainly focussed on metal-poor stars that often happen to be giants in the MW halo, we favor the giant metallicities as our generic output, $\FeH_\mathrm{phot}$. We nevertheless provide the photometric metallicities derived using the dwarf model, $\FeH_\mathrm{phot,dw}$, which should be preferred for stars that are specifically known to be dwarf stars. When the evolutionary stage of a star is known (for instance via parallax-based distances built from \Gaia; \eg \citealt{bailer-jones21}), one should of course use the most appropriate photometric metallicity values.

For a given star of de-reddened magnitudes ($CaHK_0$, $G_0$, $G_\mathrm{BP,0}$, $G_\mathrm{RP,0}$) and associated Gaussian uncertainties (not including uncertainties in the de-reddening), the probability distribution function (PDF) on any photometric metallicity estimate is determined through 100 Monte Carlo samplings of the magnitude uncertainties. $\FeH_\mathrm{phot}$ and $\FeH_\mathrm{phot,dw}$ are redetermined for each sampling of the uncertainties. Even though the uncertainties on the magnitudes are assumed to be symmetrical, the resulting PDFs on the photometric metallicities can be very asymmetrical as the narrow-band color does not vary linearly with decreasing metallicity. This can be appreciated in Figure \ref{fig_colorspace}, in which two stars with the same metallicity difference are much closer together if they are metal-poor. For this reason, we do not quantify the PDF with a simple dispersion but instead provide the values of the 16th, 50th, and 84th percentiles of a given Monte Carlo PDF. While we usually use the metallicities calculated directly from the favored values of the magnitudes, the median values (50th percentile) are also systematically reported in the catalogues. Additionally, we report \texttt{mcfrac}, the fraction of Monte Carlo samples for each star that falls within the limits of the metallicity grid.

\subsection{Applicability of the model}
Given the main science goals of the Pristine survey, which focus on metal-poor stars, most of the effort on the model is placed in the region $\FeH_\mathrm{phot}<-1.0$ to avoid being driven and dominated by the much more numerous stars above this metallicity cut at the expense of the metal-poor region. A direct consequence is that, while we do provide the photometric metallicities of stars with $\FeH_\mathrm{phot}>-1.0$ and the Pristine model provides reasonable results in this range (see the next section), these metallicities may be biased in unexpected ways (\eg from the increased impact of the gravity on the depth of the Ca H \& K lines, changes in the mean [$\alpha$/Fe] abundance, or non-LTE and 3D effects on the line strengths). Alternative \FeH\ estimates based on the \Gaia\ BP/RP information and trained, in particular, on large spectroscopic surveys (APOGEE, LAMOST) that are vastly dominated by stars with $\FeH>-1.0$ may be more appropriate to explore this regime \citep[\eg][]{andrae23,zhang23}. We stress that our results are in particular optimized to study the low-metallicity regime ($\FeH< -1.0$) of FGK stars and that it performs best for stars with $4,000 \lta \teff \lta 6,000$\,K.
 
\section{The photometric metallicity catalogues}
\label{sec:met_catalogues}
\begin{table*}
\begin{center}
  \caption{Description of the columns of the Pristine-\Gaia\ synthetic metallicity catalogue. \label{table:FeHphot_CaHKsyn}}
  \begin{tabular}{@{}lp{10cm}ll@{}}
Column & Description & Unit & Type\\
\hline
\texttt{source\_id} & \Gaia\ DR3 \texttt{source\_id} & --- & \texttt{longint}\\
\texttt{RA} & \Gaia\ DR3 right ascension & ICRS (J2016) & \texttt{float}\\
\texttt{Dec} & \Gaia\ DR3 declination & ICRS (J2016) & \texttt{float}\\
\texttt{E(B-V)} & \citet{schlegel98} extinction value & mag & \texttt{float} \\
\texttt{G\_0} & de-reddened \Gaia\ $G$ magnitude & mag & \texttt{float}\\
\texttt{d\_G} & $\delta G$, uncertainty on the \Gaia\ $G$ magnitude & mag & \texttt{float}\\
\texttt{BP\_0} & de-reddened \Gaia\ $G_{BP}$ magnitude & mag & \texttt{float}\\
\texttt{d\_BP} & $\delta G_{BP}$, uncertainty on the \Gaia\ $G_{BP}$ magnitude & mag & \texttt{float}\\
\texttt{RP\_0} & de-reddened \Gaia\ $G_{RP}$ magnitude & mag & \texttt{float}\\
\texttt{d\_RP} & $\delta G_{RP}$, uncertainty on the \Gaia\ $G_{RP}$ magnitude & mag & \texttt{float}\\
\texttt{CaHK\_0} & de-reddened Pristine-like $CaHK_\mathrm{syn}$ magnitude & mag & \texttt{float}\\
\texttt{d\_CaHK\_syn} & $\delta CaHK_\mathrm{syn}$, uncertainty on the Pristine-like $CaHK_\mathrm{syn}$ magnitude & mag & \texttt{float}\\
\hline
\texttt{FeH\_CaHKsyn} & Photometric metallicity $\FeH_\mathrm{phot}$ based on the $CaHK_\mathrm{syn}$ magnitude and using the (favored) giant model & dex & \texttt{float}\\
\texttt{FeH\_CaHKsyn\_16th} & Lower bound of the 68\% confidence interval on \texttt{FeH\_CaHKsyn} based on the Monte Carlo sampling of the magnitude uncertainties & dex & \texttt{float}\\
\texttt{FeH\_CaHKsyn\_50th} & Median value of \texttt{FeH\_CaHKsyn} based on the Monte Carlo sampling of the magnitude uncertainties & dex & \texttt{float}\\
\texttt{FeH\_CaHKsyn\_84th} & Upper bound of the 68\% confidence interval on \texttt{FeH\_CaHKsyn} based on the Monte Carlo sampling of the magnitude uncertainties & dex & \texttt{float}\\
\texttt{mcfrac\_CaHKsyn} & Fraction of the Monte Carlo samples of the magnitude uncertainties that fall outside the model color-color space & --- & \texttt{float}\\
\texttt{G\_0\_dw} & Same as \texttt{G\_0} but for the dwarf model & mag & \texttt{float}\\
\texttt{BP\_0\_dw} & Same as \texttt{BP\_0} but for the dwarf model & mag & \texttt{float}\\
\texttt{RP\_0\_dw} & Same as \texttt{RP\_0} but for the dwarf model & mag & \texttt{float}\\
\texttt{CaHK\_0\_dw} & de-reddened Pristine-like $CaHK_\mathrm{syn}$ magnitude & mag & \texttt{float}\\
\texttt{FeH\_CaHKsyn\_dw} & Same as \texttt{FeH\_CaHKsyn\_dw} but for the dwarf model & dex & \texttt{float}\\
\texttt{FeH\_CaHKsyn\_dw\_16th} & Same as \texttt{FeH\_CaHKsyn\_dw\_16th} but for the dwarf model & dex & \texttt{float}\\
\texttt{FeH\_CaHKsyn\_dw\_50th} & Same as \texttt{FeH\_CaHKsyn\_dw\_50th} but for the dwarf model & dex & \texttt{float}\\
\texttt{FeH\_CaHKsyn\_dw\_84th} & Same as \texttt{FeH\_CaHKsyn\_dw\_84th} but for the dwarf model & dex & \texttt{float}\\
\texttt{mcfrac\_CaHKsyn\_dw} & Same as \texttt{mcfrac\_CaHKsyn} but for the dwarf model & --- & \texttt{float}\\
\hline
\texttt{Pvar} & Probability for the source to be variable, as defined in Section~\ref{sec:variability} & --- & \texttt{float}\\
\texttt{RUWE} & \Gaia\ DR3 Renormalised Unit Weight Error & --- & \texttt{float}\\
\texttt{Cstar} & \Gaia\ DR3 corrected flux excess, $C^*$, as defined in equation~6 of \citet{riello21} & mag & \texttt{float}\\
\texttt{Cstar\_1sigma} & Normalized standard deviation of $C^*$ for the $G$ magnitude of this source, as defined in equation~18 of \citet{riello21} & --- & \texttt{float}\\
\end{tabular}
\end{center}
\end{table*}

\begin{table*}
\begin{center}
  \caption{Description of the columns of the Pristine DR1 metallicity catalogue. \label{table:FeHphot_Pr}}
  \begin{tabular}{@{}lp{10cm}ll@{}}
Column & Description & Unit & Type\\
\hline
\texttt{source\_id} & \Gaia\ DR3 \texttt{source\_id} & --- & \texttt{longint}\\
\texttt{RA} & \Gaia\ DR3 right ascension & ICRS (J2016) & \texttt{float}\\
\texttt{Dec} & \Gaia\ DR3 declination & ICRS (J2016) & \texttt{float}\\
\texttt{E(B-V)} & \citet{schlegel98} extinction value & mag & \texttt{float} \\
\texttt{merged\_CASU\_flag} & morphology and quality flag\footnotemark[10] & --- & \texttt{int}\\
\texttt{G\_0} & de-reddened \Gaia\ $G$ magnitude & mag & \texttt{float}\\
\texttt{d\_G} & $\delta G$, uncertainty on the \Gaia\ $G$ magnitude & mag & \texttt{float}\\
\texttt{BP\_0} & de-reddened \Gaia\ $G_{BP}$ magnitude & mag & \texttt{float}\\
\texttt{d\_BP} & $\delta G_{BP}$, uncertainty on the \Gaia\ $G_{BP}$ magnitude & mag & \texttt{float}\\
\texttt{RP\_0} & de-reddened \Gaia\ $G_{RP}$ magnitude & mag & \texttt{float}\\
\texttt{d\_RP} & $\delta G_{RP}$, uncertainty on the \Gaia\ $G_{RP}$ magnitude & mag & \texttt{float}\\
\texttt{CaHK\_0} & de-reddened Pristine $CaHK$ magnitude & mag & \texttt{float}\\
\texttt{d\_CaHK} & $\delta CaHK$, uncertainty on the Pristine $CaHK$ magnitude & mag & \texttt{float}\\
\hline
\texttt{FeH\_Pristine} & Photometric metallicity $\FeH_\mathrm{phot}$ based on the Pristine $CaHK$ magnitudes and using the (favored) giant model & dex & \texttt{float}\\
\texttt{FeH\_Pristine\_16th} & Lower bound of the 68\% confidence interval on \texttt{FeH\_Pristine} based on the Monte Carlo sampling of the magnitude uncertainties & dex & \texttt{float}\\
\texttt{FeH\_Pristine\_50th} & Median value of \texttt{FeH\_Pristine} based on the Monte Carlo sampling of the magnitude uncertainties & dex & \texttt{float}\\
\texttt{FeH\_Pristine\_84th} & Upper bound of the 68\% confidence interval on \texttt{FeH\_Pristine} based on the Monte Carlo sampling of the magnitude uncertainties & dex & \texttt{float}\\
\texttt{mcfrac\_Pristine} & Fraction of the Monte Carlo samples of the magnitude uncertainties that fall outside the model color-color space & --- & \texttt{float}\\
\texttt{G\_0\_dw} & Same as \texttt{G\_0} but for the dwarf model & mag & \texttt{float}\\
\texttt{BP\_0\_dw} & Same as \texttt{BP\_0} but for the dwarf model & mag & \texttt{float}\\
\texttt{RP\_0\_dw} & Same as \texttt{RP\_0} but for the dwarf model & mag & \texttt{float}\\
\texttt{CaHK\_0\_dw} & de-reddened Pristine-like $CaHK_\mathrm{syn}$ magnitude & mag & \texttt{float}\\
\texttt{FeH\_Pristine\_dw} & Same as \texttt{FeH\_Pristine\_dw} but for the dwarf model & dex & \texttt{float}\\
\texttt{FeH\_Pristine\_dw\_16th} & Same as \texttt{FeH\_Pristine\_dw\_16th} but for the dwarf model & dex & \texttt{float}\\
\texttt{FeH\_Pristine\_dw\_50th} & Same as \texttt{FeH\_Pristine\_dw\_50th} but for the dwarf model & dex & \texttt{float}\\
\texttt{FeH\_Pristine\_dw\_84th} & Same as \texttt{FeH\_Pristine\_dw\_84th} but for the dwarf model & dex & \texttt{float}\\
\texttt{mcfrac\_Pristine\_dw} & Same as \texttt{mcfrac\_Pristine} but for the dwarf model & --- & \texttt{float}\\
\hline
\texttt{Pvar} & Probability for the source to be variable, as defined in Section~\ref{sec:variability} & --- & \texttt{float}\\
\texttt{RUWE} & \Gaia\ DR3 Renormalised Unit Weight Error & --- & \texttt{float}\\
\texttt{Cstar} & \Gaia\ DR3 corrected flux excess, $C^*$, as defined in equation~6 of \citet{riello21} & mag & \texttt{float}\\
\texttt{Cstar\_1sigma} & Normalized standard deviation of $C^*$ for the $G$ magnitude of this source, as defined in equation~18 of \citet{riello21} & --- & \texttt{float}\\
\end{tabular}
\end{center}
\end{table*}

\begin{figure*}
\begin{center}
\includegraphics[width=\hsize]{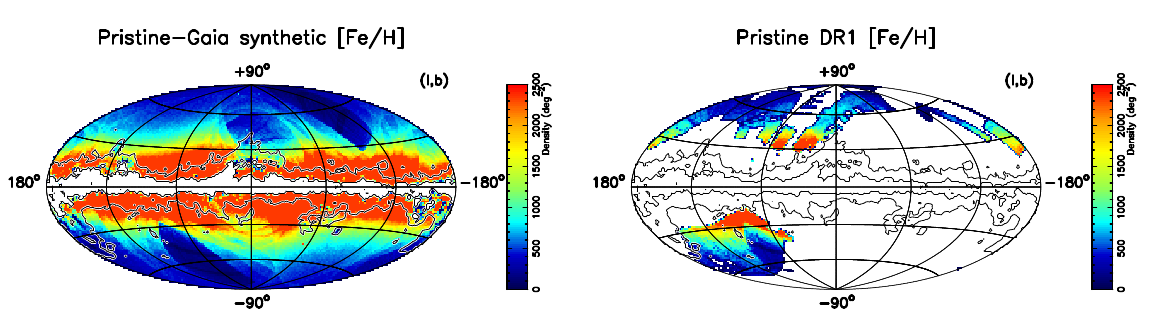}
\caption{Density of sources with photometric metallicities in the Pristine-\Gaia\ synthetic catalogue (left-hand panel) and the DR1 catalogue (right-hand panel). The black contours are lines with $E(B-V)=0.15$ and 0.5. By construction, regions with extinction $E(B-V)>0.5$ are removed from the photometric metallicity catalogues and produce the white regions in the left-hand map. The impact of the \Gaia\ scanning law on the S/N of the BP/RP information and, consequently, on that of the $CaHK_\mathrm{syn}$ magnitudes, is clearly visible; it is responsible for most of the irregular features in this map. The map of Pristine metallicities is more limited on the sky but also denser, owing to the higher S/N Pristine data: some stars with BP/RP information do not have high enough S/N to make it through the enforced $\delta CaHK_\mathrm{syn}<0.1$ cut but have a high-enough S/N in Pristine to generate a Pristine metallicity.\label{fig:maps}}
\end{center}
\end{figure*}

Both photometric catalogues presented in Sections~\ref{sec:CaHKsyn} and~\ref{sec:Pristine} are pushed through the model described in Section~\ref{sec:model} to generate the Pristine-\Gaia\ synthetic and the Pristine DR1 photometric metallicity catalogues, respectively. The content of the resulting catalogues are presented in Tables~\ref{table:FeHphot_CaHKsyn} and~\ref{table:FeHphot_Pr}. The sky density of sources in both photometric metallicity catalogues is displayed in Figure~\ref{fig:maps}.

\subsection{Building the catalogues}
Both datasets are processed in the exact same way and, with Pristine now calibrated onto the $CaHK_\mathrm{syn}$ photometric system, are expected to yield complementary catalogues of $\FeH_\mathrm{phot}$ that are on the same scale. In addition to the selection in the color-color space of Figure~\ref{fig_colorspace} that was discussed in the previous section\footnote{These are: $0.5<(G_\mathrm{BP} - G_\mathrm{RP})_0<1.5$, $[(CaHK-G)_0-2.5(G_\mathrm{BP}-G_\mathrm{RP})]_0<-0.6$, and, at most, 0.2\,mag above the ``no-metal'' line in the Figure.}, we further implement two cuts to the photometric catalogues before pushing them through the model. A cut on the $CaHK$ photometric uncertainties, $\delta CaHK_\mathrm{syn}<0.1$ or $\delta CaHK<0.1$ and a cut on the extinction, $E(B-V)<0.5$, remove noisy photometric measurements and stars that may be plagued by systematics that stem from large extinction corrections. We note that the uncertainties on the photometric metallicities are often asymmetric and larger on the metal-poor side. The catalogues list the photometric metallicities calculated for the favored magnitude values (which we favor), along with the median metallicity obtained from sampling the magnitude uncertainties and the limits of the central 68\% confidence interval, $\FeH_\mathrm{phot,16th}$ and $\FeH_\mathrm{phot,84th}$. In what follows, we approximate the photometric metallicity uncertainty, $\delta\FeH_\mathrm{phot} = 0.5(\FeH_\mathrm{phot,84th}-\FeH_\mathrm{phot,16th}$), as half the range of metallicity spanned by the central 68\% confidence interval.

\begin{figure*}
\begin{center}
\includegraphics[width=\hsize]{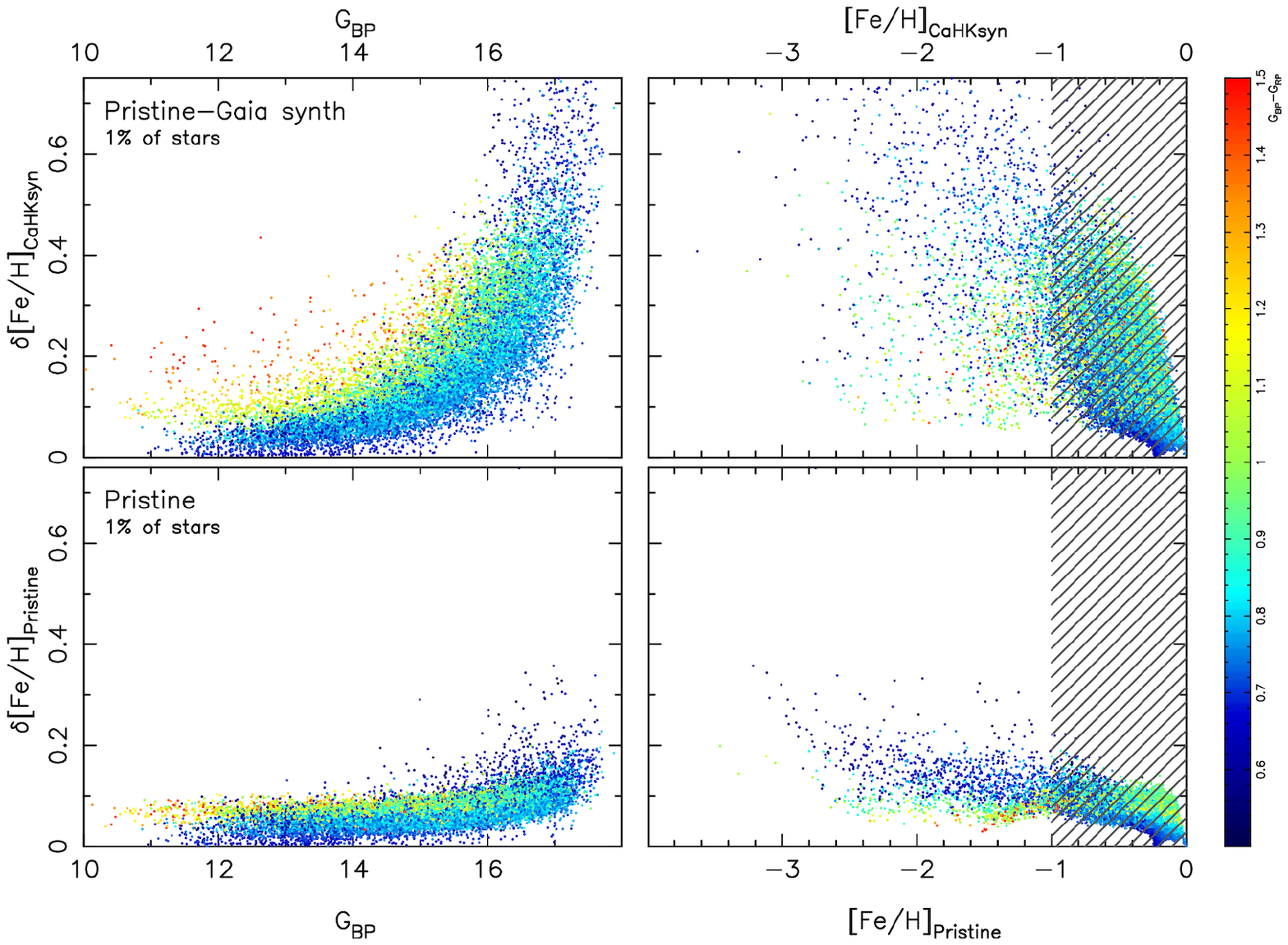}
\caption{Mean metallicity uncertainties as a function of the magnitude of a star or its photometric metallicities for a random 1\% of stars in the Pristine-\Gaia\ synthetic catalogue (top) and the Pristine DR1 catalogue (bottom). Dots are color-coded by the $G_\mathrm{BP}-G_\mathrm{RP}$ color of a star. The dashed regions corresponds to the regime for which metallicities might be hampered by systematics due to a stronger impact of the gravity of a star on its derived metallicity (see Section~\ref{sec:model}). The uncertainties are a complex function of the magnitude of a star, its color, and its metallicity because of the nontrivial mapping of metallicities on the color-color space shown in Figure~\ref{fig_colorspace}. The metallicities based on Pristine magnitudes show much lower uncertainties owing to their significantly higher S/N.\label{fig:d_FeH}}
\end{center}
\end{figure*}

Figure~\ref{fig:d_FeH} shows the typical uncertainties attached to the photometric metallicities in the two catalogues. As expected, the metallicities based on the Pristine $CaHK$ data have significantly smaller uncertainties than those determined from the \Gaia-based $CaHK_\mathrm{syn}$ magnitudes. The photometric metallicity uncertainties for the latter can be quite significant, especially in the lowest metallicity regimes and/or for bluer stars, or for magnitudes fainter than $G_\mathrm{BP}\sim16.0$. The bottom-right panel of the figure, which shows the Pristine photometric metallicity uncertainties as a function of the Pristine photometric metallicities, is quite structured and much more so than its equivalent for the Pristine-\Gaia\ synthetic photometric metallicities. This effect is due to the complex translation of magnitude uncertainties into photometric metallicity uncertainties. In the case of the Pristine-\Gaia\ synthetic metallicities, the much larger uncertainties on $\delta CaHK_\mathrm{syn}$ means that these dominate the model grid effects.

In addition to the complex impact of color on the derived photometric metallicities and associated uncertainties (at fixed magnitude, redder stars have a lower S/N in the Ca H \& K region but, at the same time, a fixed difference in photometric metallicities corresponds to a larger difference in $CaHK$), these uncertainties also scale with the photometric metallicity since iso-metallicity lines are closer together in the color-color space of Figure~\ref{fig_colorspace} for decreasing metallicities. Any sharp quality cut on the photometric metallicity uncertainties, therefore, is more detrimental to more metal-poor stars, at the risk of removing the most metal-poor stars. On the other hand, not including any quality cut on the photometric metallicity uncertainties would include a large number of the much more numerous metal-rich stars, scattered into the realm of the rare stars of lowest metallicities. This effect is of course worse for the lower quality Pristine-\Gaia\ synthetic catalogue. Keeping these complications in mind and depending on the science goal, we alternatively use quality cuts of $\delta\FeH_\mathrm{phot}<0.3$\,dex and $\delta\FeH_\mathrm{phot}<0.5$\,dex in what follows.

\subsection{A comparison of the two Pristine catalogues}
\begin{figure*}
\begin{center}
\includegraphics[width=\hsize]{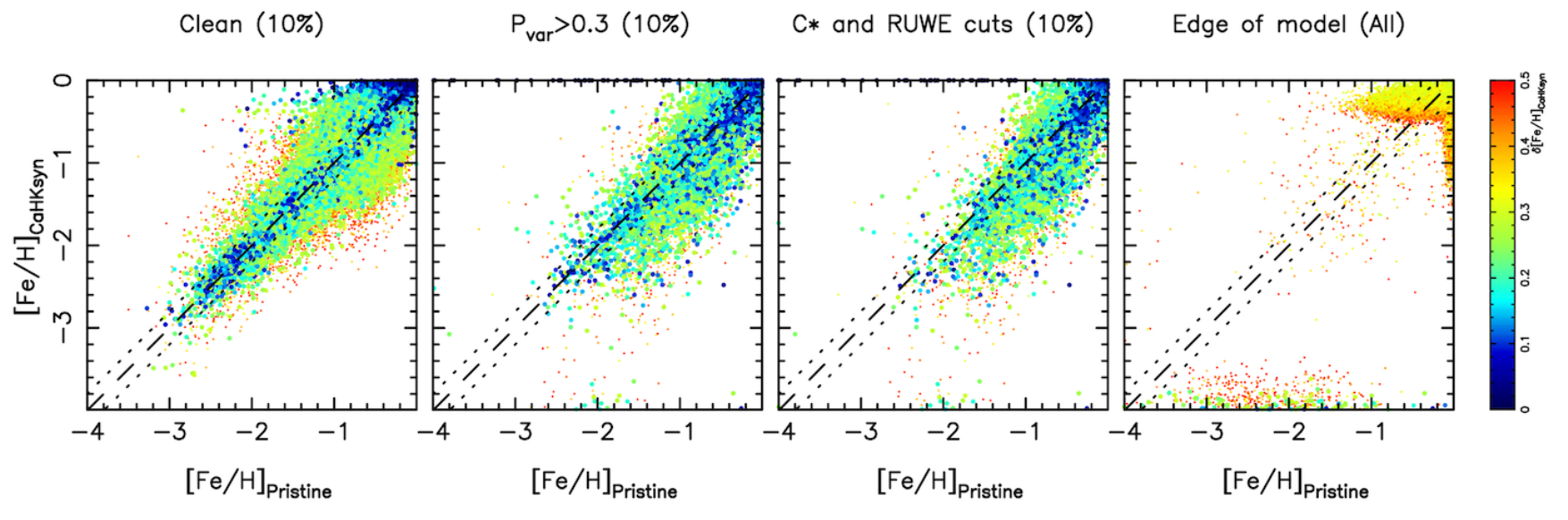}
\caption{Comparison of the photometric metallicities in the Pristine-\Gaia\ synthetic and the Pristine DR1 catalogues for a random 10\% of the overlap between the two catalogues. Points are colored by the uncertainties on the photometric metallicities from the Pristine-\Gaia\ synthetic catalogue and small and large dots correspond to stars with a loose and tight quality cut on the metallicity uncertainties, respectively ($\delta\FeH_\mathrm{phot}<0.5$ and $0.3$). The dashed lines correspond to the one-to-one lines and the dotted lines show offsets of 0.2\,dex from them. As expected, the metallicities derived from the two sets of magnitudes are consistent since they are on the same photometric system. The next three panels show the impact of quality cuts to clean the sample by removing stars with a probability of being variable ($P_\mathrm{var}>0.3$; second panel), flagged as potentially problematic in \Gaia\ (RUWE$>1.4$ and $|C^*| > \sigma_{C^*}$; third panel), or at the edge of the grid (fourth panel; here we shown all the catalogue stars that follow this criterion). The different quality cuts remove stars that do not behave as expected.\label{comparison}}
\end{center}
\end{figure*}

To better understand the limitations of the more noisy Pristine-\Gaia\ synthetic catalogue, we can use stars that have both Pristine-\Gaia\ synthetic and Pristine metallicities. Figure~\ref{comparison} presents this comparison for a randomly selected subsample of stars in common between the two photometric metallicity catalogues. As expected because the Pristine $CaHK$ magnitudes are now calibrated using the $CaHK_\mathrm{syn}$ magnitudes, the metallicities calculated from the two sets of magnitudes are consistent with each other and follow the one-to-one line in the left-hand panel of the figure. The uncertainties on the Pristine-\Gaia\ synthetic metallicities, even though large, do not show any obvious bias when restricting the sample to stars with $\delta\FeH_\mathrm{phot}<0.3$\,dex (large dots) and $\delta\FeH_\mathrm{phot}<0.5$\,dex (small dots). In this panel, we have applied different quality cuts and we show in the other three panels the stars that were removed by those.

The second panel of the figure shows stars that appear variable according to our model (Section~\ref{sec:variability}; $P_\mathrm{var}>0.3$). As expected, variability negatively affects both metallicity catalogues, with a larger distribution than in the first panel. The impact of variability is present in both catalogues, but the effect is likely worse for the Pristine catalogue than for the Pristine-\Gaia\ synthetic catalogue because the Pristine $CaHK$ photometry of a star was taken at a different time from the broad-band \Gaia\ magnitudes. In the past, this has led to contamination of the Pristine EMP-candidate sample when variability information was not available \citep{lombardo23}. We strongly encourage applying a similar cut on $P_\mathrm{var}$ when using the catalogues.

The third panel of Figure~\ref{comparison} shows the impact of cleaning the data with two cuts based on the \Gaia\ diagnostics: $\textrm{RUWE}>1.4$ and $|C^*| > 1\times \sigma_{C^*}$, as defined by \citet[equations 6 and 18]{riello21}. There is a significant overlap between stars removed with these cuts and the variability cut.

Finally, the fourth panel of the Figure shows the impact of removing stars that are at the edge of the metallicity model: $\FeH_\mathrm{phot}=-4.0$ or 0.0, $\FeH_\mathrm{phot,down}=-4.0$ or 0.0, or a large fraction of Monte Carlo samplings are at the edge of the grid ($\texttt{mcfrac}<0.8$). In particular at the most metal-poor edge of the grid, this removes a small number of objects from the Pristine-\Gaia\ synthetic catalogue that have random metallicities compared to the higher S/N Pristine DR1 metallicities. Of course, this latter cut may have the consequence of removing true UMP stars that fall very close to the no-metals line in Figure~\ref{fig_colorspace}. In the right-most panel of Figure~\ref{comparison}, there are clearly a small number of objects removed by this cut that are EMP candidates in the higher quality Pristine catalogue. It should therefore be used with caution when searching for or working on such extreme stars.

\subsection{Best practices when using the photometric metallicity catalogues}
\label{sec:qualcuts}

Given the checks shown in the previous subsection and, in particular, in Figure~\ref{comparison}, we recommend a list of quality cuts when using the two catalogues. Except if otherwise specified, these cuts are implemented throughout the rest of the paper:

\begin{enumerate}
\item $\FeH_\mathrm{phot}<0.0$ and $\FeH_\mathrm{phot}>-4.0$, as well as a fraction of the Monte Carlo samplings of the magnitude uncertainties that are within the color range of the metallicity model more than 80\% of the time ($\texttt{mcfrac}>0.8$). These cuts remove stars located at the edge of the metallicity grid\footnote{However, bear in mind that this restriction (as well as the $\FeH_\mathrm{phot}>-4.0$ restriction above) may in fact remove some true UMP stars that would be located near the ``no-metals'' line in the color-color space, or scattered upwards of it by noise in the photometry.}.
\item Photometric uncertainties $\delta\FeH_\mathrm{phot}<0.5$\,dex or, depending on the science case, $\delta\FeH_\mathrm{phot}<0.3$\,dex\footnote{We note, however, that the confidence intervals are in fact asymmetrical and that, in some cases, this cut may be too crude.}.
\item $P_\mathrm{var}<0.3$ to remove stars that are potentially variable\footnote{This cut could be loosened at the bright end, for which the S/N of the \Gaia\ magnitudes is so high that even tiny variations of the magnitude of a star would lead to a high $P_\mathrm{var}$ while not significantly affecting $\FeH_\mathrm{phot}$.}.
\item \Gaia\ quality cuts $\textrm{RUWE}<1.4$\footnote{\citet{yuan15} has shown that unresolved binaries can lead to stars in binary systems being assigned a more metal-poor photometric metallicity than their true metallicity. \citet{xu22} recommends using the stricter $\textrm{RUWE}<1.1$ for analyses where this is expected to be an issue.} and $|C^*| < 1\times \sigma_{C^*}$, with $C^*$ the corrected flux excess and $\sigma_{C^*}$ the width of its distribution at a given magnitude, as defined in equations~6 and 18 of \citet{riello21}\footnote{This $C^*$ restriction as stated here is quite strict and likely removes some good stars. For a more inclusive selection, one might opt for 2 or $3\sigma_{C^*}$ rather than $1\sigma_{C^*}$.}.
\item \texttt{CASU\_flag=-1} or \texttt{CASU\_flag=-2} for stars observed by Pristine to ensure that the objects were not identified as extended objects by the photometric pipeline (likely because of the difficulty to build a good PSF; see sub-section~\ref{sec:Pristine_phot}) and that potentially different individual detections of the source did not have conflicting \texttt{CASU\_flags} when they were merged (see Sub-section~\ref{sec:merging}).
\end{enumerate}

Finally, we reiterate that most of the effort in the metallicity model was placed in the low metallicity regime ($\FeH_\mathrm{phot}<-1.0$). While we show below that the photometric metallicities above this limit still behave well, studies specifically focussing on $\FeH_\mathrm{phot}>-1.0$ may wish to use other catalogues that are more specifically designed for this regime.

\begin{table*}
\begin{center}
\footnotesize
  \caption{Numbers of candidate stars below a given metallicity limit in the two catalogues. \label{table:numbers}}
  \begin{tabular}{@{}l|rrrrr@{}}
  \hline
Quality cut & Number of stars & $\FeH_\mathrm{phot}<-1.0$ & $\FeH_\mathrm{phot}<-2.0$ & $\FeH_\mathrm{phot}<-3.0$ \\
     \hline
Pristine-\Gaia\ synthetic catalogue\\
     \hline
$E(B-V)<0.5$ \& $\delta\FeH_\mathrm{phot}<0.5$ & 31,605,960 & 1,927,650 & 177,488 & 6,331\\
$E(B-V)<0.3$ \& $\delta\FeH_\mathrm{phot}<0.5$ & 27,882,616 & 1,774,800 & 166,143 & 5,770\\
$E(B-V)<0.3$ \& $\delta\FeH_\mathrm{phot}<0.3$ & 23,254,626 & 829,320 & 84,384 & 1,585\\
     \hline
Pristine DR1 catalogue\\
     \hline
$E(B-V)<0.5$ \& $\delta\FeH_\mathrm{phot}<0.5$ & 3,812,595 & 480,602 & 57,786 & 1,801\\
$E(B-V)<0.3$ \& $\delta\FeH_\mathrm{phot}<0.5$ & 3,810,843 & 480,263 & 57,752 & 1,798\\
$E(B-V)<0.3$ \& $\delta\FeH_\mathrm{phot}<0.3$ & 3,783,243 & 469,066 & 53,636 & 795\\
\end{tabular}
\end{center}
\tablefoot{All quality cuts listed in sub-section~\ref{sec:qualcuts} are applied to the catalogues to yield these numbers. Additional cuts on the $E(B-V)$ extinction values and the photometric uncertainties $\delta\FeH_\mathrm{phot}$ are listed in the first column. }
\end{table*}

Applying these various quality cuts results in 31,605,960 stars with reliable photometric metallicities in the Pristine-\Gaia\ synthetic catalogue and 1,927,650 of these correspond to metal-poor star candidates ($\FeH_\mathrm{phot}<-1.0$). Adopting various cuts on extinction ($E(B-V)<0.5$ or, more strictly, $<0.3$) and $\delta\FeH_\mathrm{phot}$ ($<0.5$ or, more robustly, $<0.3$) yields the numbers listed in Table~\ref{table:numbers}. In particular, restricting to a high-quality sample, the catalogue lists 84,384 VMP star candidates and 1,585 EMP star candidates. From our own follow-up of Pristine selected EMP star candidates, we expect that most of those stars truly have $\FeH<-2.5$ and that $>20$\% of EMP star candidates indeed have $\FeH<-3.0$ when followed up with spectroscopy (\citealt{youakim17,aguado19} and see also \citealt{caffau17,caffau20,venn20,kielty21,lardo21,lucchesi22,sestito23}), although it is important to remove the contamination from potentially variable stars \citep{bonifacio19,lombardo23}. The Pristine DR1 catalogue is overall smaller, with 3,783,243 good-quality photometric metallicities, among which 469,066 metal-poor star candidates, 53,636 VMP star candidates, and 795 EMP star candidates, but contains many fainter targets.

\subsection{Internal and external comparisons with metallicity catalogues}

\begin{figure*}
\begin{center}
\includegraphics[width=0.66\hsize,angle=0]{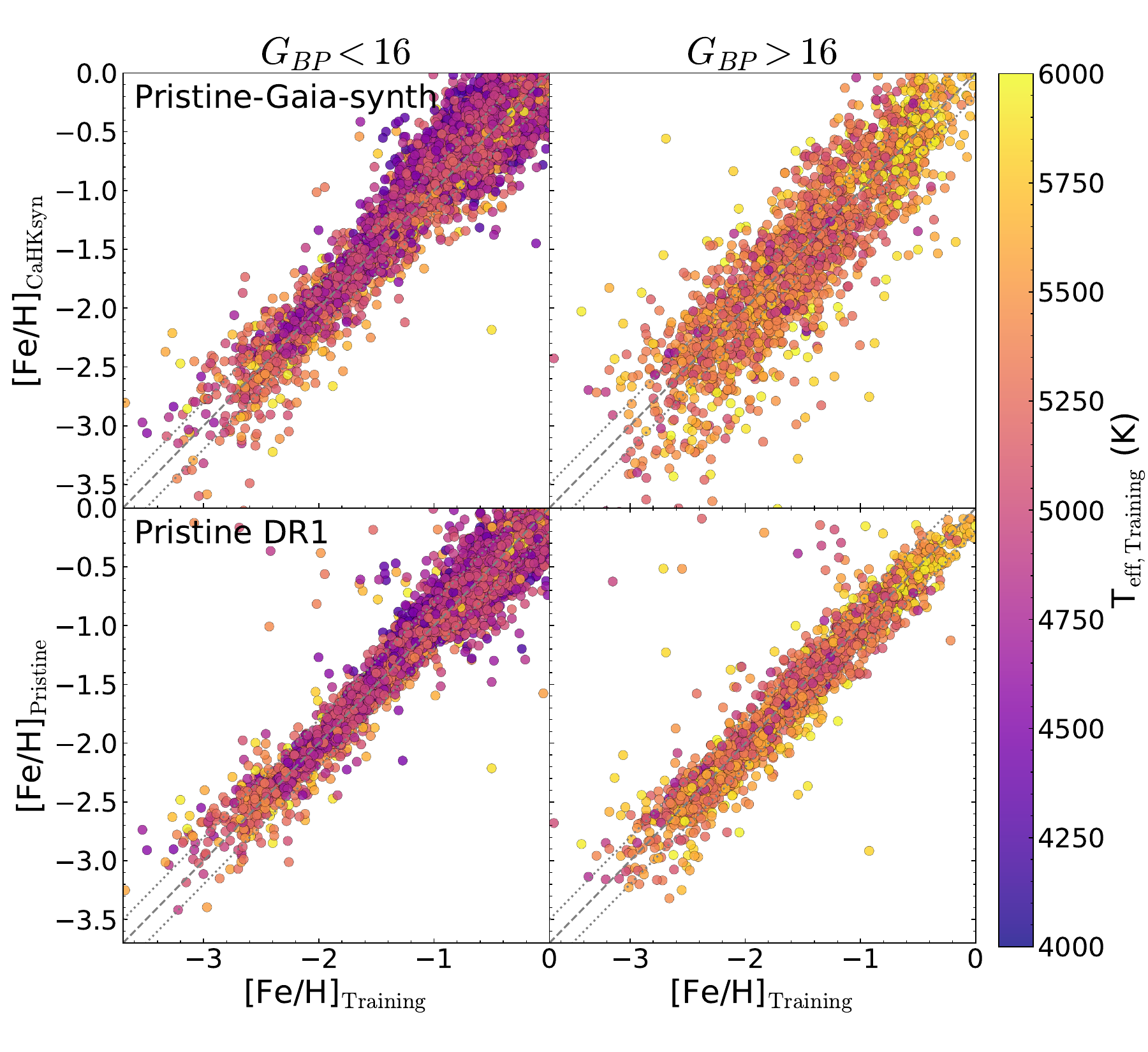}
\includegraphics[width=0.33\hsize,angle=0]{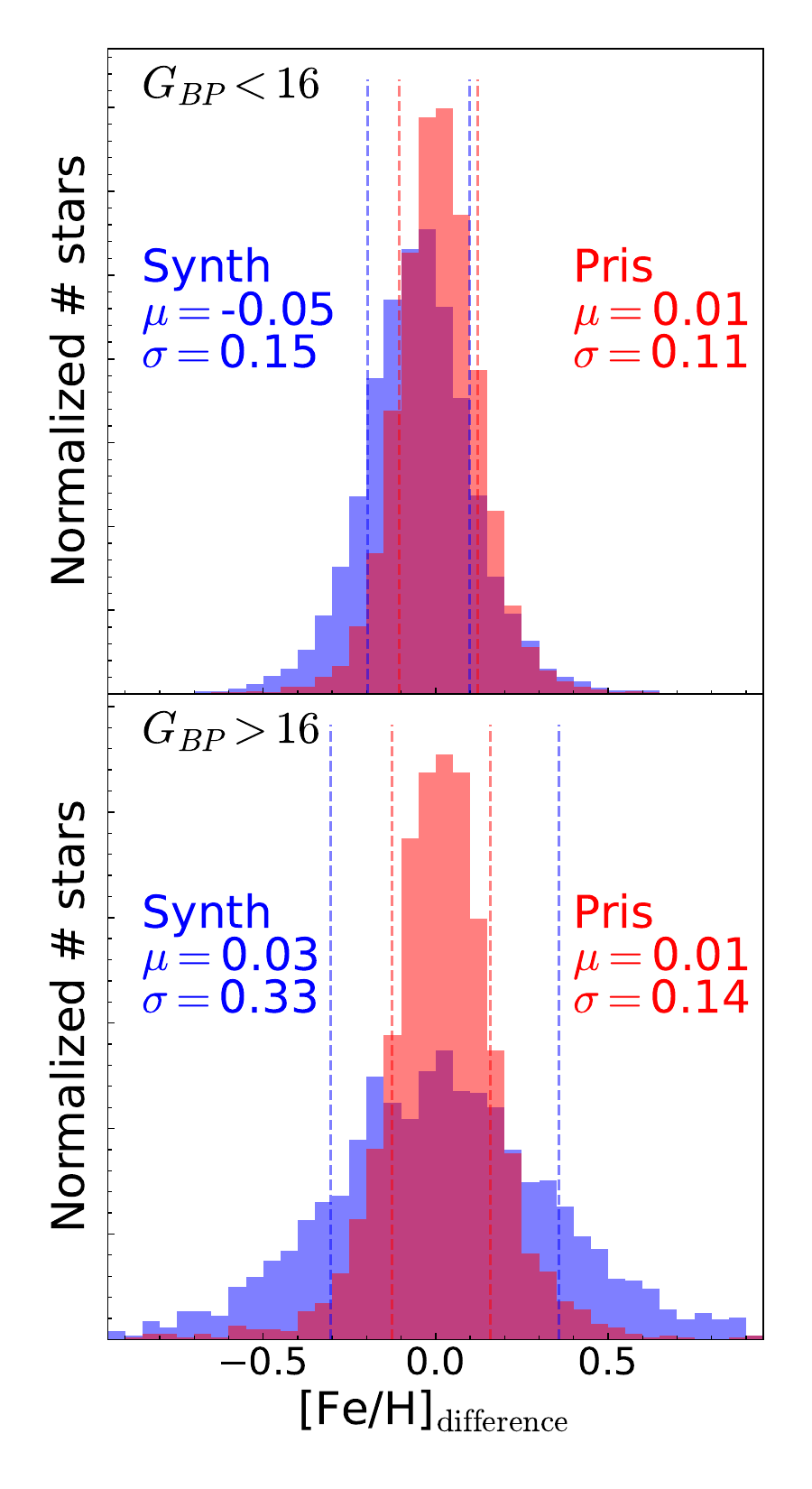} 
\caption{Comparisons of the Pristine-\Gaia\ synthetic (top row of the left-hand panels) and Pristine DR1 (bottom row of the left-hand panels) photometric metallicities with the spectroscopic metallicities of the cleaned giant star training set (see text for details). The right-hand panels show the histograms of metallicity differences for the two catalogues, with the Pristine-\Gaia\ synthetic catalogue in blue and Pristine DR1 in red. The catalogues are split as a function of $G_\mathrm{BP}$ magnitude, into the bright end ($G_\mathrm{BP}<16$; top) and the fainter end ($G_\mathrm{BP}<16$; bottom). At brighter magnitudes than this limit, both catalogues are similarly competitive, as can be appreciated from the confidence interval of the metallicity differences in the right-hand panels (16th to 84th percentiles; vertical dashed lines) and the average of these percentile offsets quoted as $\sigma$ in the two panels. At fainter magnitudes, the Pristine DR1 catalogue continues to remain very competitive while the Pristine-\Gaia\ synthetic catalogue suffers from the lower signal-to-noise of the $CaHK_\mathrm{syn}$ magnitudes, leading to a wider distribution of metallicity differences. \label{fig:int_comparison}}
\end{center}
\end{figure*}

\subsubsection{Internal comparisons}
To test the quality of the resulting photometric metallicities, Figure~\ref{fig:int_comparison} presents a comparison of the calculated photometric metallicities with the spectroscopic metallicities of stars from the training set that was used to build the $(CaHK,G,G_\mathrm{BP},G_\mathrm{RP})\rightarrow\FeH_\mathrm{phot}$ model. Beyond the cleaning of the training sample already mentioned in Section~\ref{sec:model}, we apply the additional quality criteria of Section \ref{sec:qualcuts} with the exception that we place not cut on the $\delta\FeH_\mathrm{phot}$ uncertainty. We restrict this analysis to giant stars (with $\logg<  3.9$) that are cooler than 6,000\,K. We also leave out the sample from \citet{li18b} from this specific comparison, because that particular data set shows significantly more scatter. By construction, there should be good agreement between the photometric and spectroscopic metallicities for this sample and it is indeed the case. The left-hand panels of the figure show that, at the bright end ($G_\mathrm{BP}<16.0$, where the uncertainties on the $CaHK_\mathrm{syn}$ uncertainties and photometric metallicity uncertainties remain reasonably small; cf. Figures~\ref{d_CaHK} and~\ref{fig:d_FeH}), both input $CaHK$ catalogues yield very similar results. The histograms in the right-hand panels show the distribution of the differences between the photometric and spectroscopic metallicities. Their confidence intervals, bounded by their 16th and 84th percentiles (vertical lines) correspond to average uncertainties of 0.11 and 0.15\,dex, even when including the regime with $\FeH_\mathrm{phot}>-1.0$, for which the model is not specifically designed. For fainter magnitudes, however, the \Gaia-Pristine synthetic metallicities become significantly less accurate than their Pristine counterpart, with uncertainties that reach 0.33\,dex (vs. 0.14\,dex), as expected from the higher S/N of the Pristine photometry. The medians of the distributions of metallicity differences inform us on the photometric metallicity bias, which remains small in all cases ($<0.05$\,dex).

\begin{figure}
\begin{center}
\includegraphics[width=\hsize,angle=0]{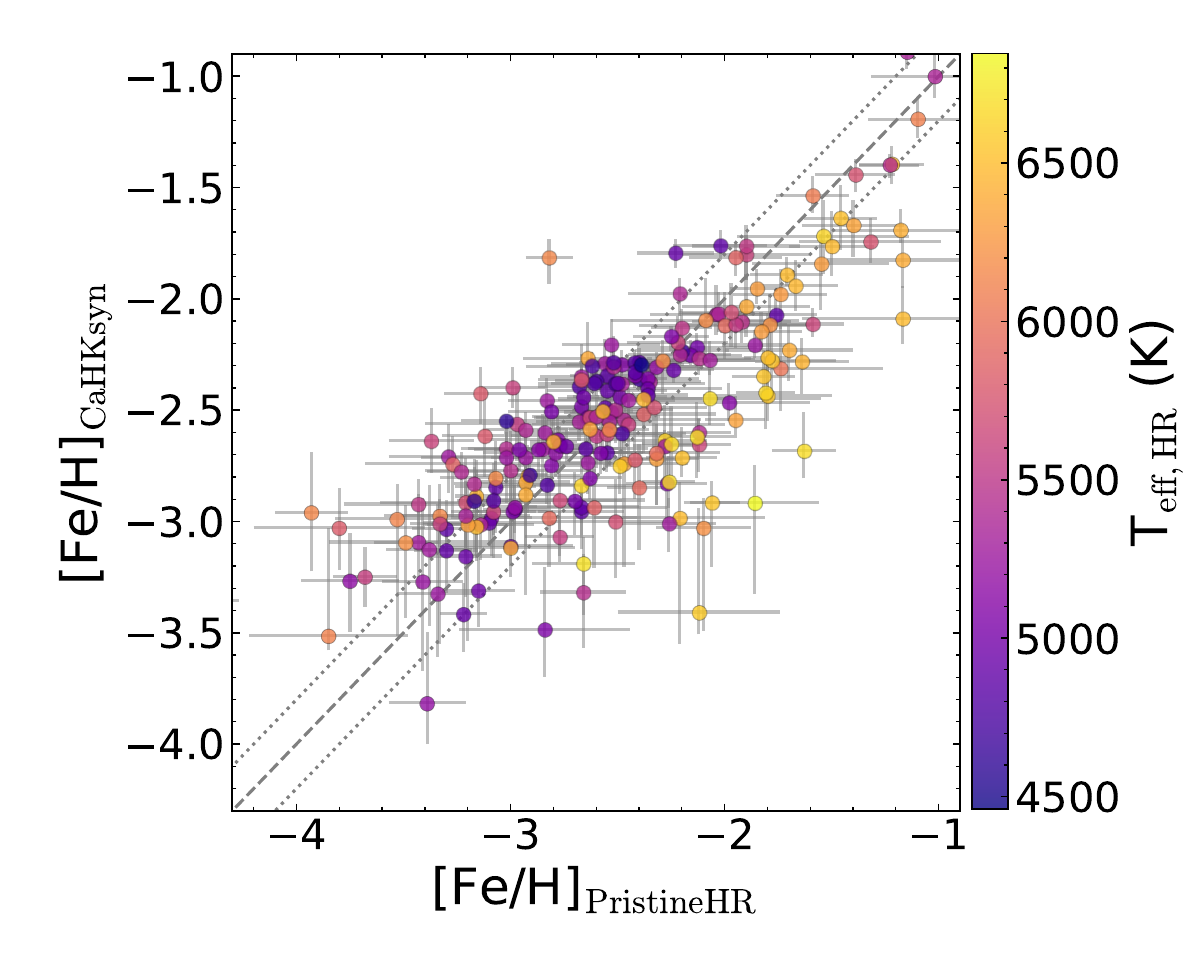}
\caption{
Comparison of the metallicities from the Pristine-\Gaia\ synthetic catalogue with the high-resolution spectroscopic metallicities accumulated by the Pristine collaboration. The photometric metallicities are in very good agreement with their spectroscopic counterparts throughout the metallicity scale, which means that the issues (variable stars, subtly saturated SDSS stars, \dots)  that led us to erroneously target stars with $\FeH_\mathrm{PristineHR}>-2.0$ as VMP candidates are now resolved. There remains an offset between the photometric and spectroscopic metallicities with the temperature of a star (the color of the dots). Stars with $\teff>6,000$\,K do tend to have lower photometric metallicities than their spectroscopic metallicities. \label{fig:Pr_comparison}}
\end{center}
\end{figure}

Since the start of the Pristine survey 8 years ago, the team has been actively pursuing spectroscopic follow-up of promising (very) metal-poor targets \citep{youakim17,caffau17,aguado19,bonifacio19,caffau20,venn20,kielty21,lardo21,lucchesi22,martin22a,yuan22b,sestito23,lombardo23}.  Figure~\ref{fig:Pr_comparison} focusses specifically on a comparison with these follow-up efforts in high-resolution as presented in \citet{caffau17,bonifacio19,venn20,kielty21,lardo21,lucchesi22,martin22a,yuan22b,lombardo23,sestito23}. Since many of these stars also feature in our training sample, this Figure does not present an independent validation of our photometric metallicities, but it is insightful to see how the photometric metallicities compare to these high-resolution results. While some of the samples followed up suffered from contamination of higher metallicity sources due to subtle saturation effects in the SDSS-photometry used at the time, or variable stars (see, in particular, \citealt{caffau17,bonifacio19,venn20,lucchesi22,lombardo23} for details), it is clear from this comparison that the current catalogues and recommended quality cuts would result in a very successful sample selection, as the more metal-rich stars are now correctly identified as such. We do, however, still see a clear trend with temperature and, as expected, the photometric metallicities are less accurate for stars with higher temperatures above $\sim$6,000\,K, a regime for which the lines of equal metallicity are closer together in our color-color space.

\subsubsection{External comparisons}
\begin{figure*}
\begin{center}
\includegraphics[width=\hsize,angle=0]{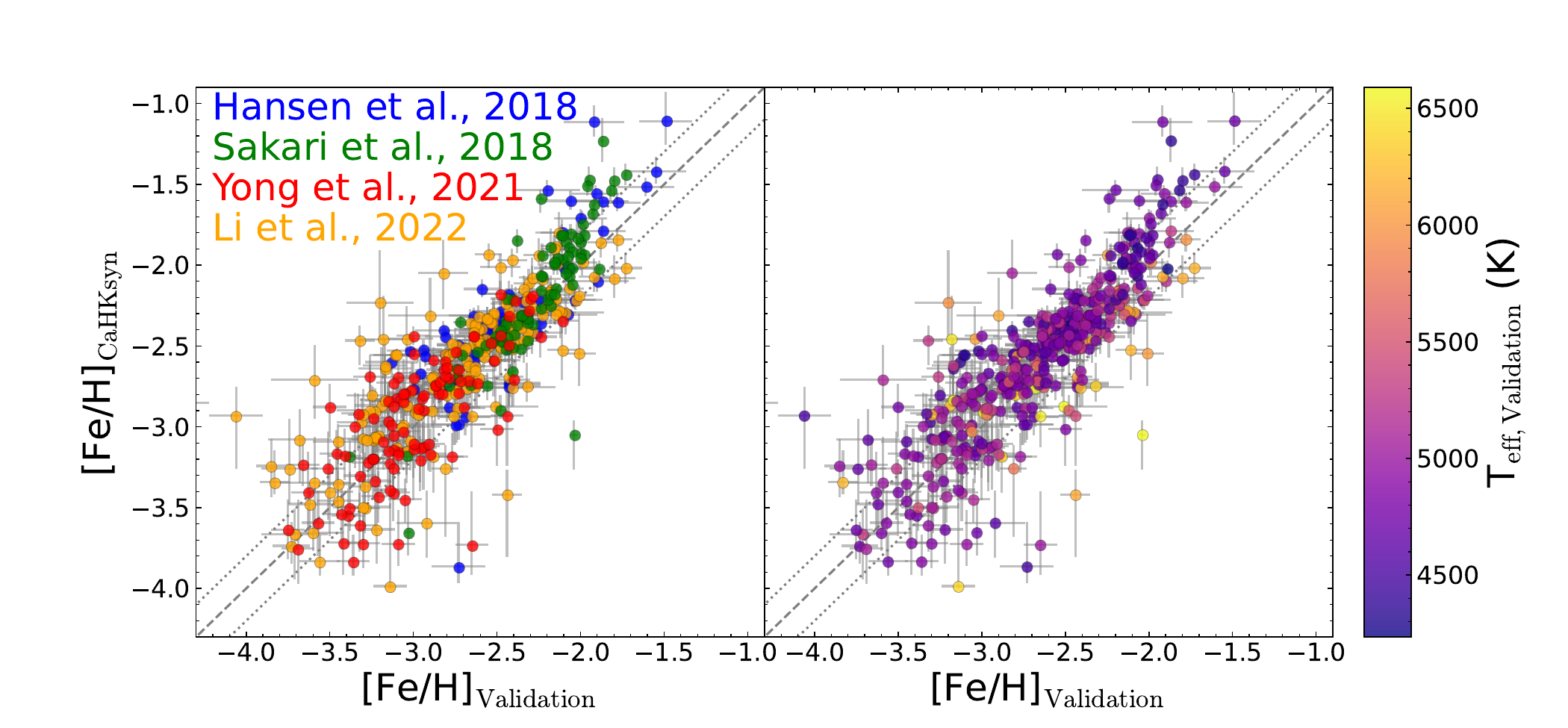}
\caption{Comparison of the Pristine-\Gaia\ synthetic photometric metallicities with metallicities from high-resolution spectroscopy samples of metal-poor stars (\citealt{hansen18,sakari18,yong21,li22b}; $R>25,000$). In the left-hand panel, color traces the origin of the spectroscopic metallicities, as listed in the label. The photometric metallicities are of high enough quality that they track the spectroscopic metallicities down to the lowest metallicities available. In the right-hand panel, the color codes the temperatures of the same stars, showing that the agreement is generally better for cooler rather than hotter stars. In both panels, the dashed line corresponds to the 1-to-1 line and the dotted lines represent offsets of 0.2\,dex. \label{fig:ext_comparison}}
\end{center}
\end{figure*}

A more interesting and powerful test of the photometric metallicities in the metal-poor regime comes from their comparison with a set of spectroscopic metallicities of metal-poor stars obtained with high-resolution spectroscopy that were not included in the training sample. These include four sets of high-resolution spectroscopic observations (\citealt{hansen18,sakari18,yong21,li22b}; $R>25,000$). All these stars are mainly located in the southern sky (where the coverage of the Pristine survey is extremely limited) or were only recently released, which explains why they were not taken into account when building the training set. They consequently provide a completely independent test of the quality of the photometric metallicities. To test the consistency of the method over a range of metallicities and/or temperatures, we deliberately chose to include in this comparison large samples of homogeneously analyzed stars, rather than a compilation of individual follow-up results from the literature. This comparison is presented in Figure~\ref{fig:ext_comparison} and shows an excellent agreement between the photometric metallicities we calculate and the high-quality spectroscopic metallicities. Despite increased uncertainties in the EMP regime ($\FeH<-3.0$), the Pristine metallicities are able to discriminate between stars of metallicities $\FeH\sim-2.0$, $\FeH\sim-3.0$ and even $\FeH\sim-3.5$. The right-hand panel of the Figure however illustrates that this capability decreases somewhat for hotter stars. Nevertheless, for most FGK stars, the catalogue of metallicities we publish here should be particularly suited to hunt for the low end of the MDF over the whole sky, just as the Pristine metallicities have been used to track such stars over the Pristine footprint \citep{starkenburg18,youakim17,aguado19,arentsen20b,caffau20,venn20,kielty21,lucchesi22,sestito23}.

\subsubsection{Comparisons with other \Gaia-based metallicity catalogues}
\begin{figure*}
\begin{center}
\includegraphics[width=0.495\hsize,angle=0]{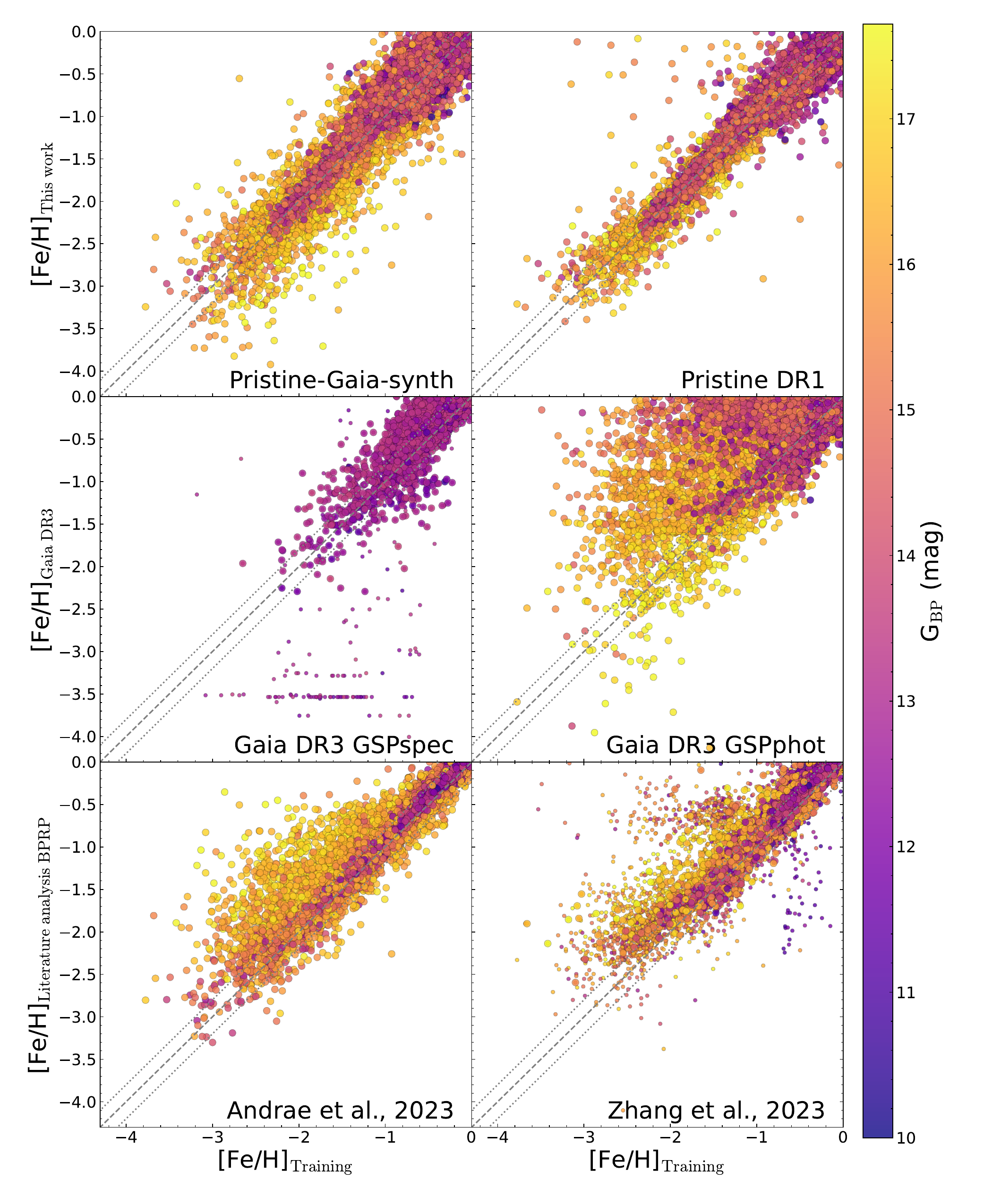}
\includegraphics[width=0.495\hsize,angle=0]{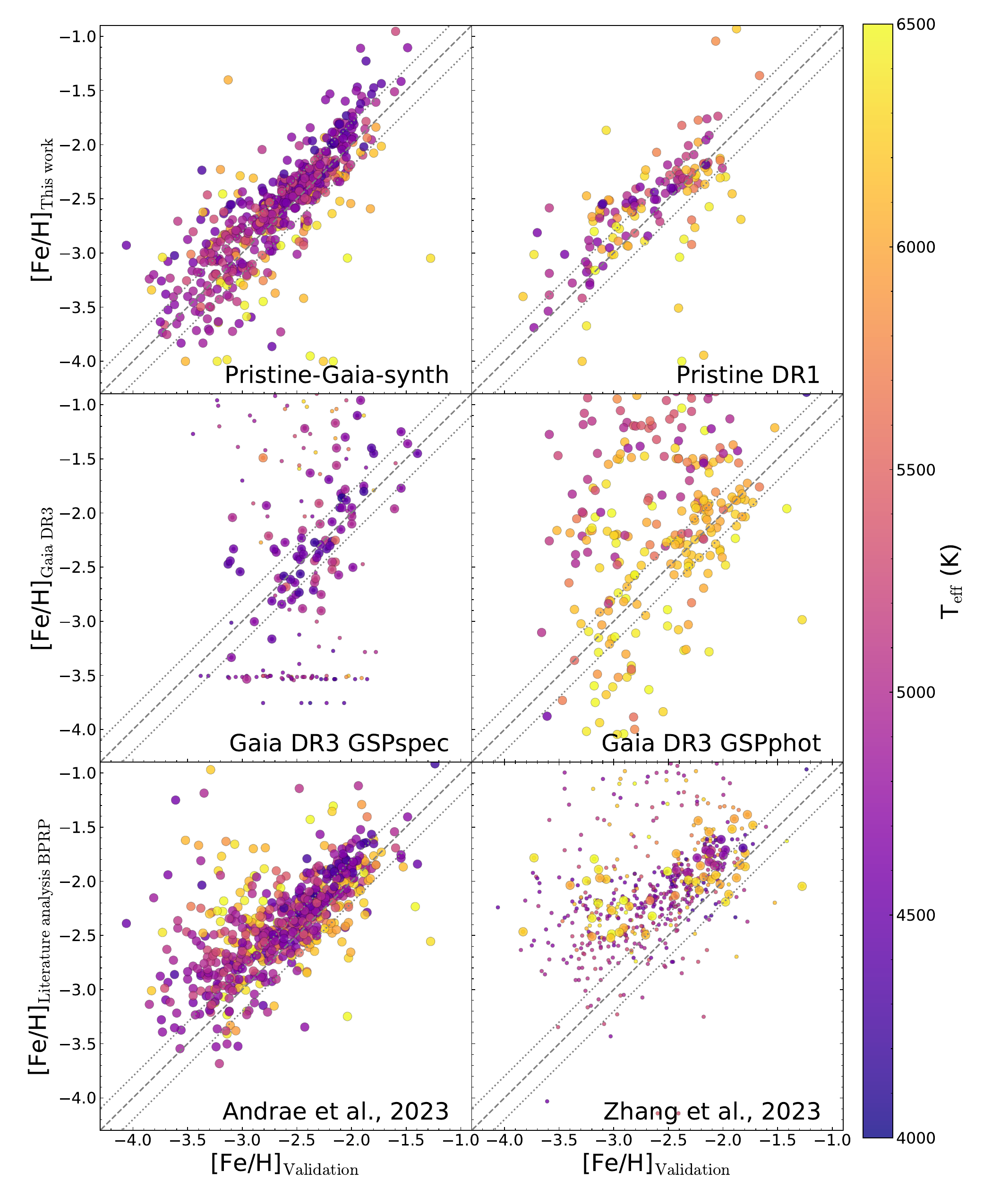} 
\caption{Comparison of various metallicity catalogues based on \Gaia\ DR3 with the giant stars in the training (left, see text for the construction of this sample) and with the validation sample at lower metallicities (right, see text for details) and their spectroscopic metallicities. After the comparison with the Pristine-\Gaia\ synthetic catalogue, and the Pristine DR1 catalogue, the panels show the comparisons for the \Gaia\ GSP-Spec metallicities based on the \Gaia\ RVS spectra \citep[][smaller circles correspond to stars flagged as unreliable in this catalogue, following the same criteria discussed in their sub-section~10.5 and represented in their Figure 26]{recio-blanco23}, the \Gaia\ GSP-Phot catalogue based on the BP/RP information \citep{andrae22}, the XGBoost algorithm \citep{andrae23} and the \citet[][smaller circles correspond to stars flagged to have unreliable metallicities]{zhang23} catalogue that use the BP/RP information and some additional non-\Gaia\ photometry. The color encodes the magnitude of a star in the left-hand set of panels and their temperature in the right-hand set of panels. The dashed line corresponds to the 1-to-1 line and the dotted lines represent offsets of 0.2\,dex. \label{fig:comparison_other_catalogues}}
\end{center}
\end{figure*}

Finally, Figure~\ref{fig:comparison_other_catalogues} shows a succinct comparison of current \Gaia-based metallicity catalogues with the giant stars in our training set (left-hand set of panels) and the validation sample presented in the previous sub-section (right-hand set of panels). In particular, we show a comparison of the spectroscopic metallicities of those two sets with the metallicities of catalogues from:
\begin{itemize}
\item the Pristine-\Gaia\ synthetic and Pristine DR1 photometric metallicity catalogues, as already shown above (top rows);
\item \citet{recio-blanco23}, based on the \Gaia\ Radial Velocity Spectrometer (RVS) spectra around the Calcium triplet region, with metallicities derived using the General Stellar Parametriser-spectroscopy (GSP-Spec), using the recommended set of flags for low metallicity stars\footnote{We use the set of flags described in their Sub-section 10.5 and in the caption of their Figure~26.} (middle left-hand panel);
\item \citet[]{andrae22}, with metallicities derived using the General Stellar Parameterizer from Photometry (GSP-Phot) applied to the \Gaia\ astrometry, photometry, and BP/RP spectrophotometry (middle right-hand panel);
\item \citet[v2.1 catalogue]{andrae23}, based on a data-driven algorithm, XGBoost, applied to the BP/RP coefficients (excluding the region of the Ca H\&K region because of low S/N) and AllWISE magnitudes, trained on a sample of APOGEE data complemented by a small number of VMP stars (bottom left-hand panel);
\item \citet{zhang23}, with a catalogue of stellar parameters (including metallicities) obtained by forward modeling the stellar type, extinction, and parallax of stars onto the space of XP spectra and near-infrared 2MASS and WISE photometry, using the information of the LAMOST survey for training (bottom right-hand panel).
\end{itemize}

As mentioned above, the top row of panels in the left-hand set is a beneficial comparison for the Pristine metallicities since it shows a comparison with the sample that was used to build the Pristine model. While this comparison is only including stars on the giant branch (with $\logg$ $< 3.9$) and $\teff < 6,000$\,K, a comparison with the turn-off, or main-sequence stars is shown in the appendix (see Figures~\ref{fig:dwarfcomp} and \ref{fig:dwarfcomp2}). We note that, generally, the performance is better for giants and moreover decreases toward hotter stars. The top row of panels in the right-end set, however, provides an external comparison with the validation sample that was not used to build any of the models, except the \citet{andrae23} model that was partially trained on the \citet{li22b} sample that constitutes about a quarter of the validation set. In all panels, the color of the data points encodes the magnitude of a star (left-hand set of panels) or its $\teff$ (right-hand set of panels).

These comparisons are very informative and show the strengths and weaknesses of the different metallicity data sets that can be quite complementary and should be used preferentially for different science goals. Stars with GSP-Spec metallicities are all bright since they require RVS spectra\footnote{Even though \Gaia\ RVS spectra are observed down to $G\sim15$, not all of these have high signal-to-noise. GSP-Spec analyzed spectra with signal-to-noise ratios $>20$ yield atmospheric parameters and the corresponding stars typically have $G<$12--13.}. Because of that, they do not contain a lot of metal-poor stars. Applying the strict flags listed in \citet{recio-blanco23} is absolutely essential to clean up the data in the VMP regime and to prevent being dominated by spurious metallicity values (small points in the corresponding panel). Despite this clean-up, there remains stars with GSP-Spec metallicities that deviate from those of the high-resolution validation sample\footnote{At least in part because GSP-Spec metallicities have large uncertainties in the VMP regime.}. The strength of these spectra however lies in the additional information available for them, including, of course, the radial velocities that provide full 6-dimensional phase-space information, from the rich multi-dimensional abundance information, and from the high-quality \logg, \teff, and further information that they can provide \cite[\eg][]{kordopatis23}.

GSP-Phot, XGBoost, and the \citet{zhang23} metallicities go significantly deeper as they were derived using the BP/RP information and are more directly comparable to the two Pristine metallicity catalogues. The GSP-Phot metallicities are definitely challenged and include a sizable number of stars with erroneously high metallicities. This is particularly problematic when comparing with the validation data set in the VMP regime (corresponding bottom panel), where strong GSP-Phot systematics lead to metallicities that cover the full range.

The XGBoost \citep{andrae23} and \citet{zhang23} catalogues, which are both based on the BP/RP coefficients but also supplemented by information external to \Gaia, provide metallicities that are much better behaved over the full metallicity range, especially at the bright end (magenta points in the right-hand set of panels). However, in the VMP regime, even though the metallicities of both catalogues are significantly better behaved than GSP-Phot values, a non-negligible fraction of stars are systematically assigned more metal-rich photometric metallicities compared to their spectroscopic counterpart (as reported by both \citealt{andrae23} and \citealt{zhang23}). These stars are still often assigned metallicities below $\FeH_\mathrm{phot}<-2.0$, but this would limit the ability to build a pure and complete sample of stars with $\FeH<-2.5$ or $<-3.0$. This behavior is somewhat worse for the \citet{zhang23} catalogues, possibly because \citet{andrae23} took the extra step of complementing their APOGEE training sample with a specific set of VMP stars. A similar overestimation of the metallicities is also visible for $\FeH_\mathrm{spec}<-1.0$ for fainter stars when comparing with the training set and could indicate that this bias is related to the S/N of the spectral features, either because a star is VMP, or because it is faint. Both catalogues do work well for bright and/or fairly metal-rich stars (despite a small number of catastrophic failures in the \citealt{zhang23} catalogue), the latter of which constitute most of the MW stars. Both samples also have the advantage of providing metallicities in all regions of the MW: using near-infrared photometry in conjunction with the BP/RP coefficients and training on APOGEE or LAMOST stars that cover a wide range of extinction values helps XGBoost and the \citet{zhang23} algorithm to separate the impact of extinction on the input features.

None of these four external catalogues were specifically built to treat VMP stars. We may therefore expect that the Pristine-\Gaia\ synthetic and the Pristine DR1 catalogues handle the very low-metallicity range better and this is indeed what we can see in the top rows of panels in Figure~\ref{fig:comparison_other_catalogues}. As already discussed above, the Pristine metallicities are well aligned with the spectroscopic metallicities, despite a small tendency for them to be underestimated in the lowest metallicity regime. In the $\FeH>-1.5$ regime, we note that the XGBoost metallicities show less scatter than the Pristine metallicities at the bright end and, at the metal-rich end, the Pristine metallicities are clearly suboptimal compared to the XGBoost or most of the \citet{zhang23} values. A comparison of the plots for the two Pristine catalogues emphasizes what we already mentioned above: the Pristine DR1 catalogue has higher S/N than the Pristine-\Gaia\ synthetic catalogue (left-hand set of panels) but the latter covers a significantly larger footprint and allows for a more thorough comparison, in particular with the high-resolution validation set (right-hand set of panels).

\subsection{Impact of quality cuts on the search for UMP stars}
\begin{figure}
\centering
\includegraphics[width=1.0\hsize]{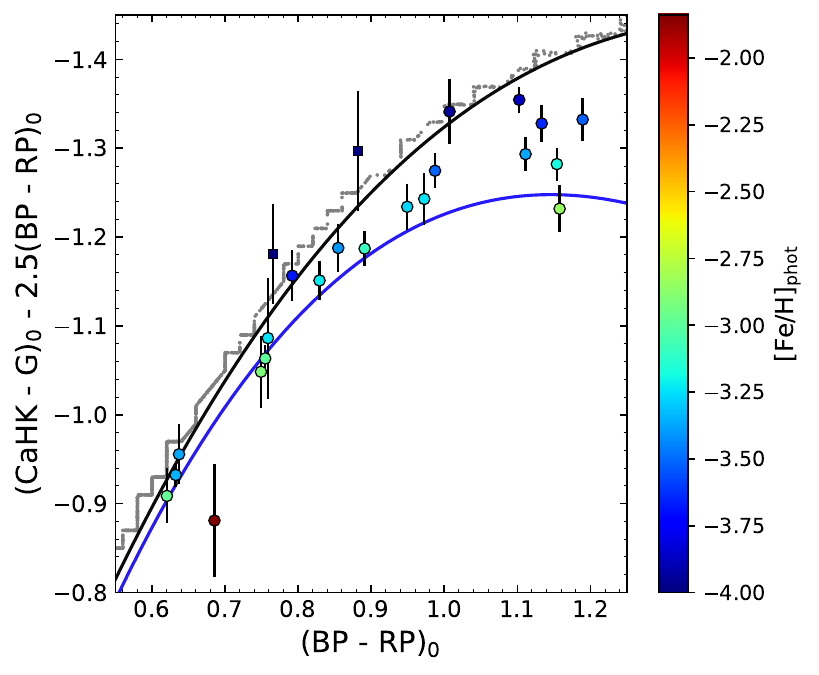}
\caption{Distribution in the Pristine color-color space of the 23 known UMPs in common with the Pristine-\Gaia synthetic catalogue. Stars are color-coded by their photometric metallicities $\FeH_\mathrm{phot}$. The blue and black lines are polynomial fits to the expected location of stars with $\FeH=-3.0$ (blue) and $\FeH=-4.0$ (black) based on the spectral synthesis described in section~\ref{sec:ccspace} and also represented by star symbols in Figure~\ref{fig_colorspace}. The gray points correspond to stars with $-4.0<\FeH_\mathrm{phot}<-3.99$ and highlight the location of the region with $\FeH_\mathrm{phot}=-4.0$, as assigned by the Pristine model. Only two stars, represented by squares, are above this region and technically rejected by the $\FeH_\mathrm{phot}>-4.0$ cut recommended in sub-section~\ref{sec:qualcuts}. Because these stars have large uncertainties, requesting only $\FeH_\mathrm{phot,84th}>-4.0$ would select them back the sample of high-quality stars.\label{fig:ump}} 
\end{figure}

The search for the most metal-poor stars and, in particular, UMP stars is one of the key uses for the metallicity catalogues. However, the dearth of good calibrators and models at those metallicities led us to fix the lowest metallicities output by the model described in section~\ref{sec:model} to $\FeH_\mathrm{phot}=-4.0$. In addition, this metallicity limit is assigned to all stars within 0.15\,mag of the $\FeH=-4.0$ model in the color-color space of Figure~\ref{fig_colorspace} and we consequently recommend not using stars with $\FeH_\mathrm{phot}=-4.0$ for a generic use of the catalogue. One may therefore naturally wonder whether the proposed quality cuts negatively impact the search for the most metal-poor stars by simply flagging them out.

To explore this question, we cross-match the list of 42 UMP stars provided by \citet{sestito19} with the Pristine-\Gaia synthetic metallicity catalogue. There are 23 stars in common between the two catalogues, with the other stars being too faint to have a metallicity in the Pristine-\Gaia synthetic catalogue. Figure~\ref{fig:ump} shows the distribution of those 23 stars in the Pristine color-color space. Reassuringly, almost all stars are within the region $-4.0<\FeH_\mathrm{phot}<-3.0$. Within this extremely low-metallicity region, they are not specifically located at the lower metallicity edge of the model (the upper edge in the figure). This could be due to the large amount of carbon present in these stars that could increase the absorption in the CaHK filter leading to photometric metallicities that are biased high (see sub-section~\ref{sec:carbon} below). In particular, this is the case for the most problematic star, SDSS~J081554.26+472947.5, that appears in red in the figure, with $\FeH_\mathrm{phot} = -1.84^{+0.39}_{-0.57}$ in the Pristine-\Gaia synthetic catalogue. This star is also present in the Pristine DR1 catalogue, with a much more consistent $\FeH=-3.64^{+0.46}_{-0.36}$, which means that its erroneous metallicity in the Pristine-\Gaia synthetic catalogue is driven by its large $CaHK_\mathrm{syn}$ uncertainties, $\delta CaHK_\mathrm{syn}=0.06$, and its blue color, for which the iso-metallicity lines are closer in the Pristine color-color space.

While there is a handful of true UMP stars that are located above the $\FeH=-4.0$ theoretical model (black line), most of these are below the $\FeH_\mathrm{phot}=-4.0$ threshold, as defined by the model given the pixelization of the color-color space (gray line; see sub-section~\label{sec:algorithm}). Most of these stars would not be remove by the $\FeH_\mathrm{phot}>-4.0$ recommended above. There remains two genuine UMP stars, represented by square symbols in the figure and that both have spectroscopic $\FeH<-5.0$, that are located above this line and would be rejected by the recommended cut. These two stars also have large $CaHK$ uncertainties, as can be seen from their large vertical error bars and they are being pushed in the $\FeH_\mathrm{phot}=-4.0$ region may simply be a consequence of these large uncertainties. Nevertheless, a catalogue user that aims at isolating candidate UMP stars may wish to loosen the recommended cut to $\FeH_\mathrm{phot,84th}>-4.0$. This change would prevent the removal of the two egregious UMPs.

\subsection{Impact of extinction and latitude}
\begin{figure}
\centering
\includegraphics[width=1.0\hsize]{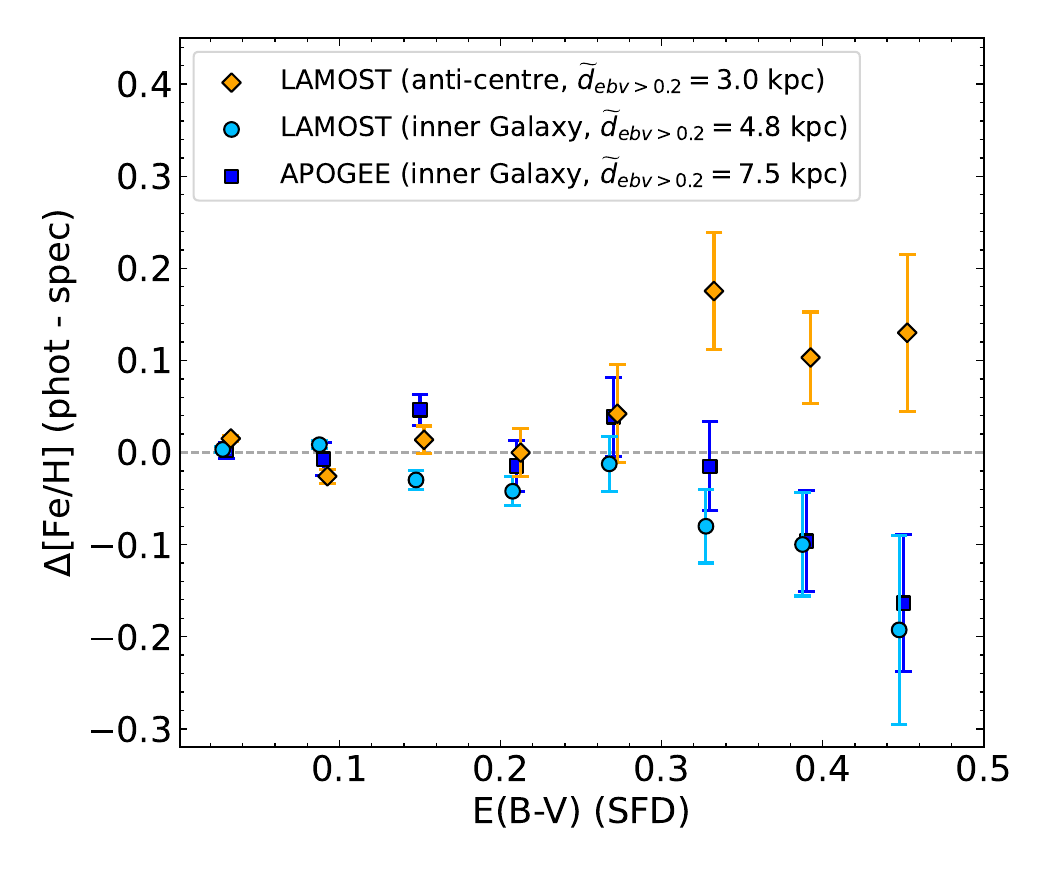}
\caption{Median difference between Pristine-\Gaia synthetic photometric metallicities and APOGEE/LAMOST spectroscopic metallicities as a function of the extinction values $E(B-V)$ from the map of \citet[][\ie SFD]{schlegel98} used to deredden the magnitudes. The samples are split in Galactic longitude to investigate spatial trends, with APOGEE having mostly stars in the inner Galaxy and the LAMOST sample covering two spatial regions: toward the inner Galaxy or the anticenter. The median distance of high extinction stars with $E(B-V) > 0.2$ in each sample are given in the legend. Error bars correspond to $\sigma/\sqrt{N}$, with $\sigma$ the standard deviation between $\FeH_\mathrm{phot}$ and $\FeH_\mathrm{spec}$ in a given bin after clipping 3-sigma outliers, and $N$ the number of stars in this bin.\label{fig:extcomp}} 
\end{figure}

Even though the largest extinction regions, with $E(B-V)>0.5$, are excluded from this study, we stress that, as we adopt an integrated extinction map that has its limitations \citep{schlegel98}, much care should be taken for stars in high extinction regions that are listed in the catalogue. To test possible systematics with $E(B-V)$, we compare our Pristine-\Gaia synthetic photometric metallicities with spectroscopic metallicities from APOGEE DR17\footnote{\url{https://www.sdss4.org/dr17/}} \citep{abdurrouf22} and LAMOST DR8\footnote{\url{https://www.lamost.org/dr8/}} \citep{zhao12}, which both have many stars across a range of extinction. We select giant stars with spectroscopic [Fe/H]~$<-1.0$ passing basic spectroscopic quality criteria, remove globular cluster stars from the sample, and only keep stars with $\delta\FeH_\mathrm{phot}<0.3$\,dex. To investigate spatial trends, we limit APOGEE to the inner Galaxy ($|l| < 45\deg$) and split the LAMOST sample in two: $l < 90\deg$ (inner Galaxy) and $l > 90\deg$ (anticenter). The result is shown in Fig.~\ref{fig:extcomp}. Up until $E(B-V) = 0.3$, the comparison between photometric and spectroscopic metallicities is excellent, to within $\sim 0.05$\,dex. Between $0.3 < E(B-V) < 0.5$, systematic offsets begin to appear, but, on average, they remain $<0.2$\,dex. There is an opposite trend for the inner Galaxy and the anticenter, which might be related to systematics in the reddening map and/or to different distance distributions for stars in the sample (nearby stars might be over-corrected and distant stars might be under-corrected for extinction when using an integrated extinction map). In summary, our photometric metallicities appear to be good for $E(B-V) < 0.3$, while for $0.3 < E(B-V) < 0.5$ some biases appear to be creeping in.\footnote{In their study to calibrate the J-PLUS photometry from the \Gaia XP data, \citet{lopez-sanjuan24} find a potential trend of their calibration with extinction, which they venture could be the sign of a degradation of the calibration of the XP data in regions of high extinction. It could explain some of the systematic deviations we measure when the extinction increases.}

Additionally, comparisons of our results with the spectroscopic samples reveal a larger fraction of metal-rich contaminants at low latitudes, close to the Galactic plane, even at lower extinction values in the \citet{schlegel98} maps. We recommend additional caution in these regions. As they are overwhelmingly dominated by metal-rich stars, any source of contamination from those in the metal-poor regime, even small, can become problematic.

\subsection{Impact of carbon}
\label{sec:carbon}

The presence of strong carbon bands in the spectrum of a star can significantly bias the estimated photometric metallicity as carbon absorption in the wavelength range of the CaHK filter are interpreted as stronger Ca H \& K lines. This is of particular importance in the VMP regime because a sizable fraction of VMP and EMP stars are expected to be carbon-enhanced ($[\textrm{C}/\textrm{Fe}] > 0.7$), even though different data sets yield different conclusions on the exact value of this fraction \citep[\eg][]{arentsen22}. From the Pristine spectroscopic follow-up presented in \citet{aguado19}, \citet{arentsen22} showed that selecting EMP candidates from the Pristine photometric metallicities leads to a dearth of cool ($T_{\rm eff} < 5500$\,K) Carbon Enhanced Metal-Poor (CEMP) stars (see also \citealt{caffau20}). To investigate the effect of this bias, we compare the metallicities from the Pristine-\Gaia\ synthetic catalogue with the spectroscopic metallicities from the SAGA database \citep{suda08}. As SAGA is a compilation of different sources the selection function may well be biased with regards to carbon, but for comparative purposes it suits our needs well. We use the recommended version of the data set from April 7, 2021 and limit the comparison to stars with $-4.0<\FeH_\mathrm{SAGA}<-1.0$ to match the recommended range for the photometric metallicities. The catalogues are crossmatched with a search radius of $1''$ and [M/H] values are used for stars without $\FeH$ values in SAGA. 

\begin{figure}
\begin{center}
\includegraphics[width=1.05\hsize,angle=0]{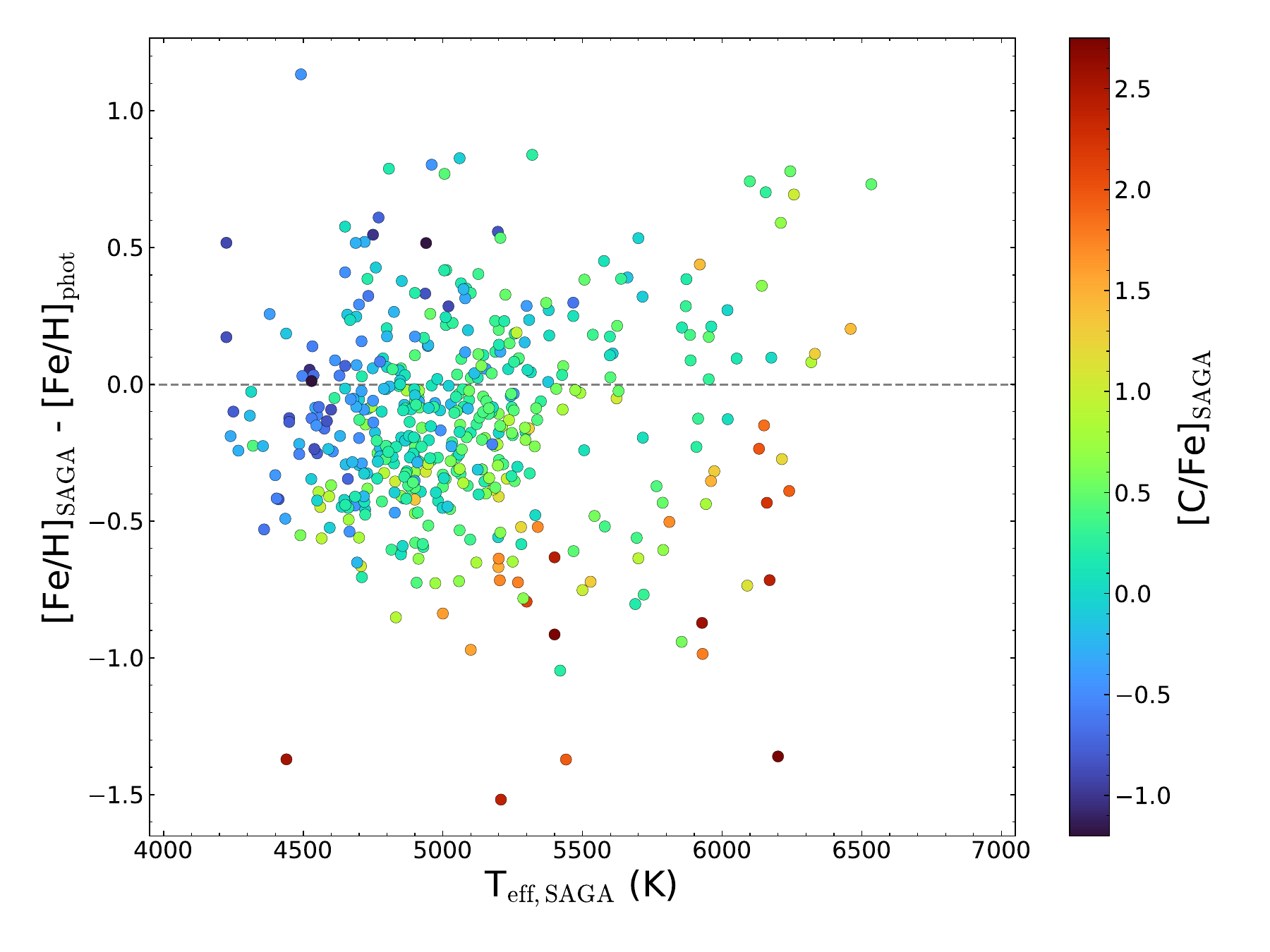}
\caption{Impact of the carbon abundance of a star on the derived Pristine photometric metallicities. The difference between the photometric metallicities from the Pristine-\Gaia\ synthetic catalogue and the spectroscopic metallicities from the SAGA database is plotted against the spectroscopic effective temperature and color-coded by the carbon abundance, [C/Fe]. Stars that are strongly carbon-enhanced show a clear bias and have overestimated Pristine metallicities, especially for lower $\teff$. \label{fig:carbon}}
\end{center}
\end{figure}

Figure \ref{fig:carbon} shows the difference in metallicity between Pristine and SAGA, color-coded by their carbon abundance, [C/Fe], as listed in the SAGA database. It is clear that stars with large carbon abundances have photometric metallicities that are significantly overestimated. In the cool regime, for stars with $T_{\rm eff} < 5500$\,K, most photometric metallicities are overestimated for stars with [C/Fe]$ > 0.7$, with a mean discrepancy of 0.70\,dex. For the hotter stars with $T_\mathrm{eff} > 5500$\,K the mean deviation is noticeably smaller at 0.33\,dex. The temperature dependance of these discrepancies are in good agreement with the bias found by \citet{arentsen22}. Because of this, we caution against drawing conclusions about the fraction of stars with significant carbon enhancement in samples derived from the Pristine catalogues. A more comprehensive analysis of the causes of this bias is under study (M. Montelius et al., in prep.).

\section{Science tests}
\label{sec:tests}
\subsection{Science test 1: The metallicity of globular clusters}
\begin{figure*}
\begin{center}
\includegraphics[width=0.9\hsize]{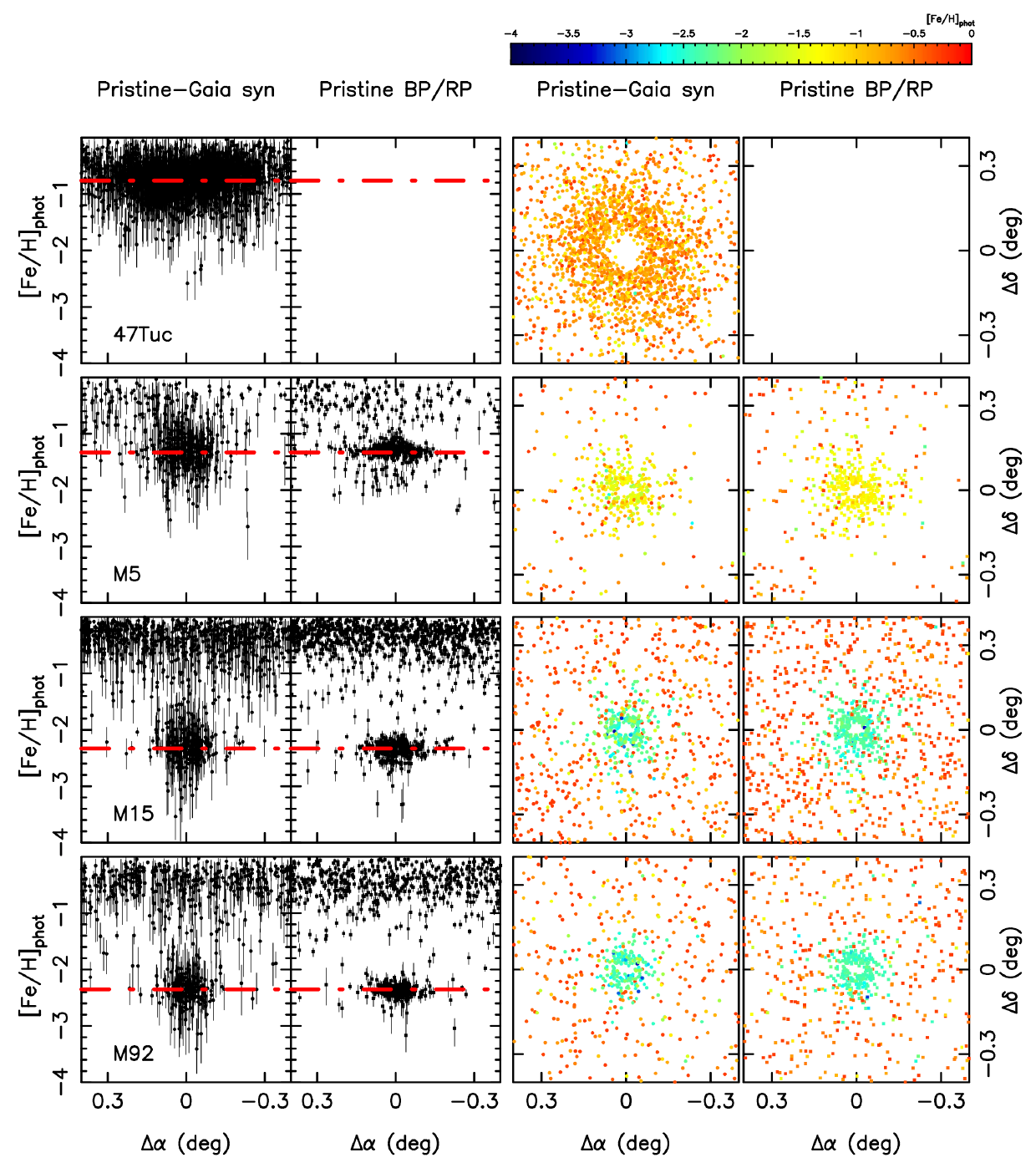}
\caption{Metallicity of stars in both the Pristine-\Gaia\ synthetic and Pristine DR1 catalogues around four globular clusters that span a wide metallicity range. From top to bottom, 47\,Tuc, M\,5, M\,15, and M\,92. The two left-most columns show the distribution of metallicities of stars in both catalogues as a function of right ascension. They are in good agreement with the known metallicity of the clusters, indicated by the red dot-dashed lines \citep{carretta09}. The two right-most columns of panels show the distribution of stars around the clusters, color-coded by metallicity. With the metallicity information, likely members of the clusters are very easy to pick out among the field contamination.\label{fig:GCs}}
\end{center}
\end{figure*}

An illustration of both the quality and the scientific usefulness of the photometric metallicities we provide is presented in Figure~\ref{fig:GCs}. The figure shows the photometric metallicities for all stars in the vicinity of four different MW globular clusters (GCs) that have been chosen to span a wide range in metallicity. Here, we use stars with $\delta\FeH_\mathrm{phot}<0.5$\,dex. Close to the center of the clusters, crowding becomes an issue and the holes in the data are produces by the imposed quality cuts (subsection~\ref{sec:qualcuts}), especially the \Gaia\ quality cuts. Apart from these central regions without data, the member stars of metal-poor GCs are very easy to isolate from the foreground contamination of MW stars and the resulting photometric metallicities are all in excellent agreement with their literature metallicities, despite spanning almost 2\,dex (the red lines in the two first columns of the figure; 47\,Tuc, $\FeH=-0.76\pm0.02$; M\,5, $\FeH=-1.33\pm0.02$; M\,15, $\FeH=-2.33\pm0.02$; M\,92, $\FeH=-2.35\pm0.05$; \citealt{carretta09}). In the case of the two most metal-poor clusters, M\,15 and M\,92, it becomes trivial to separate metal-rich, MW contaminants from member stars and opens up an exciting discovery space for ``extra-tidal'' cluster member stars, or cluster stellar stream stars for those that are disrupting (\eg M\,92; \citealt{thomas20}). When a cluster appears in both metallicity catalogues, the higher signal-to-noise of the Pristine photometry, combined with the $\delta CaHK<0.1$ quality cut to calculate $\FeH_\mathrm{phot}$, translate to significantly more stars with Pristine DR1 metallicities.

To further explore the accuracy of our photometric metallicities, we systematically examine the whole population of MW GCs. Using catalogues of known GCs \citep{harris96,baumgardt21}, we select stars around GCs that also have \Gaia\ color-magnitude and proper motion information compatible with the properties of the clusters. We then extract photometric metallicities from the Pristine-\Gaia\ synthetic and Pristine DR1 catalogues for those stars, once again using the quality cuts presented in Section~\ref{sec:qualcuts}, except that we use a stricter $\delta\FeH_\mathrm{phot}<0.3$\,dex. For those clusters with a well-populated red giant branch (more than 100 stars above the sub-giant branch) we compute the mean difference between the Pristine metallicities and the literature value by fitting a Gaussian to that distribution.

\begin{figure*}
\begin{center}
\includegraphics[width=\hsize]{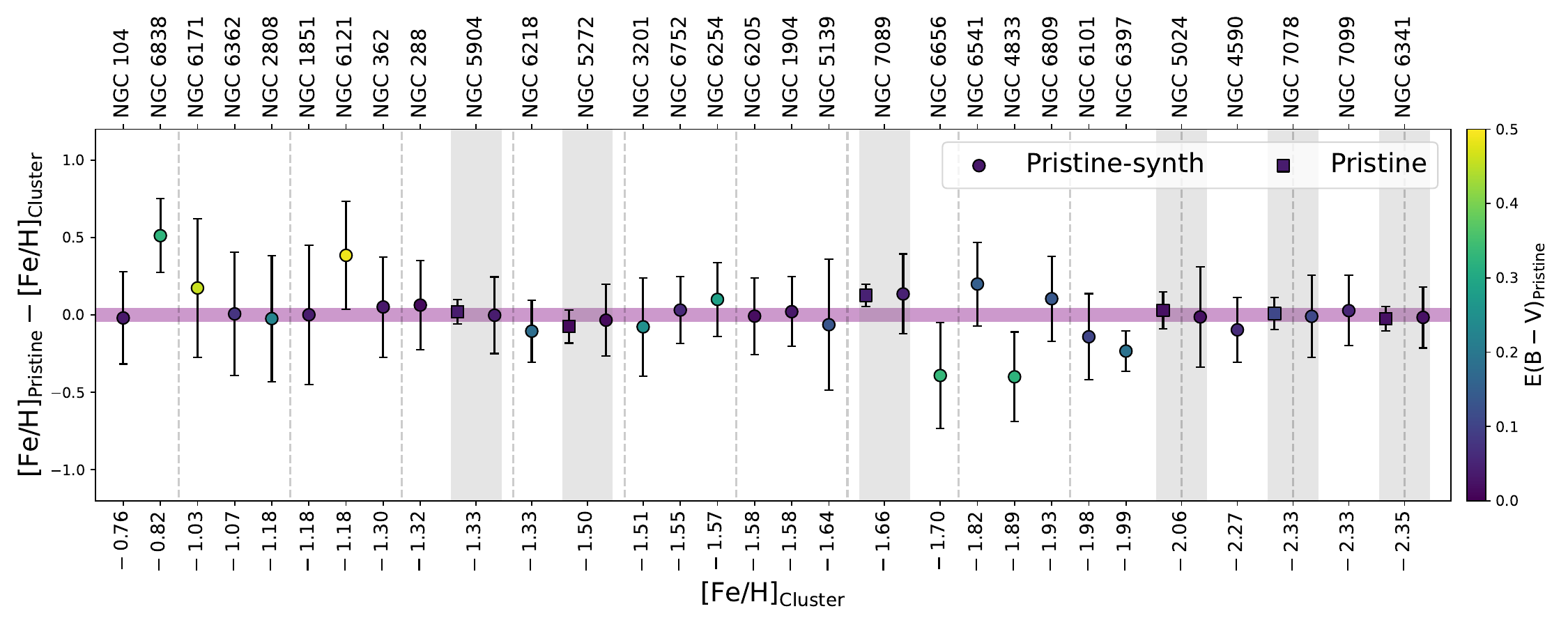}
\caption{Difference between the mean metallicity of 35 GCs with at least 100 stars in the Pristine-\Gaia\ synthetic (circles) and Pristine DR1 (squares) catalogues, compared to the literature values \citep{carretta09}, over a broad range of metallicities. The gray bands highlight clusters that are present in both catalogues. Cluster names and metallicities are listed, and the symbols are color-codded by the $E(B-V)$ value toward each GC. The expected zero purple line and the vertical dashed lines are highlighted for visualization purposes.\label{fig:GCs_all}}
\end{center}
\end{figure*}

In total, we can compare results from the photometric metallicity catalogues and literature metallicities \citep{carretta09} for 35 clusters that cover the entire metallicity distribution of MW GCs. The results are shown in Figure~\ref{fig:GCs_all} and highlight a very good agreement over the full metallicity range. There may be the hint of a mild overestimation of the Pristine metallicities at the metal-rich end. However, some of these differences also appear driven by the redenning (coded by the color of the symbols in the figure). More highly reddened clusters tend to yield mean Pristine DR1 and Pristine-\Gaia\ synthetic metallicities that deviate more significantly from the literature values. Nevertheless, this comparison with MW GCs emphasizes the very good quality of the Pristine photometric metallicities that were assembled without any information on the metallicity of the GC clusters.

\subsection{Science test 2: Slicing the MW by metallicity}

\begin{figure*}
\begin{center}
\includegraphics[width=0.95\hsize]{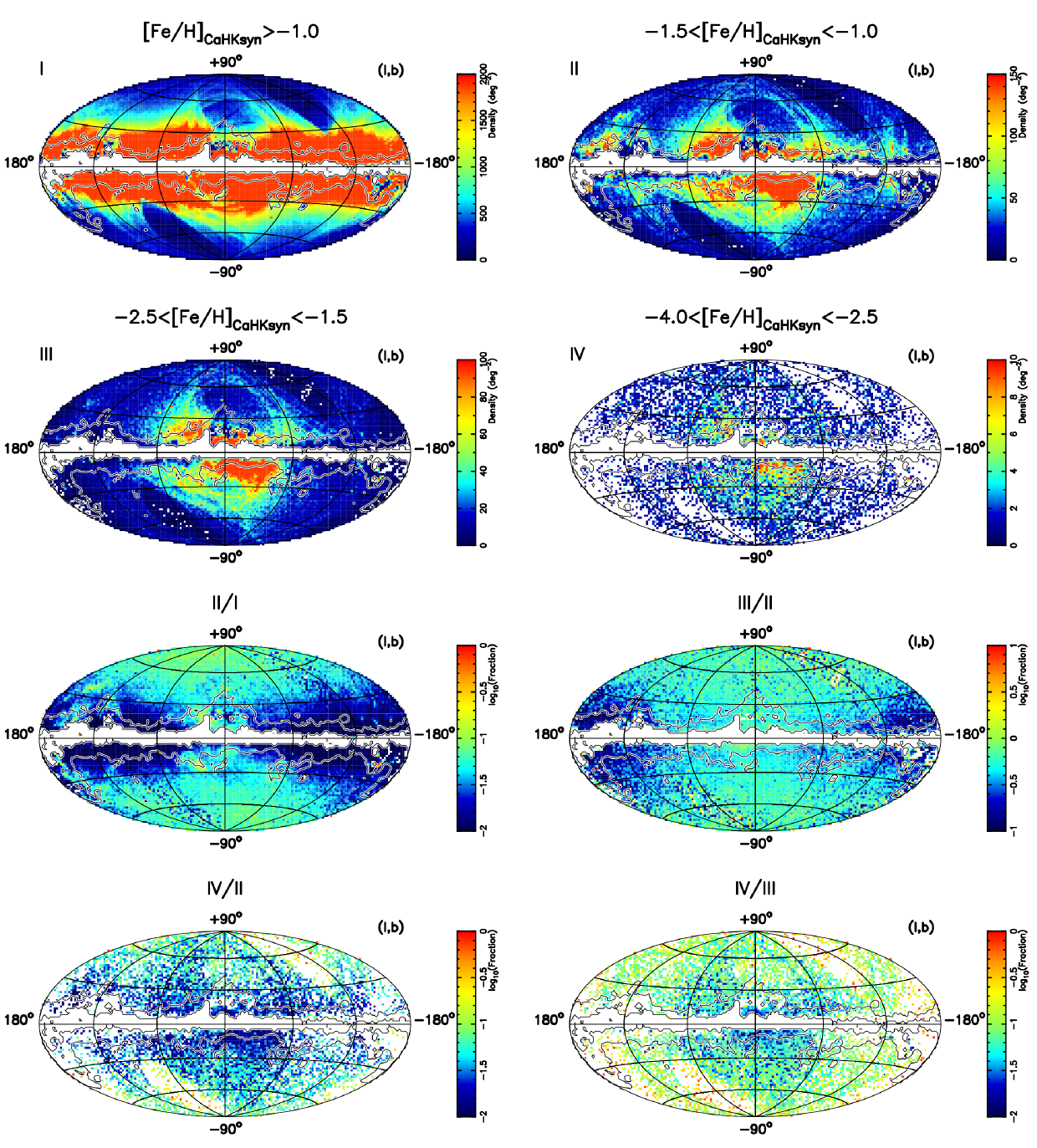}
\caption{Density maps for different slices of Pristine-\Gaia\ synthetic metallicities, projected in Galactic coordinates. The contours correspond to $E(B-V)=0.5$ and $E(B-V)=0.15$. From top-left to bottom-right, the four top panels display density for metallicity ranges $\FeH_\mathrm{CaHKsyn}>-1.0$ (panel I), $-1.5<\FeH_\mathrm{CaHKsyn}<-1.0$ (panel II), $-2.5<\FeH_\mathrm{CaHKsyn}<-1.5$ (panel III), and $-4.0<\FeH_\mathrm{CaHKsyn}<-2.5$ (panel IV). To overcome the inhomogeneous survey and highlight the transition in the shape of the MW, the bottom four panel shows ratios of density maps. From top-left to bottom-right, II/I, III/II, IV/II, and IV/III.\label{fig:maps_metslices}}
\end{center}
\end{figure*}

A second illustration of the power of the all-sky catalogue of photometric metallicities we publish is presented in Figure~\ref{fig:maps_metslices}, where we show density maps of MW stars for different metallicity slices. With the aim for these maps to be all-sky, we only use the Pristine-\Gaia\ synthetic catalogue and include stars with $\delta\FeH_\mathrm{phot}<0.3$\,dex, along with the different quality cuts described in subsection~\ref{sec:qualcuts}. We further add a restriction on the parallax to remove nearby stars ($\varpi+\delta\varpi<1''$). Changes in the shape of the MW\footnote{The absence of the Magellanic Clouds from these maps stems from the relative brightness of the Pristine-\Gaia\ synthetic catalogue. Being distant and very red, giant branch stars from the Clouds are effectively removed by the $\delta CaHK<0.1$ cut. The \citet{schlegel98} dust map is also not very reliable toward the central regions of the two satellite galaxies and, in this region, the de-reddened magnitudes should be used with caution.} with metallicity are directly evident, with the most metal-rich stars ($\FeH>-1.0$, map I) exhibiting a clear domination of the disk that transitions to the rounder distribution of the halo as more and more metal-poor samples are explored with $-1.5<\FeH<-1.0$ (map II), $-2.5<\FeH<-1.5$ (map III), and finally $-4.0<\FeH<-2.5$ (map IV).

Some of the features in those maps are however directly related to the \Gaia\ scanning law \citep{cantat-gaudin23} and it can be difficult to visualize the exact shape of the MW stars intersected by the complex volume probed by the catalogue. We therefore also show maps of density ratios in the bottom four panels of the figure. These have the advantage of exhibiting fewer artifacts and more clearly highlight the differences in the morphology of the MW in the different metallicity slices.

With the mean metallicity of the halo being $\FeH\sim-1.5$ \citep[\eg][]{ryan91,carney96,ivezic08,schork09,beers12,ibata17,youakim20}, the ratio of the two most metal-rich maps, II and I, clearly highlights the transition from the disk-dominated metallicity regime to the rounder halo-dominated regime. Interestingly, the III-to-II density ratio map implies that, overall, the density of stars with $-1.5<\FeH<-1.0$ and $-2.5<\FeH<-1.5$ are fairly similar, except in the anticenter direction that is clearly more metal-rich. This is possibly due to the numerous stellar features in this region, globally referred to as the Galactic Anticenter Stellar Structure \citep[GASS; \eg][]{newberg02,ibata03,slater14,morganson15} and now thought to be perturbed disk material sent to high latitudes \citep[\eg][]{bergemann18,laporte20,li21}.

Focussing on the low metallicity end of the halo with the IV/II map and, especially, the IV/III map (bottom row of the Figure) highlights that the relative density of the most metal-poor stars is higher toward the outer MW than toward the central Galactic regions (in the bottom-right panel of the figure, the directions toward the inner galaxy has lower density ratios that other direction). This is reminiscent of the findings of \citet{starkenburg17a} and \citet{el-badry18} that lines of sight toward the inner galaxy are privileged directions to find the oldest stars but that they suffer from the domineering presence of in situ stars formed as or after the halo of the MW assembled. Additional information, such as kinematics and detailed chemical abundances, are essential to sift through these stars to isolate samples of accreted stars that trace the assembly of the proto-MW \citep[\eg][]{arentsen20a,belokurov22,rix22}

More detailed information on the ``mono-abundance'' relative shape of the MW is not directly measurable from these maps that probe an irregular volume of the Galaxy around the Sun's position. However, the regular shape of the maps in Figure~\ref{fig:maps_metslices} implies that it should be straightforward to parametrize the shape of the MW as a function of metallicity. Provided, of course, that the very complex selection function imparted onto the data by the \Gaia\ scanning law, the choices made by the \Gaia\ consortium as to which star would be included in the catalogue of BP/RP coefficients and, of course, our own color and S/N cuts. This will be the focus of future work.

\section{Summary}
\label{sec:summary}

In this paper, we have taken advantage of the recently released \Gaia\ DR3 spectrophotometric BP/RP information to build a catalogue of synthetic $CaHK_\mathrm{syn}$ magnitudes mimicking the narrow-band photometry of the Pristine survey that, centered on the calcium H \& K lines, is sensitive to the metallicity of a star. We seize this opportunity to reprocess the full $\sim$11,500 images of the Pristine survey gathered since 2015 and present the current state of the survey that now covers more than 6,500\,deg$^2$. In particular, we use the \Gaia-based synthetic observations to refine the calibration of the Pristine photometry with a neural-net based algorithm, \texttt{PhotCalib}. We show that the new photometric catalogue is significantly flatter than before, better accounting for systematics across the field of view, and accurate at the 13\,mmag level (from a previously underestimated $\sim$40\,mmag with the previous calibration). The Pristine model that derives the photometric metallicity of a star from $CaHK$ magnitudes and broadband magnitudes is updated to rely solely on \Gaia\ broadband information ($G$, $G_\mathrm{BP}$, $G_\mathrm{RP}$). As part of this new model, we also present a more reliable way to iteratively deal with the extinction correction that is challenging with the \Gaia\ broadband magnitudes and now folds in the metallicity of a star. Finally, as photometrically variable sources are a significant source of spurious metallicities, we build a probabilistic variability model based on the photometric uncertainties of the \Gaia\ sources.

Pushing both the Pristine $CaHK$ magnitudes and the BP/RP-based synthetic $CaHK_\mathrm{syn}$ magnitudes through the Pristine model, we build and distribute two catalogues of photometric metallicities for reliable stars ($E(B-V)<0.5$ and $\delta CaHK<0.1$ or $\delta CaHK_\mathrm{syn}<0.1$): the Pristine-\Gaia\ synthetic catalogue and the Pristine DR1 catalogue for stars in common between Pristine and the BP/RP catalogue of \Gaia\ DR3. The latter serves as the first data release of the Pristine survey. We show that both catalogues can be used to build reliable samples of metal-poor stars and are particularly well suited to track VMP and EMP stars. The two catalogues are also complementary: the Pristine-\Gaia\ synthetic catalogue provides photometric metallicities over most of the sky but the Pristine DR1 catalogue (which is, by definition, limited to the Pristine footprint) contains significantly higher quality metallicities that go significantly fainter.

We conclude by showing test cases of how both catalogues of photometric metallicities can very efficiently determine the metallicity of MW GCs and make it easy to isolate extra-tidal cluster stars when the cluster's metallicity separates from the metallicity of MW field (more metal-rich) stars. Finally, we use the all-sky catalogue of Pristine-\Gaia\ synthetic photometric metallicities to display the change of morphology of the MW with metallicity, from a flat, disk-like distribution at high metallicity to a spheroidal halo as the metallicity decreases.

The release of the BP/RP information in the \Gaia\ DR3 catalogue truly opens a new age in the chemo-dynamical decomposition of the metal-poor MW, over the full MDF. And we can but look forward to the \Gaia DR4 data that will provide BP/RP information at higher S/N and for fainter stars.

\textbf{Author contributions: }NFM and ES are co-PIs of the Pristine survey, designed the project and led the writing of the paper. NFM handled the data reduction and ES built the metallicity model, the training set (with KV's help) and worked on the comparisons with external data sets. ZY took care of the calibration. MF coordinated the CaHKsyn calculations with FDA. AA-A developed the extinction calculations with HZ's help. AA-A, AE, and AV updated the code of the metallicity model to its public version. FG, MM, and SR developed the comparisons with globular clusters, carbon-rich stars, and UMP stars. RA, MB, and PM provided their \Gaia expertise. All other authors are active members of the Pristine collaboration and took part in the writing and editing of the paper.

\begin{acknowledgements}
We are deeply grateful to the CFHT staff for performing the Pristine observations in queue mode, for their reactivity in adapting the schedule, and for answering our questions during the data-reduction process. We are also particularly grateful to the \Gaia\ consortium for providing data of such wonderful quality and variety to the astronomy community.

NFM, ZY, and RAI gratefully acknowledge support from the French National Research Agency (ANR) funded project ``Pristine'' (ANR-18-CE31-0017) along with funding from the European Research Council (ERC) under the European Unions Horizon 2020 research and innovation programme (grant agreement No. 834148). ES, MB, and MM acknowledge funding through VIDI grant ``Pushing Galactic Archaeology to its limits'' (with project number VI.Vidi.193.093) which is funded by the Dutch Research Council (NWO). AA acknowledges support from the Herchel Smith Fellowship at the University of Cambridge and a Fitzwilliam College research fellowship supported by the Isaac Newton Trust.
M.B. and P.M. acknowledge the support to activities related to the ESA/\Gaia\ mission by the Italian Space Agency (ASI) through contract 2018-24-HH.0 and its addendum 2018-24-HH.1-2022 to the National Institute for Astrophysics (INAF). 
EFA acknowledge support from the Agencia Estatal de Investigaci\'on del Ministerio de Ciencia e Innovaci\'on (AEI-MCINN) under grant ``At the forefront of Galactic Archaeology: evolution of the luminous and dark matter components of the Milky Way and Local Group dwarf galaxies in the \Gaia\ era'' with reference PID2020-118778GB-I00/10.13039/501100011033. EFA also acknowleges support from the `Mar\'ia Zambrano' fellowship from the Universidad de La Laguna. JIGH and CAP acknowledge financial support from the Spanish Ministry of Science and Innovation (MICINN) project PID2020-117493GB-I00. We benefited from the International Space Science Institute (ISSI) in Bern, CH, thanks to the funding of the team ``Pristine''. This research was supported by the International Space Science Institute (ISSI) in Bern, through ISSI International Team project 540 (The Early Milky Way). DA also acknowledges financial support from the Spanish Ministry of Science and Innovation (MICINN) under the 2021 Ram\'on y Cajal program MICINN RYC2021-032609. G.B. acknowledges support from the Agencia Estatal de Investigaci\'on del Ministerio de Ciencia en Innovaci\'on (AEI-MICIN) and the European Regional Development Fund (ERDF) under grant number AYA2017-89076-P, the AEI under grant number CEX2019-000920-S and the AEI-MICIN under grant number PID2020-118778GB-I00/10.13039/501100011033. 

Based on observations obtained with MegaPrime/MegaCam, a joint project of CFHT and CEA/DAPNIA, at the Canada-France-Hawaii Telescope (CFHT) which is operated by the National Research Council (NRC) of Canada, the Institut National des Sciences de l'Univers of the Centre National de la Recherche Scientifique of France, and the University of Hawaii.

This work has made use of data from the European Space Agency (ESA) mission \Gaia\ (\url{https://www.cosmos.esa.int/gaia}), processed by the \Gaia\ Data Processing and Analysis Consortium (DPAC, \url{https://www.cosmos.esa.int/web/gaia/dpac/consortium}). Funding for the DPAC has been provided by national institutions, in particular the institutions participating in the \Gaia\ Multilateral Agreement. 

Funding for the Sloan Digital Sky Survey IV has been provided by the Alfred P. Sloan Foundation, the U.S. Department of Energy Office of Science, and the Participating Institutions. SDSS-IV acknowledges support and resources from the Center for High Performance Computing  at the University of Utah. The SDSS website is www.sdss.org. SDSS-IV is managed by the Astrophysical Research Consortium for the Participating Institutions of the SDSS Collaboration including the Brazilian Participation Group, the Carnegie Institution for Science, Carnegie Mellon University, Center for Astrophysics | Harvard \& Smithsonian, the Chilean Participation Group, the French Participation Group, Instituto de Astrof\'isica de Canarias, The Johns Hopkins University, Kavli Institute for the Physics and Mathematics of the Universe (IPMU) / University of Tokyo, the Korean Participation Group, Lawrence Berkeley National Laboratory, Leibniz Institut f\"ur Astrophysik Potsdam (AIP),  Max-Planck-Institut f\"ur Astronomie (MPIA Heidelberg), Max-Planck-Institut f\"ur Astrophysik (MPA Garching), Max-Planck-Institut f\"ur Extraterrestrische Physik (MPE), National Astronomical Observatories of China, New Mexico State University, New York University, University of Notre Dame, Observat\'ario Nacional / MCTI, The Ohio State University, Pennsylvania State University, Shanghai Astronomical Observatory, United Kingdom Participation Group, Universidad Nacional Aut\'onoma de M\'exico, University of Arizona, University of Colorado Boulder, University of Oxford, University of Portsmouth, University of Utah, University of Virginia, University of Washington, University of Wisconsin, Vanderbilt University, and Yale University.

Guoshoujing Telescope (the Large Sky Area Multi-Object Fiber Spectroscopic Telescope LAMOST) is a National Major Scientific Project built by the Chinese Academy of Sciences. Funding for the project has been provided by the National Development and Reform Commission. LAMOST is operated and managed by the National Astronomical Observatories, Chinese Academy of Sciences.

\end{acknowledgements}

%
%

\bibliographystyle{aa}



\begin{appendix}

%
%
%

\section{Extinction coefficients}
Tables~\ref{tab:extinction_giant} and~\ref{tab:extinction_dwarf} list the coefficients of the polynomial relations used to fit the synthetic photometry from \texttt{dustapprox}, as described in section~\ref{sec:extinction} and equation~\ref{eq:kf}.

\begin{table}[h]
\begin{center}
\fontsize{7}{7}\selectfont
\caption{Polynomial coefficients for $k_\mathrm{f}$ (Equation~\ref{eq:kf}) for giants. Only coefficients larger than 0.001 are listed. \label{tab:extinction_giant}}
\begin{tabular}{l l c c c c c c c c c|} 
 &  factor    & $G$ & $G_{BP}$ & $G_{RP}$ & $CaHK$ & $A_0/A_V$ \\ [0.5ex] 
 \hline
  $a_{0}$ &   1 & $-$0.0897 &  0.3441 &  0.4009 &  1.5409 &  1.1753 \\
  $a_{1}$ &   $T$ &  1.9108 &  1.4676 &  0.5364 &  0.0377 & $-$0.5199 \\
  $a_{2}$ &   $A_0$ &  0.0307 &  0.0302 & $-$0.0059 &      &  0.0045 \\
  $a_{3}$ &   \FeH &  0.0113 &  0.0128 & $-$0.0027 & $-$0.0023 &  0.0059 \\
  $a_{4}$ &   $T$$^2$ & $-$1.2696 & $-$0.9191 & $-$0.3920 & $-$0.0252 &  0.3903 \\
  $a_{6}$ &   $\FeH^2$ & $-$0.0077 & $-$0.0087 &      &      &  0.0020 \\
  $a_{7}$ &   $T$ $A_0$ & $-$0.0921 & $-$0.0778 & $-$0.0019 &      &      \\
  $a_{8}$ &   $T$ \FeH & $-$0.0547 & $-$0.0728 &  0.0042 &  0.0028 & $-$0.0012 \\
  $a_{9}$ &   $A_0$ \FeH &  0.0023 &  0.0025 &      &      &      \\
 $a_{10}$ &  $T$$^3$ &  0.3183 &  0.2122 &  0.1005 &  0.0060 & $-$0.1017 \\
 $a_{13}$ &  $T$$^2$ $A_0$ &  0.0284 &  0.0274 &      &      &      \\
 $a_{15}$ &  $T$$^2$ \FeH &  0.0300 &  0.0405 & $-$0.0011 &      & $-$0.0017 \\
 $a_{16}$ &  $T$ \FeH$^2$ &  0.0035 &  0.0038 &      &      & $-$0.0011 \\
  \hline
 $\sigma$ &   &  0.0017 &  0.0019 &  0.0005 &  0.0001 &  0.0005 \\
 \hline
\end{tabular}
\vspace{0.05cm}
\end{center}
\end{table}

\begin{table}[h]
\begin{center}
\fontsize{7}{7}\selectfont
\caption{Polynomial coefficients for $k_\mathrm{f}$ (Equation~\ref{eq:kf}) for dwarfs. Only coefficients larger than 0.001 are listed. \label{tab:extinction_dwarf}}
\begin{tabular}{l l c c c c c c c c c|} 
 &  factor    & $G$ & $G_{BP}$ & $G_{RP}$ & $CaHK$ & $A_0/A_V$ \\ [0.5ex] 
 \hline
  $a_{0}$ &  1 &  0.2554 &  0.6765 &  0.4620 &  1.5524 &  1.0985 \\
  $a_{1}$ &  $T$ &  0.9525 &  0.6622 &  0.3391 &  0.0091 & $-$0.2832 \\
  $a_{2}$ &  $A_0$ &  0.0113 &  0.0062 & $-$0.0061 &      &  0.0042 \\
  $a_{3}$ &  \FeH & $-$0.0312 &  0.0107 & $-$0.0214 &      &  0.0076 \\
  $a_{4}$ &  $T$$^2$ & $-$0.4075 & $-$0.2720 & $-$0.1963 & $-$0.0022 &  0.1637 \\
  $a_{5}$ &  $A_0$$^2$ &  0.0017 &  0.0011 &      &      &      \\
  $a_{6}$ &  $\FeH^2$ & $-$0.0065 & $-$0.0027 & $-$0.0021 &      &      \\
  $a_{7}$ &  $T$ $A_0$ & $-$0.0601 & $-$0.0381 & $-$0.0013 &      &      \\
  $a_{8}$ &  $T$ \FeH &  0.0195 & $-$0.0475 &  0.0305 &      & $-$0.0071 \\
  $a_{9}$ &  $A_0$ \FeH &  0.0024 &  0.0019 &      &      &      \\
 $a_{10}$ & $T$$^3$ &  0.0657 &  0.0393 &  0.0395 &      & $-$0.0328 \\
 $a_{13}$ & $T$$^2$ $A_0$ &  0.0158 &  0.0113 &      &      &      \\
 $a_{15}$ & $T$$^2$ \FeH &      &  0.0229 & $-$0.0100 &      &  0.0016 \\
 $a_{16}$ & $T$ \FeH$^2$ &  0.0026 &      &  0.0015 &      &      \\
  \hline
 $\sigma$ &   &  0.0022 &  0.0019 &  0.0012 &  0.0001 &  0.0008 \\
 \hline
\end{tabular}
\vspace{0.05cm}
\end{center}
\end{table}

\section{Comparison of photometric metallicities with the training set for dwarf stars}
\begin{figure}
\begin{center}
\includegraphics[width=\hsize]{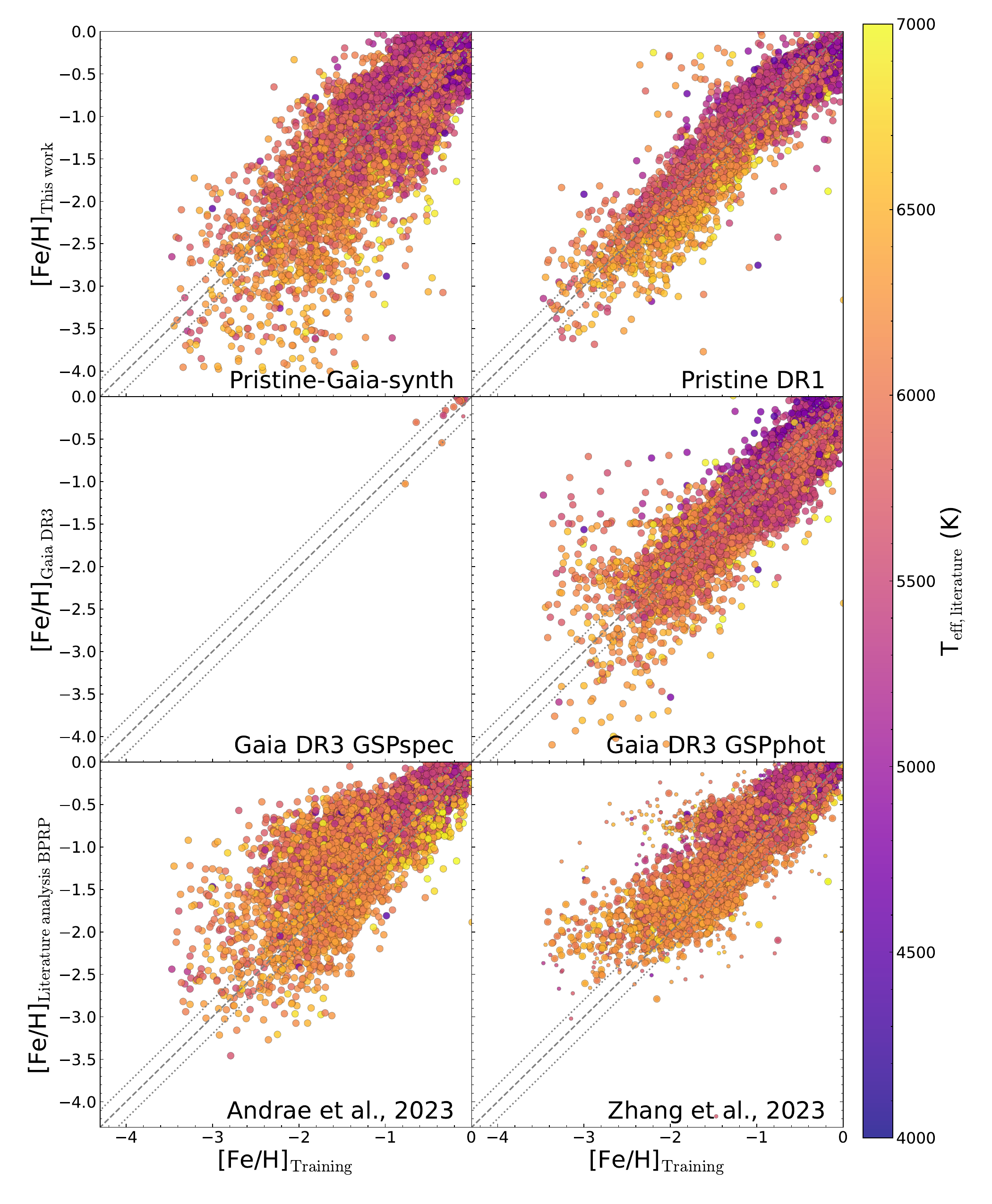}
\caption{Comparison of various metallicity catalogues based on \Gaia\ DR3 with the stars in the training sample with $\logg > 3.9$ corresponding to the main-sequence stars. After the comparison with the Pristine-\Gaia\ synthetic catalogue, and the Pristine DR1 catalogue, the panels show the comparisons for the \Gaia\ GSP-Spec metallicities based on the \Gaia\ RVS spectra \citep[][smaller circles correspond to stars flagged as unreliable in this catalogue, following the same criteria as in their Figure 26]{recio-blanco23}, the \Gaia\ GSP-Phot catalogue based on the BP/RP information \citep{andrae22}, the XGBoost algorithm that uses the BP/RP information and some additional non-\Gaia\ photometry \citep{andrae23}, the BP/RP-based metallicities from \citet[][smaller circles are flagged results]{zhang23}. In all panels, the color encodes the temperature of the star from the training sample. The dashed line corresponds to the 1-to-1 line and the dotted lines represent offsets of 0.2\,dex. \label{fig:dwarfcomp}}
\end{center}
\end{figure}

\begin{figure}
\begin{center}
\includegraphics[width=\hsize]{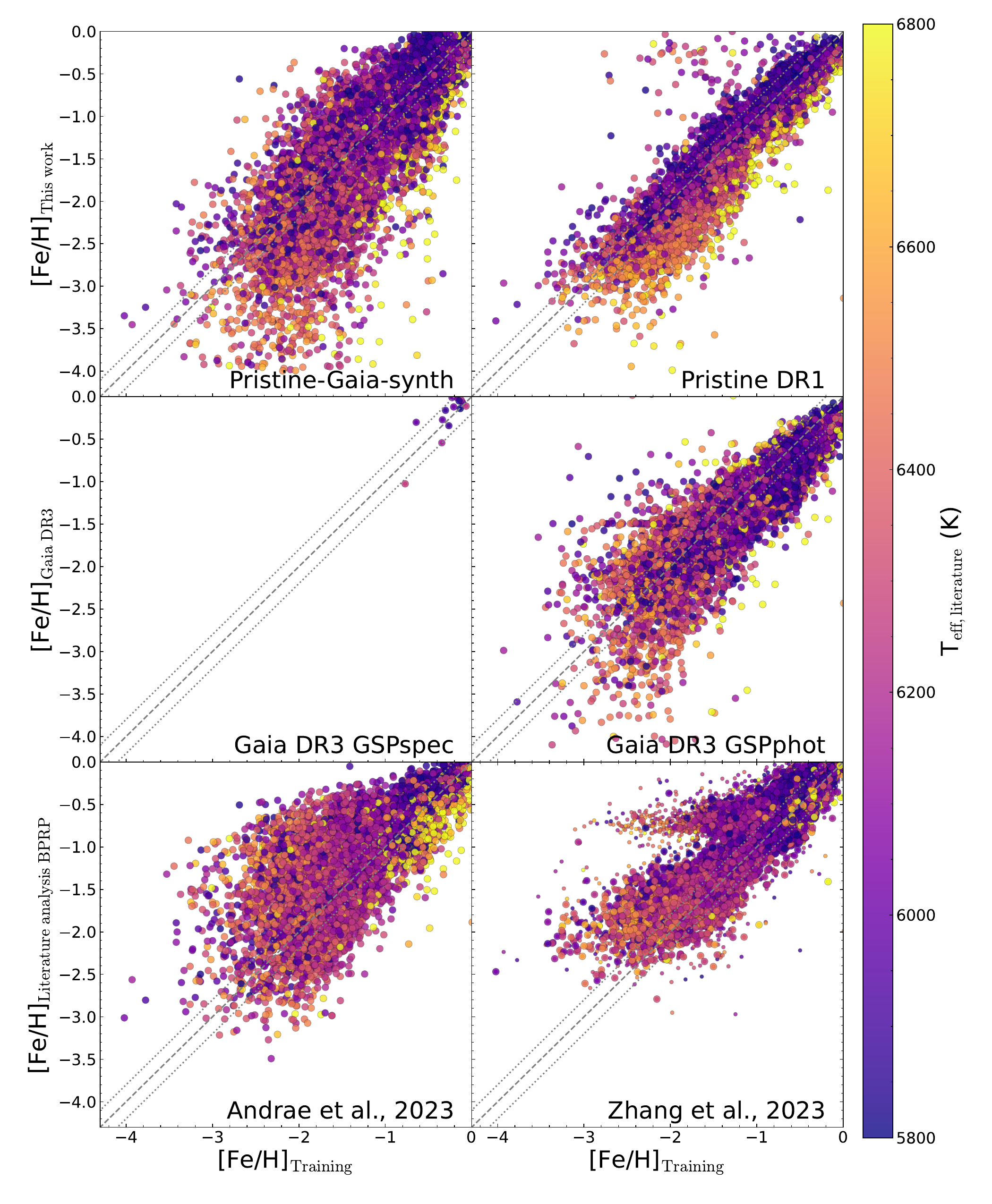}
\caption{Similar to Figure \ref{fig:dwarfcomp}, but for training sample turn-off stars only (defined as $\teff > 5800$ K, regardless of their $\logg$). A clear trend with temperature can be seen in the Pristine-DR1 catalogue, even in this relatively small temperature range. Our photometric metallicities estimates degrade for hotter stars. \label{fig:dwarfcomp2}}
\end{center}
\end{figure}

Figures~\ref{fig:dwarfcomp} and~\ref{fig:dwarfcomp2} are the same as Figure~\ref{fig:comparison_other_catalogues}, but for the dwarf and main-sequence turnoff stars of the training sample, respectively, and color-coded by the temperature of a star. This comparison with the training sample shows somewhat poorer quality than for the giant stars but the degrading performance stems mainly from the main-sequence turnoff stars. This is understandable as these fairly hot stars have narrow spectral Ca H \& K features that translate into a loss of metallicity resolution for a fixed photometric uncertainty. As shown in Figure~\ref{fig_colorspace}, a fixed difference in $\FeH_\mathrm{phot}$ is significantly harder to measure as it corresponds to a smaller $CaHK$ difference for these blue stars than it is for a redder star. It is interesting to note on the other hand that the \citet{andrae22} GSP-Phot values have less catastrophic outliers when compared to our dwarf training sample than for the giants, possibly because of the additional information brought by the parallax in the GSP-Phot analysis. 

\section{Comparison of the variability model with the \Gaia\ variability classes}\label{app:dwarfs}
\begin{figure*}
\begin{center}
\includegraphics[width=\hsize]{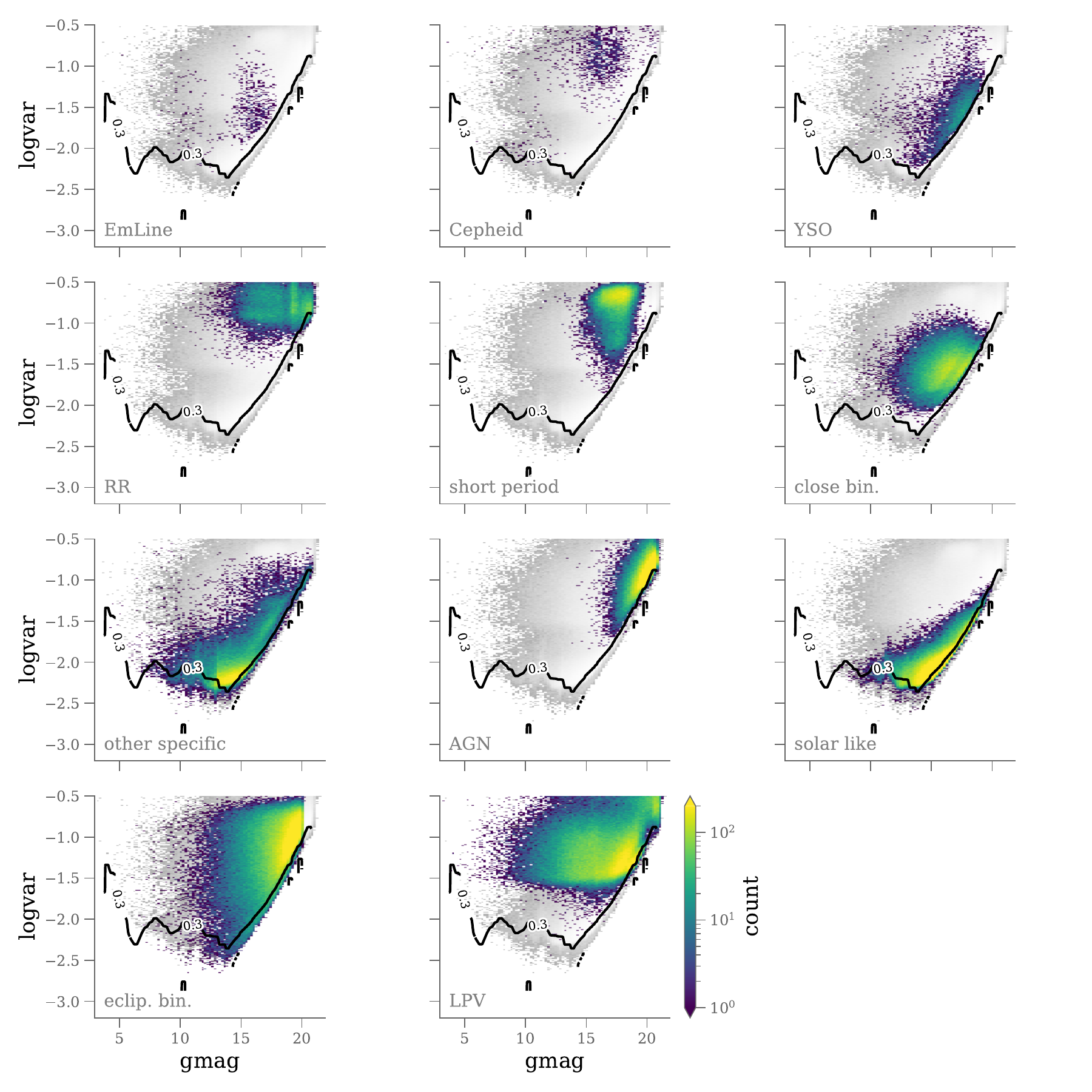}
\caption{Distribution of the main \Gaia\ DR3 variable classes, as defined by \citet{gavras23}, in the variability space of Figure~\ref{fig_gvar_vs_gmag} used to define $P_\mathrm{var}$. All panels show the distribution of the variables identified in \Gaia\ DR3 (\texttt{phot\_variable\_flag = 'VARIABLE'}) in gray and the density distribution of a given class of variables, as labeled in each panel. We only show classes with at least 5,000 identified sources and we overlay our $P_\mathrm{var}=0.3$ threshold for comparison. The proposed cut effectively rejects the most obvious classes of variables.\label{fig:varclass}}
\end{center}
\end{figure*}

Figure~\ref{fig:varclass} presents the distribution of the different classes of variable sources in \Gaia\ \citet{gavras23} in the variability space of Figure~\ref{fig_gvar_vs_gmag}. The proposed quality cut that keeps stars with $P_\mathrm{var}<0.3$ does indeed remove most of these identified variable stars, irrespective of the variable class.

\end{appendix}

\end{document}